\documentclass[aps,twocolumn,showpacs,floatfix,amssymb]{revtex4}
\usepackage{graphicx}
\begin{document}

\title{Spin fluctuations and high-temperature superconductivity
 in cuprates }

\author{ Nikolay M. Plakida$^{a,b}$ }
\affiliation{ $^a$Joint Institute for Nuclear Research, 141980
Dubna, Russia\\
 $^{b}$Max-Planck-Institut f\"{u}r Physik komplexer Systeme,
D-01187, Dresden,  Germany}

\date{\today}

\begin{abstract}
To describe the cuprate superconductors, models of strongly correlated electronic systems, such as the Hubbard or $t$-$J$ models, are commonly employed. To study   these models,  projected (Hubbard)  operators have to be used.  Due to the unconventional commutation relations for the Hubbard operators, a specific kinematical interaction of electrons with spin and charge fluctuations emerges.
The interaction is induced by the intraband hopping with a coupling parameter of the order of the kinetic energy of electrons $\, W $  which is much larger than the antiferromagnetic exchange interaction $J$ induced by the interband hopping.
This  review  presents a consistent microscopic theory of spin excitations and superconductivity for cuprates where these  interactions are taken into account  within the Hubbard operator technique. The low-energy spin excitations are considered for the $t$--$J$ model, while the electronic properties are studied using the two-subband extended Hubbard model where the intersite Coulomb repulsion $V$ and electron-phonon interaction are taken into account.
\end{abstract}


\maketitle

%To be published in Physica C

%------------------------------------
\section{Introduction}
\label{sec:1}

Discovery of high-temperature superconductivity (HTSC) in cuprates by Bednorz and
M\"uller~\cite{Bednorz86} caused an unprecedented  scientific activity in
study of this extraordinary phenomenon (see, e.g., Refs.~\cite{Schrieffer07,Plakida10}). In recent years intensive experimental investigations have presented detailed information concerning  unconventional physical properties of cuprates. However, theoretical studies of  various microscopical models have not yet brought about a commonly accepted theory of superconductivity in cuprates. Two most frequently discussed mechanisms of HTSC    are  the conventional electron-phonon pairing (see, e.g., Refs.~\cite{Kulic04,Maksimov10}) and  the spin-fluctuation mediated pairing~\cite{Scalapino95,Scalapino12} in the framework of phenomenological spin-fermion models (see, e.g., Refs.~\cite{Monthoux94b,Moriya00,Chubukov04,Abanov08}). A weak intensity of spin fluctuations at optimal doping found in the magnetic inelastic neutron scattering experiments~\cite{Bourges98} is the main argument against the spin-fluctuation pairing mechanism. However,  recent magnetic resonant inelastic x-ray scattering (RIXS) experiments have revealed that paramagnon antiferromagnetic (AF) excitations persist with a similar dispersion and comparable intensity in a large family of cuprate superconductors, even in the overdoped region (see, e.g., Refs.~\cite{LeTacon11,LeTacon13,Dean13,Dean14,Minola15}). These experiments
prove that spin fluctuations have sufficient strength to mediate HTSC in cuprates and to explain various physical properties of cuprate materials such as, the ``kink'' phenomenon observed in the electronic spectrum by  the angle-resolved
photoemission spectroscopy~\cite{Kordyuk10}, the optical  and dc
conductivity~\cite{Vladimirov12}, etc.   Therefore, we can assume  that the spin-fluctuation mediated pairing plays the major role while the electron-phonon pairing is less important. This assumption is supported by observation of the weak isotope effect in optimally doped samples (see reviews ~\cite{Zhao01,Khasanov04,Muller07}).
\par
The main problem in a theoretical study of the cuprate superconductors is that strong electron correlations there  precludes from application of the conventional Fermi-liquid approach in  description of their  electronic structure~\cite{Fulde95}. They are Mott-Hubbard  (more accurately, charge-transfer) insulators  where the conduction band splits into two Hubbard subbands where the projected (Hubbard) electronic operators should be used. Recently we have developed  a microscopic theory of spin excitations and superconductivity for cuprates within the Hubbard operator (HO) technique ~\cite{Plakida10,Vladimirov09,Vladimirov11,Plakida06,Plakida07,Plakida13,Plakida14}.
We demonstrate an important role of the  kinematical interaction for the  HOs induced by the non-Fermy commutation relations for electronic operators, both for studying the spin excitations and electron spectrum. In this review we present the results of these investigations.

First, in Sec.~\ref{sec:2},  we explain how the kinematical
interaction  appears  in the  Hubbard model in the limit of
strong electron correlations.  Then in Sec.~\ref{sec:3} we
formulate a theory of spin excitations  within the
relaxation-function approach  for calculation of the dynamical
spin susceptibility (DSS) in the normal and superconducting states
~\cite{Vladimirov09,Vladimirov11}. The spin dynamics at
arbitrary frequencies and wave vectors is studied for various
temperatures and  hole doping. In the superconducting state, the DSS reveals the magnetic resonance mode  at the AF wave vector ${\bf Q} = \pi(1,1)$ at low
doping. We show that it can be  observed even above superconducting  $T_c$ due to its weak damping.
This is explained by an involvement of spin excitations in the decay
process besides the particle-hole continuum usually considered in the
random-phase-type approximation where the resonance mode can appear
only below $T_c$.

In the second part of the paper, in Secs.~\ref{sec:4}~and~\ref{sec:5}, a theory of the electronic spectrum and superconductivity is presented within the Mori-type projection technique for the single-particle electronic Green functions (GFs)
\cite{Plakida07,Plakida13,Plakida14}. We consider the extended Hubbard model where the intersite Coulomb repulsion and the electron-phonon interaction are taken into account.  An exact  Dyson equation for the normal and
anomalous (pair) GFs  is derived which is solved in the
self-consistent Born approximation (SCBA) for the self-energy. We found the $d$-wave
pairing with high-$T_c$ mediated by the kinematical spin-fluctuation interaction which can be suppressed only for a large intersite Coulomb repulsion  $\,V > W$.
Isotope effect  on $T_c$ induced by electron-phonon interaction is weak at optimal doping and increases at low doping. We emphasize that the kinematical interaction is absent in the spin-fermion models and is lost in the slave-boson (-fermion) models treated in the mean-field approximation (MFA). In conclusion we summarize our results.

\section{Kinematical interaction in the Hubbard model}
\label{sec:2}

We consider the extended  Hubbard model~\cite{Hubbard63}  on a square lattice
\begin{eqnarray}
H &= &\sum_{i \neq j, \sigma} \, t_{ij} \, a_{i\sigma}^{\dag}
a_{j\sigma} - \mu \, \sum_{i} N_{i}
\nonumber \\
& + & (U/2) \,\sum_{i} N_{i \sigma}  N_{i \bar\sigma} +
(1/2)\,\sum_{ i \neq j} V_{ij}\, N_{i} N_{j},
 \label{1}
\end{eqnarray}
where $t_{i,j}$ is the single-electron hopping parameter, $a^{\dag}_{i\sigma}$ and
$a_{i\sigma}$ are the Fermi creation and annihilation  operators for  electrons with spin $\sigma/2 \;(\sigma = \pm 1), \, \bar\sigma = -\sigma) $ on the
lattice site $i$, $U$ is the on-site Coulomb interaction (CI) and the $V_{ij}$ is the
intersite CI. $ \, N_{i} =\sum_{\sigma} N_{i\sigma}, \,
N_{i\sigma}=a^{\dag}_{i\sigma}a_{i\sigma} $ is the number operator and $\mu$ is the
chemical potential.
\par
In the strong correlation limit, $U\gg |t_{i,j}|$, the  Fermi operators $a^\dag_{i\sigma}, \,
a_{i\sigma}$ in (\ref{1}) fail to describe single-particle electron excitations in the system and  the HOs~\cite{Hubbard65}, referring to the singly occupied  and doubly occupied   subbands, should be used: $a^\dag_{i\sigma} =  a^\dag_{i\sigma}(1- N_{i \bar\sigma})  + a^\dag_{i\sigma} N_{i \bar\sigma} \equiv X_{i}^{\sigma
0}  +  \sigma \, X_{i}^{2\bar\sigma}$. In terms of the HOs the model
(\ref{1}) reads
\begin{eqnarray}
 H &= & \varepsilon_1\sum_{i,\sigma}X_{i}^{\sigma \sigma}
  + \varepsilon_2\sum_{i}X_{i}^{22}
+  \frac{1}{2} \sum_{i\neq j}\,V_{ij} N_i N_j
\nonumber \\
& + &\sum_{i\neq j,\sigma}\, t_{ij}\,\bigl\{ X_{i}^{\sigma
0}X_{j}^{0\sigma}
  + X_{i}^{2 \sigma}X_{j}^{\sigma 2}
 \nonumber \\
& + & \sigma \,(X_{i}^{2\bar\sigma}X_{j}^{0 \sigma} + {\rm
H.c.})\bigr\},
 \label{2}
\end{eqnarray}
where $\varepsilon_1 = - \mu$ is the single-particle energy  and $\varepsilon_2 =  U - 2
\mu $ is the two-particle energy. The matrix HOs $X_{i}^{\alpha\beta} =
|i\alpha\rangle\langle i\beta|$ describes  transition from the state $|i,\beta\rangle$ to
the state $|i,\alpha\rangle$ on a lattice site $i$ taking into account four  possible
states for holes: an empty state $(\alpha, \beta =0) $, a singly occupied hole state
$(\alpha, \beta = \sigma)$, and a doubly occupied hole state $(\alpha, \beta = 2) $. The number operator and the spin operators in terms of the HOs are defined as
\begin{eqnarray}
  N_i &=& \sum_{\sigma} X_{i}^{\sigma \sigma} + 2 X_{i}^{22},
\label{3}\\
S_{i}^{\sigma} & = & X_{i}^{\sigma\bar\sigma} ,\quad
 S_{i}^{z} =  (\sigma/2) \,[ X_{i}^{\sigma \sigma}  -
  X_{i}^{\bar\sigma \bar\sigma}] .
\label{4}
\end{eqnarray}
To compare our results with cuprates, we consider the hole-doped case where the chemical potential $\mu$ is determined by the equation for the average hole  occupation number
\begin{equation}
  n =  1 + \delta = \langle \, N_i \rangle .
    \label{3a}
\end{equation}
Here  $\langle \ldots \rangle$ denotes the statistical average
with the Hamiltonian (\ref{2}).
\par
From the multiplication rule for the HOs, $\, X_{i}^{\alpha\beta} X_{i}^{\gamma\delta} =
\delta_{\beta\gamma} X_{i}^{\alpha\delta}$, it follows their commutation relations
\begin{equation}
\left[X_{i}^{\alpha\beta}, X_{j}^{\gamma\delta}\right]_{\pm}=
\delta_{ij}\left(\delta_{\beta\gamma}X_{i}^{\alpha\delta}\pm
\delta_{\delta\alpha}X_{i}^{\gamma\beta}\right)\, ,
 \label{5}
\end{equation}
with the upper  sign  for the Fermi-type operators (such as
$X_{i}^{0\sigma}$) and the lower sign for the Bose-type operators
(such as  the number  or  spin operators). The HOs obey the
completeness relation
\begin{equation}
 X_{i}^{00} +
 \sum_{\sigma} X_{i}^{\sigma\sigma}  + X_{i}^{22} = 1,
 \label{5a}
\end{equation}
which rigorously preserves the local constraint  that only one quantum
state $\alpha$ can be occupied on any lattice site $i$.

The unconventional commutation relations (\ref{5}) for HOs result
in the  kinematical interaction. The term was introduced by Dyson
in a general theory of spin-wave interactions ~\cite{Dyson56}. It
was pointed out that the spin-wave creation $\, b_i^{\dag}=
S_{i}^{-} $ and annihilation   $\, b_i =S_{i}^{+} \,$ operators
obey the commutation relations
\begin{equation}
b_i b_j^{\dag} -  b_j^{\dag} b_i =
  \delta_{i,j}\,(1- 2\, b_i^{\dag} b_i ),
 \label{6}
\end{equation}
which show that they are Bose operators on different lattice sites
and Fermi operators on the same lattice site. The last term $\,
 b_i^{\dag} b_i\,$  in Eq.~(\ref{6}) just brings about   the
kinematical interaction which may be small for low density of spin
excitations, $\, b_i^{\dag} b_i  \ll 1\,$.

Similar commutation relations  can be written for the electron
creation $X_{i}^{\sigma 0}\, $ and annihilation $
X_{j}^{0\sigma}\,$ operators,
\begin{eqnarray}
&& X_{i}^{0\sigma} X_{j}^{\sigma 0}+ X_{j}^{\sigma
0}X_{i}^{0\sigma}
 = \delta_{i,j}(1 - X_{i}^{\bar\sigma \bar\sigma} - X_i^{22})
 \nonumber \\
 && = \delta_{ij}(1 - N_{i \sigma} /2 + \sigma S_i^z),
 \label{7}
\end{eqnarray}
where the relation (\ref{5a}) was used. Therefore, the HOs for electrons  in the singly occupied subband can be
considered as Fermi operators on different lattice sites but on
the same lattice site electron scattering on charge  $N_{i
\sigma}$ and spin $S_i^\alpha$ fluctuations occurs which
determines the kinematical interaction. This results in dressing
of the electron hopping by spin and charge fluctuations
with the coupling constant of the order of the hopping parameter
$\, t$.

\section{Spin excitation spectrum}
\label{sec:3}

To study low-energy excitations such as spin excitations in the limit of strong
correlations, $U \gg t$, we can consider only one subband, e.g.,  the singly occupied
subband for electrons taking into account virtual hopping to the second  subband by the
exchange interaction $J_{ij} = 4\, t_{ij}^2/ U $. This results in the $t$-$J$ model which
in terms of the HOs  reads
\begin{eqnarray}
H &=& - \sum_{i \neq j,\sigma}t_{ij}X_{i}^{\sigma 0}X_{j}^{0\sigma}
 - \mu \sum_{i \sigma} X_{i}^{\sigma \sigma}
\nonumber \\
 &  + &\frac{1}{4} \sum_{i \neq j,\sigma} J_{ij}
\left(X_i^{\sigma\bar{\sigma}}X_j^{\bar{\sigma}\sigma}  -
   X_i^{\sigma\sigma}X_j^{\bar{\sigma}\bar{\sigma}}\right) .
\label{b1}
\end{eqnarray}
In the $t$-$J$ model the doubly-occupied states are neglected and therefore the number
operator (\ref{3}) and the completeness relation (\ref{5a})  take the forms:
$\,n_{i}=\sum_{\sigma} \,X_{i}^{\sigma\sigma} $ and  $ X_{i}^{00} + \sum_{\sigma}
X_{i}^{\sigma\sigma} = 1$, respectively. The chemical potential $\mu$ is  determined from
the equation for the average electron density
\begin{equation}
n = \sum_{ \sigma} \langle X_{i}^{\sigma \sigma} \rangle =  1 - \delta,
 \label{b1a}
\end{equation}
where $ \delta = \langle X_{i}^{00} \rangle$ is the hole concentration and $\langle \ldots
\rangle$ denotes the statistical average with the Hamiltonian (\ref{b1}).
\par
Various approaches have been used to study the spin dynamics within  the $t$–-$J$ model
(\ref{b1})  (for a review see, e.g., Refs.~\cite{Izyumov97} and ~\cite{Plakida06}). In
particular, in the slave boson or fermion methods,  a local constraint prohibiting a
double occupancy of any quantum state is difficult to treat rigorously. An application of
a  special diagram technique for HOs in the $t$--$J$ model results in a complicated
analytical expression for the DSS~\cite{Izyumov90}. Studies of finite clusters by
numerical methods were important in elucidating static and dynamic spin interactions,
though they have limited energy and momentum resolutions (see, e.g.,
Refs.~\cite{Dagotto94,Jaklic00,Eder95}).
\par
To overcome this complexity, we apply the projection Mori-type technique elaborated for
the two-time thermodynamic GFs~\cite{Zubarev60,Plakida73,Tserkovnikov81,Plakida11}. In
this method an exact representation for the self-energy (or polarization operator) can be
derived which, when evaluated in the mode-coupling approximation (MCA), yields physically
reasonable results even for strongly interacting systems.

We use  the spin-rotation-invariant relaxation-function theory for the DSS to calculate
the static properties in the generalized mean-field approximation (GMFA) similarly to
Ref.~\cite{Winterfeldt98} and the dynamic spin-fluctuation spectra using the MCA  for the
force-force correlation functions. Thereby, we capture both the local and itinerant
character of charge carriers in a consistent way. In calculating the static properties,
in particular the static susceptibility and spin-excitation spectrum, we pay  particular
attention to a proper description of AF short-range order (SRO) and its implications on
the spin dynamics. For the undoped case described by the Heisenberg model our results are
similar to those in Refs.~\cite{Winterfeldt97} and ~\cite{Winterfeldt99}. For a finite
doping our theory yields a reasonable agreement with available exact diagonalization (ED)
data and neutron scattering experiments.

A similar approach based on the Mori projection technique for the single-particle
electron GF and spin GF has been used in
Refs.~\cite{Sega03,Prelovsek04,Sherman03,Sherman04,Sherman06}. The magnetic resonance
mode observed in the superconducting state was studied within the memory-function
approach in Refs.~\cite{Sherman03a,Sega03,Sega06,Prelovsek06}.

First we present the results for the DSS in the normal state obtained in
Ref.~\cite{Vladimirov09} and then we consider the theory of the magnetic resonance mode
in the superconducting state developed in Ref.~\cite{Vladimirov11}.

\subsection{Relaxation-function theory}
\label{sec:3a}

In Ref.~\cite{Vladimirov05},  applying the Mori-type projection technique~\cite{Mori65},
elaborated for the relaxation function, we have derived an exact representation  for the
DSS $\, \chi({\bf q}, \omega) \,$ related to the retarded commutator GF~\cite{Zubarev60},
\begin{equation}
\chi({\bf q}, \omega) = -\langle \!\langle {S}_{\bf q}^{+}| {S}_{-\bf q}^{-} \rangle
\!\rangle_{\omega} =  \frac{m({\bf q})} { \omega_{\bf q}^2 +\omega \, \Sigma({\bf
q},\omega) - \omega^2 } ,
 \label{b2}
\end{equation}
where $m({\bf q})=\langle [i\dot{S}^{+}_{\bf q}, S_{-\bf q}^{-}]\rangle  = \langle [\,
[{S}^{+}_{\bf q}, H], \, S_{-\bf q}^{-}]\rangle$, and $\omega_{\bf q}$ is the
spin-excitation spectrum in the GMFA. The self-energy is given by the many-particle
Kubo-Mori relaxation function
\begin{equation}
\Sigma({\bf q},\omega)=[1/m({\bf q})]\, (\!( - \ddot{S}_{\bf q}^{+}\,| - \ddot{S}_{-\bf
q}^{-})\!)_{\omega}^{(\rm pp)}\, ,
 \label{b3}
\end{equation}
where    $- \ddot{S}_{\bf q}^{\pm} = [\, [{S}_{\bf q}^{\pm},H], \,H] $ (for details see
Ref.~\cite{Vladimirov05}). The Kubo-Mori relaxation function and the scalar product are
defined as (see, e.g.,~Ref.~\cite{Tserkovnikov81})
\begin{equation}
(( A | B ))_{\omega}= - i \int_{0}^{\infty} dt  e^{i\omega t}
  (A(t), B),
   \label{b2a}
\end{equation}
and
\begin{equation}
 (A(t), B) = \int_{0}^{\beta}d\lambda
   \langle A(t-i\lambda) B \rangle,
  \quad \beta = 1/k_{\rm B} T ,
\label{b2ab}
\end{equation}
respectively.  The ``proper part''  {\rm(pp)} of the relaxation function (\ref{b3}) does
not contain parts connected by a single zero-order relaxation function  which corresponds
to the projected time evolution in the original Mori projection technique~\cite{Mori65}.
The spin-excitation spectrum is given by the spectral function defined by the imaginary
part of the DSS (\ref{b2}),
\begin{eqnarray}
\chi''({\bf q}, \omega)  =  \frac{- \omega \, \Sigma{''}({\bf q},\omega)\; m({\bf q})}
{[\omega^2 -  \omega_{\bf q}^2 - \omega \, \Sigma{'}({\bf q},\omega)]^2  + [\omega \,
\Sigma{''}({\bf q},\omega)]^2} \, ,
 \label{b4}
\end{eqnarray}
where $\, \Sigma({\bf q},\omega +i0^+)= \Sigma{'}({\bf q},\omega) + i \Sigma{''}({\bf
q},\omega)$  and $\,\Sigma{'}({\bf q},\omega) = - \Sigma{'}({\bf q}, -\omega)$ is  the
real and $ \Sigma{''}({\bf q},\omega) =  \Sigma{''}({\bf q}, - \omega) < 0$ is the
imaginary parts of the self-energy.\\

{\bf Static properties.} The general representation of the DSS~(\ref {b2}) determines the
static susceptibility $\, \chi_{\bf q} = \chi({\bf q}, 0) \,$ by the equation
\begin{equation}
\chi_{\bf q} = ({S}_{\bf q}^{+},S_{-{\bf q}}^{-}) =  m({\bf q}) /\omega_{\bf q}^2 \, .
 \label{b5}
\end{equation}
To calculate  the spin-excitation spectrum $\omega_{\bf q}$ we use the equality for
$m({\bf q})= \langle [\, [{S}^{+}_{\bf q}, H], \, S_{-\bf q}^{-}]\rangle $:
\begin{equation}
m({\bf q}) = (-\ddot{S}_{\bf q}^{+},S_{-{\bf q}}^{-}) =
 \omega_{\bf q}^2 \, ({S}_{\bf
q}^{+},S_{-{\bf q}}^{-}),
 \label{b6}
\end{equation}
where the correlation function $(-\ddot{S}_{\bf q}^{+},S_{-{\bf q}}^{-})$ is evaluated in the GMFA  by a decoupling procedure  in the site representation   as described in
Ref.~\cite{Vladimirov09}. This procedure is equivalent to the MCA for the equal-time
correlation functions. This results in the spin-excitation spectrum
\begin{eqnarray}
\omega_{\bf q}^2 &=& 8t^2\lambda_1(1-\gamma_{\bf q})(1-n-F_{2,0}-2F_{1,1})
\nonumber \\
 &+ & 4J^2(1-\gamma_{\bf q})
\big[\, \lambda_2\frac{n}{2}-\alpha_1C_{1,0}(4\gamma_{\bf q}+1)
\nonumber\\
&+&\alpha_2(2C_{1,1}+C_{2,0})\, \big],
 \label{b7}
\end{eqnarray}
where $t$ and $J$ are the hopping parameter and the exchange interaction for the nearest
neighbors, respectively, and $\gamma_{\bf q}=(1/2)\,(\cos q_x + \cos q_y)$ (we take the
lattice spacing $a$ to be unity). The static electron and spin correlation functions are
defined as
\begin{eqnarray}
  F_{n,m} \equiv F_{\bf R} &= &\langle X_{\bf 0}^{\sigma 0}\,
X_{\bf R}^{0\sigma}\rangle = \frac{1}{N}\sum_{{\bf q} }\,F_{\bf q}
 {\rm e}^{i {\bf q R}},
\label{b8a}\\
   C_{n, m}  \equiv C_{\bf R} &=& \langle S^-_{\bf 0} \,
   S^+_{\bf R} \rangle = \frac{1}{N}\sum_{{\bf q} }\,C_{\bf q}
 {\rm e}^{i {\bf q R}} ,
 \label{b8b}
\end{eqnarray}
where ${\bf R} = n {\bf a}_x + m {\bf a}_y \,$ and  $C_{\bf q} = \langle S^+_{\bf q}
S^-_{-\bf q}  \rangle $. The direct calculation of $m({\bf q})= \langle [\, [{S}^{+}_{\bf
q}, H], \, S_{-\bf q}^{-}]\rangle $ yields
\begin{equation}
m({\bf q})= -8 \, (1-\gamma_{\bf q})\, \left[ t \, F_{1,0}+ J \, C_{1,0} \right].
 \label{b9}
\end{equation}
\par
The AF correlation length $\xi$ may be calculated  by expanding the static susceptibility
in the neighborhood of the AF wave vector ${\bf Q} = \pi(1,1)$, $\chi_{\bf Q + \bf k}=
\chi_{\bf Q}/(1+\xi^2\, k^2)$~\cite{Shimahara91,Winterfeldt97}. We get
\begin{equation}
\xi^2=\frac{8J^2\alpha_1|C_{1,0}|}{\omega_{\bf Q}^2}.
 \label{b10}
\end{equation}
The critical behavior of the model  (\ref{b1}) is reflected by the divergence of
$\chi_{\bf Q}$ and $\xi$ as $T \rightarrow 0$, i.e., by $\omega_{\bf Q}(T = 0) = 0$. In
the phase with AF long-range order (LRO) which, in two dimensions, may occur at $T =0$
only, the correlation function $C_{\bf R}$ (\ref{b8b})  is written
as~\cite{Shimahara91,Winterfeldt97}
\begin{equation}
C_{\bf R} = \frac{1}{N}\sum_{{\bf q} \neq {\bf Q}}\,C_{\bf q}
 {\rm e}^{i {\bf q R}} + C\,{\rm e}^{i {\bf Q R}} .
 \label{b8c}
\end{equation}
The condensation part $C$ determines  the staggered magnetization $\, |m| = |\langle
S_i^z\rangle| \,$ which in the spin-rotation-invariant form is given by
\begin{equation}
m^2 = \frac{3}{2N}\sum_{\bf R} \,C_{\bf R}
 {\rm e}^{- i {\bf Q R}} = \frac{3}{2} \, C.
 \label{b8m}
\end{equation}
The  GMFA spectrum (\ref {b7}) is calculated self-consistently by using this spectrum for
the static correlation function (\ref{b8b}),
\begin{equation}
C_{\bf q} = \frac{m({\bf q})}{ 2 \, \omega_{\bf q}}\,\coth \frac{\beta \, \omega_{\bf
q}}{2} \,  .
  \label{b11}
\end{equation}
The decoupling parameters $\alpha_1, \alpha_2$ and $ \lambda_1, \lambda_2$ in Eq.~(\ref
{b7}) take into account the vertex renormalization for the spin-spin and electron-spin
interaction, respectively, as explained in Ref.~\cite{Vladimirov09}. In particular, the
parameters $\alpha_1, \alpha_2$ are evaluated from the results for the Heisenberg model
at $\delta =0$ and are kept fixed for $\delta \neq 0$. The parameters $ \lambda_1,
\lambda_2$ are calculated from the sum rule $C_{0,0} = \langle S^+_0 S^-_0\rangle  =
(1/2) (1- \delta)$ with a fixed ratio $\lambda_1/\lambda_2 = 0.378$. In the
superconducting state, the electron correlation function $\,F_{\bf R} \,$ is calculated
by the spectral function for electrons  in the superconducting state. The variation of
the parameters $ \lambda_1, \lambda_2$ below the superconducting transition  is
negligibly small and practically has no influence on the spectrum $\omega_{\bf q} $.
\par
Thus, the  static susceptibility (\ref {b5}) is explicitly
determined by Eqs.~(\ref {b7}) and (\ref {b9}).\\

{\bf Self-energy.} The self-energy  (\ref{b3}) can be written in terms of the
corresponding time-dependent correlation functions as
\begin{eqnarray}
\Sigma({\bf q},\omega)= \frac{1}{2\pi m({\bf q})}
\int_{-\infty}^{\infty}d\omega^{\prime}\frac{e^{\beta\omega^{\prime}}-1}
{\omega^{\prime}(\omega-\omega^{\prime})}\nonumber\\
\,\int_{-\infty}^{\infty} dte^{i\omega^{\prime}t}\langle \, \ddot{S}^{-}_{- \bf q}\;
\ddot{S}^{+}_{\bf q}(t) \rangle^{\rm pp},
 \label{b12}
\end{eqnarray}
where $-\ddot{S}_{i }^{+} =[[{S}_{i}^{+},\, (H_{t} + H_{J})]\, ,
  \, (H_{t} +H_{J} )] \equiv \sum_{\alpha} F_{i}^{\alpha}$,
and $H_{t}$ and  $H_{J}$ are the hopping  and the exchange parts of the Hamiltonia~(\ref
{b1}) and $\alpha = tt,\, tJ,\, Jt, \, JJ$.  As shown in Ref.~\cite{Vladimirov09}, the
largest contribution at a finite hole doping $\delta >  0.05$  in (\ref {b12}) comes from
the spin-electron scattering, the $F_{i}^{tt}$ forces. In the undoped case, $\delta =0$,
for the Heisenberg model we should take into account the $F_{i}^{JJ}$ forces.

Using MCA for the corresponding time-depending correlation functions of the forces, the
spin-electron scattering contribution to the imaginary part of the self-energy
$\Sigma''_t({\bf q},\omega)$ in the superconducting state can be written as
\begin{eqnarray}
 && \Sigma''_t({\bf q},\omega)=
-\frac{\pi(2t)^4 (e^{\beta \omega }-1)}{ m({\bf q})\,\omega}
\int\!\!\int_{-\infty}^{\infty} d\omega_1  d\omega_2
 \label{b13} \\
&&\times\frac{1}{N^2}\sum_{{\bf q_1, q_2}} [1- n(\omega_1)]\, n(\omega + \omega_1 -
\omega_2)\,N(\omega_2)  B_{\bf q_2}(\omega_2)
\nonumber \\
&& \times \big[({\Lambda}^2_{\bf q_1, q_2, q_3} + {\Lambda}^2_{\bf q_3, q_2, q_1})\,
A^N_{\bf q_1}(\omega_1) \, A^N_{\bf q_3}(\omega + \omega_1 - \omega_2)
\nonumber\\
&&- 2 {\Lambda}_{\bf q_1, q_2, q_3} {\Lambda}_{\bf q_3, q_2, q_1}\, A^S_{{\bf q_1}
\sigma}(\omega_1)\, A^S_{{\bf q_3} \sigma}(\omega + \omega_1 - \omega_2) \big] ,
\nonumber
\end{eqnarray}
where ${\bf q} = {\bf q}_1 + {\bf q}_2 + {\bf q}_3\,$. The Fermi and Bose functions are
denoted by $n(\omega)= (e^{\beta\omega}+1)^{-1}$ and
$N(\omega)=(e^{\beta\omega}-1)^{-1}$. The vertex function $\Lambda_{\bf q_1, q_2, q_3}$
is defined by the equation
\begin{equation}
 \Lambda_{{\bf q}_1 {\bf q}_2 {\bf q}_3}
= 4(\gamma_{{\bf q}_3+{\bf q}_2}
  -  \gamma_{{\bf q}_1 })\, \gamma_{q_3}
+ \gamma_{{\bf q}_2}-\gamma_{{\bf q}_1 + {\bf q}_3} ,
 \label{b13a}
\end{equation}
where the terms linear in $\gamma_{{\bf q}}$ reflect the exclusion of terms in $
F_{i}^{tt}$ with coinciding sites.  Here we introduced the  spectral functions:
\begin{eqnarray}
 A^N_{\bf q}(\omega)& =& -(1/\pi){\rm Im} \langle\langle
X^{0\sigma}_{\bf q}|X^{\sigma0}_{\bf q}\rangle\rangle_{\omega},
\label{b14a}\\
 A^S_{{\bf q} \sigma}(\omega)& =& -(1/\pi){\rm Im} \langle\langle
X^{0\sigma}_{\bf q}|X^{0 \bar\sigma}_{-\bf q} \rangle\rangle_{\omega},
\label{b14b}\\
B_{\bf q}(\omega) &= &(1/\pi)\,\chi''({\bf q}, \omega),
 \label{b14c}
\end{eqnarray}
where $A^{N, S}_{\bf q}(\omega)$ are determined by the  retarded anticommutator GFs for electrons~\cite{Zubarev60}. In the normal state, the contribution proportional
to  the anomalous GF  $\, \langle\langle X^{0\sigma}_{\bf q}|X^{0 \bar\sigma}_{-\bf q}
\rangle\rangle_{\omega}\,$ disappears.
\par
The self-energy $\Sigma''_J({\bf q},\omega)$ for the Heisenberg model in the undoped case
reads
\begin{eqnarray}
&&\Sigma''_{J}({\bf q},\omega) = \frac{\pi \,(2\,J)^4\, (e^{\beta \omega }-1)}{2 \,
m({\bf q})\,\omega } \frac{1}{N^2}\sum_{{\bf q}_1,{\bf q}_2}
 \int\!\!\int_{-\infty}^{\infty} d \omega_1 d \omega_2
  \nonumber \\
&&\times\{\Gamma^2_{{\bf q}_1 {\bf q}_2 {\bf q}_3} + \Gamma_{{\bf q}_1 {\bf q}_2 {\bf
q}_3}\Gamma_{{\bf q}_2 {\bf q}_1 {\bf q}_3}\}
 N(\omega_1 )N(\omega_2)
 \label{b15} \\
&&\times N(\omega-\omega_1-\omega_2)B_{{\bf q}_1}(\omega_1) B_{{\bf q}_2}(\omega_2)
B_{{\bf q}_3}(\omega- \omega_1 -\omega_2) ,
 \nonumber
\end{eqnarray}
where  ${\bf q} = {\bf q}_1 + {\bf q}_2 + {\bf q}_3\,$. The vertex for the spin-spin
scattering reads
\begin{eqnarray}
\Gamma_{{\bf q}_1 {\bf q}_2 {\bf q}_3}= 4 (\gamma_{{\bf q}_3+{\bf q}_1} - \gamma_{{\bf
q}_2}) (\gamma_{{\bf q}_3}-\gamma_{{\bf q}_1})
\nonumber\\
-\gamma_{{\bf q}_1}+\gamma_{{\bf q}_3}+\gamma_{{\bf q}_2+{\bf q}_3}-\gamma_{{\bf
q}_2+{\bf q}_1}.
 \label{b15a}
\end{eqnarray}
The self-energy (\ref{b15}) describes the conventional for the Heisenberg model  decay of a spin wave into three spin-waves (see, e.g., Ref.~\cite{Manousakis91})).
Similarly,  the self-energy (\ref{b13}) describe the decay of a spin excitation with the energy $\,\omega\,$ and wave vector ${\bf q}$ into three excitations: a particle-hole
pair and a spin excitation. This process is controlled by the energy and momentum
conservation laws.
\par
In the  calculation of the self-energy  (\ref {b13}) we adopt MFA for the electron
spectral functions (\ref{b14a}) and (\ref{b14b}) which in the superconducting state can
be written as
\begin{eqnarray}
A^N_{\bf q}(\omega) & = & Q \sum_{\omega_1 = \pm E_{\bf q}}\frac{\omega_1 +
\varepsilon_{\bf q}}{2 \omega_1}
 \delta(\omega - \omega_1) \, ,
\label{b17a} \\
A^S_{{\bf q} \sigma}(\omega) &  = & Q \sum_{\omega_1 =\pm E_{\bf q}}\frac{\Delta_{{\bf q}
\sigma}} {2 \omega_1}
 \delta(\omega - \omega_1)\, .
\label{b17b}
\end{eqnarray}
Here $Q = 1 - n/2$ is the Hubbard weighting factor and the superconducting gap function
$\, \Delta_{{\bf q} \sigma} = ({\rm sgn}~\sigma)~\Delta_{\bf q}$. In the electron
spectrum $\varepsilon_{\bf q}$ we take into account only the nearest-neighbor hopping
$\,t \,$ and consider the energy dispersion  in the Hubbard-I approximation:
$\varepsilon_{\bf q} = - 4t\,Q\, \gamma_{\bf q}- \mu$. The spectrum of quasiparticles in
the superconducting state is given by the conventional formula $E_{\bf
q}=\sqrt{\varepsilon_{\bf q}^2 + \Delta_{\bf q}^2} $. For the spin-excitation  spectral
function (\ref{b14c}) we take the form:
\begin{equation}
 B_{\bf q}(\omega) = \frac{m({\bf q})}
 { 2 \,\widetilde{\omega}_{\bf q}}
 \,[ \delta(\omega - \widetilde{\omega}_{\bf q}) - \delta(\omega +
\widetilde{\omega}_{\bf q})] ,
 \label{b17c}
\end{equation}
where the spectrum of spin excitations $\widetilde{\omega}_{\bf q}$  is determined by the
pole of the DSS, $\,\widetilde{\omega}_{\bf q} = [\omega_{\bf q}^2 +
\widetilde{\omega}_{\bf q} \, {\rm Re}\Sigma({\bf q},\widetilde{\omega}_{\bf q})]^{1/2}
\,$.

In presenting numerical results,  we take the exchange interaction $\, J=0.3\,t $ and
measure all energies in units of $\, t \,$. For the superconducting state we consider the
$d$-wave gap in the form $\Delta^{(d)}_{\bf q} = (\Delta/2)(\cos q_x - \cos q_y)$ where
the temperature dependent amplitude $\Delta(T)$ follows the conventional
Bardeen-Cooper-Schrieffer theory. We mainly consider two  doping cases, $\delta=0.2$,
where  for $t = 0.31$~eV  the superconducting transition temperature is $k_{\rm B} T_{\rm
c} = 0.025 t \approx 91$~K,  and $\delta=0.09$  with $k_{\rm B} T_{\rm c} = 0.016 t
\approx  59$~K~\cite{Vladimirov11}.

\subsection{Spin excitations in  the normal state}
\label{sec:3b}

\begin{figure}
\includegraphics[width=0.35\textwidth]{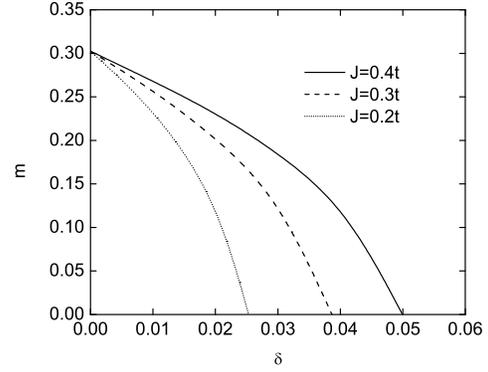}
\caption{ Staggered  magnetization as a function of doping for different values of $J/t
$.} \label{fig1}
\end{figure}

{\bf Static properties.} In the static limit, the doping dependence of the spin
correlation functions (\ref{b8b}) and the staggered magnetization $m(\delta)$ (\ref{b8m})
were calculated at zero temperature. The correlation functions $C_{1,0}, \,C_{1,1}$ and
$C_{2,1}$ show a good agreement with the ED  data of Ref.~\cite{Bonca89}. The staggered
magnetization $m(\delta)$ is plotted in Figfig.~\ref{fig1} for various values of the
exchange interaction $J/t $.

We observe a strong suppression of the AF LRO with increasing doping due to the spin-hole interaction. In the Heisenberg limit we get $m(0)= 0.303$ which agrees with the value
$m(0)= 0.3074$ found in quantum Monte Carlo simulations ~\cite{Wiese94}. At the critical doping $ \delta_c\,$ we obtain a transition from the LRO phase to a paramagnetic phase with AF SRO. It is remarkable that $ \delta_c$ is nearly proportional to $J/t$. This result agrees with that found by the cumulant approach of Ref.~\cite{Vojta96} where our $ \delta_c$ values are somewhat lower, e.g.,  $ \delta_c \simeq 0.06$ for $J/t = 0.4$.
The $ \delta_c$ values obtained are in qualitative agreement with neutron scattering
experiments on La$_{2-\delta}$Sr$_\delta$CuO$_4$ (LSCO) which reveal the vanishing of 3D LRO at $ \delta_c \simeq 0.02$~\cite{Kastner98}.

The doping dependence of the uniform static spin susceptibility $\chi = (1/2)\lim_{q
\rightarrow 0} \chi_{\bf q}$ reveals  an increase   upon doping caused by the decrease of
SRO,  i.e., of the spin stiffness against orientation along a homogeneous external
magnetic field. At large enough doping,  $\chi$ decreases with increasing $\delta $ due
to the decreasing number of spins. Therefore,  the SRO-induced maximum of $\chi$ at
$\delta_{\rm max}(T)$ appears which shifts to lower doping with increasing temperature,
since SRO effects are less pronounced at higher $T$. The doping dependence of  $\chi$, especially the maximum at $\delta_{\rm max}(T)$  found in \cite{Vladimirov09} is in accord with the ED results of Ref.~\cite{Jaklic96}.

The AF correlation length $\xi$ (\ref{b10})  was studied as a function of doping and
temperature.  The behavior of $\,\xi \,$ in the zero-temperature limit as function of
doping and $J/t$ can be explained  by considering the  staggered magnetization at $T = 0$
depicted in Fig.~\ref{fig1}. At a given value of $J/t$ and $\delta < \delta_c$, in the
limit $T \to 0$, the correlation length $\xi$ shows a divergence connected with the
closing of the AF gap $\omega_{\bf Q} \to 0$ due to appearance of AF LRO. At zero doping,
$\xi^{-1}(T)$ exhibits the known exponential decrease as $T \rightarrow
0$~\cite{Shimahara91,Winterfeldt97}. At $\delta > \delta_c$, the ground state has no AF LRO, i.e., we have $\omega_{\bf Q} > 0$, and the correlation length saturates at  $T \to 0$.  A reasonable  agreement of the temperature dependence of $\xi^{-1}(T, \delta)$ with neutron-scattering experiments on LSCO at $T \leq 600$~K,~\cite{Kastner98}  was found at
the doping $\delta = 0.04$. The doping dependence of $\xi(\delta, T)$ can be described approximately by the proportionality $\xi(\delta, T) \propto 1/\sqrt{\delta}$ which
agrees with the experimental findings~\cite{Kastner98}.
\par

{\bf Dynamic properties.} Now we present results for the spin-fluctuation spectra
provided by the imaginary part of the DSS $\chi''({\bf q}, \omega)$, Eq.~(\ref {b4}). The
shape of the spin-fluctuation spectrum depends on the ratio of the damping to the energy
of the excitation: at a small damping we have a spin-wave-like behavior, while for a
large damping the spectrum shows a broad energy distribution. So, it is interesting to
consider the doping and temperature dependence of the damping  of spin fluctuations $\,
\Gamma({\bf q},\omega) $  determined by the imaginary part of the self-energy,  $\,
\Gamma({\bf q},\omega)= - (1/2)\Sigma''({\bf q},\omega)\,$. Here we mainly consider the
damping at $\, \omega = \omega_{\bf q}, \; \Gamma_{\bf q}= \Gamma({\bf q},\omega_{\bf q})
\,$.
\par
It turns out that the major contributions to the damping are given by the diagonal terms
$\,\Sigma''_{t}({\bf q},\omega)\,$, Eq.~(\ref {b13}) and $\,\Sigma''_{J}({\bf
q},\omega)\,$, Eq.~(\ref {b15}) , while the interference terms, such  as
$\,\Sigma_{Jt,Jt}''({\bf q},\omega)\,$ appear to be much smaller and may be neglected.
That is, the damping $\,\Gamma_{\bf q} \,$ is the sum of the spin-spin scattering
contribution $\,\Gamma_{J, {\bf q}} = - (1/2 )\Sigma''_{J}({\bf q},\omega_{\bf q}) \,$
and the spin-hole scattering contribution $\,\Gamma_{t,{\bf q}} = - (1/2
)\Sigma''_{t}({\bf q},\omega_{\bf q}),\, \Gamma_{\bf q} = \Gamma_{J,{\bf q}} +
\Gamma_{t,{\bf q}} \,$. Note that in Refs.~\cite{Sega03} and ~\cite{Prelovsek04} the
partition of the damping into a spin-exchange contribution and a fermionic contribution
was suggested from the ED data.
\par
The numerical calculations of  $\,\Sigma''({\bf q},\omega)\,$ are performed for the
exchange interaction $J = 0.3t$, the value which is usually used in numerical
simulations. This affords us to compare our analytical results with finite cluster
calculations and to check the reliability of our approximations.
\begin{figure}
 \includegraphics[width=0.35\textwidth]{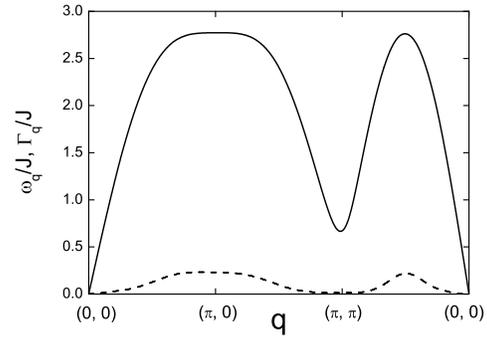}
\caption{Spectrum $\omega_{\bf q}$ (solid line) and damping $\Gamma_{\bf q}$ (dashed
line)
 in the Heisenberg limit, $\delta = 0$,  at $T=0.35 J$ .}
 \label{fig2}
\end{figure}
Let us first consider the Heisenberg limit $\delta = 0$. Figure~\ref{fig2}  shows the
spectrum of spin excitations $\omega_{\bf q}$, Eq.~(\ref {b7}),  and the damping $\,
\Gamma_ {\bf q} = \Gamma_ {J, \bf q}$. The results are similar to those obtained in
Ref.~\cite{Winterfeldt99}. In the spin-wave region, at $\,q \xi \gg 1$,  we get
well-defined quasiparticles with $\Gamma_ { \bf q} \ll \omega_{\bf q}$, as also discussed
in Ref.~\cite{Manousakis91} for the two-dimensional Heisenberg model. To compare our
results for the damping with the quantum Monte Carlo  data of Ref.~\cite{Makivic92}, we
have considered the linewidth $\Lambda_{\bf q}$ of the relaxation function $\, F({\bf
q},\omega) = 4 [\beta\omega \chi_{ \bf q}]^{-1}\chi''({\bf q},\omega) \,$  at $T = 0.35
J$, where $\Lambda_{\bf q} \simeq 2 \Gamma_{ \bf q} $, and have found a good agreement.
\par
\begin{figure}
\includegraphics[width=0.35\textwidth]{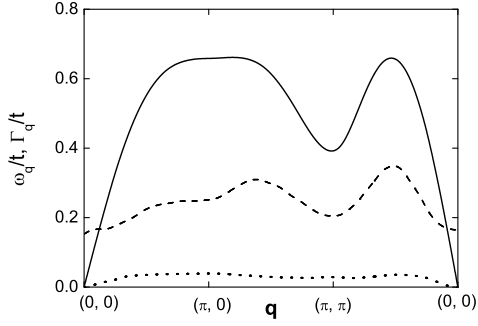}
\caption{Spectrum $\omega_{\bf q}$ (solid line), and  damping $\, \Gamma_{J, \bf q}$
(dotted line) and   $\, \Gamma_{t, \bf q}\,$ (dashed line) at $T=0.15t$ and
$\delta=0.1$.}
 \label{fig3}
\end{figure}

For non-zero  doping the  spin-hole scattering contribution $\,\Sigma''_{t}({\bf
q},\omega)\,$, Eq.~(\ref {b13}), increases rapidly with doping and temperature and
already at moderate hole concentration far exceeds the spin-spin scattering contribution
$\,\Sigma''_{J}({\bf q},\omega)\,$, Eq.~(\ref {b15}), as demonstrated in
Fig.~\ref{fig3}. Depending on ${\bf q}$, doping, and temperature, the spin excitations
may have a different character and dynamics. In particular, for the spin-spin scattering
contribution $\,\Sigma''_{J}({\bf q},\omega)\,$ we observe, in the long-wavelength limit,
$ \lim_{{\bf q} \to 0 } \Gamma_{J,{\bf q}} = 0$, as in the case of the Heisenberg limit
shown in Fig.~\ref{fig2}. This  result for the Heisenberg model was obtained by various
calculations (see, e.g., Ref.~\cite{Manousakis91}). Contrary to this behavior, the
damping $\Gamma_{t,{\bf q}}$, induced by the spin-hole scattering,  is finite in this
limit. The different behavior of $\Gamma_{J, {\bf q}}$ and $\Gamma_{t, {\bf q}}$ may be
explained by the different ${\bf q}$-dependence of the spectral functions entering
Eqs.~(\ref{b13}) and (\ref{b15}). Whereas for spin excitations the spectral function  is
proportional to $m({\bf q})/\omega_{\bf q} \sim q $ for $q \to 0$ (see Eq.~(\ref{b17c})),
for electrons it is finite in this limit (see Eq.~(\ref{b17a})). Therefore, in the limit
of $\,{\bf q} = {\bf q}_1 + {\bf q}_2 + {\bf q}_3 = 0\,$, for the spin-spin scattering
the product $m({\bf q}_1) m({\bf q}_2) m({\bf q}_3)/\omega_{{\bf q}_1}\omega_{{\bf
q}_2}\omega_{{\bf q}_3}$ gives a vanishingly small contribution to the integrals over
${\bf q}_1,{\bf q}_2$ in $\,\Sigma''_{J}({\bf q},\omega)\,$, Eq.~(\ref{b15}), while  in
the spin-hole self-energy (\ref{b13}) there is no such small factor. At low enough doping
and temperature, i.e., at small enough $\,\Gamma_{t,{\bf q}}$, we may observe
well-defined high-energy spin-wave-like  excitations with $\,q, k \gg 1/\xi \quad (k =
|{\bf q -Q }|)\,$ and $\,\Gamma_{{\bf q}} \ll \omega_{{\bf q}}\,$ propagating in AF SRO.
\par
\begin{figure}
\includegraphics[width=0.35\textwidth]{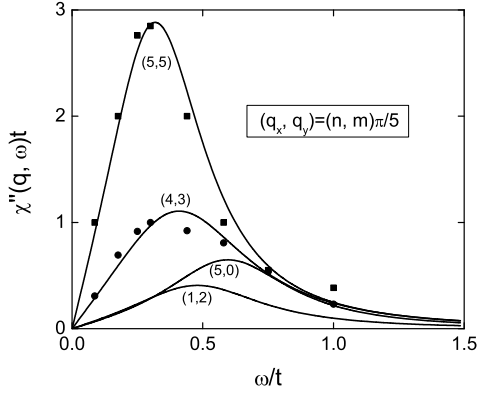}
\caption{Spectral function $\chi''({\bf q}, \omega)$ for various wave vectors at
$T=0.15t$ and $\delta=0.1$ in comparison with  ED data (filled symbols,
Ref~\cite{Prelovsek04}).}
 \label{fig4}
\end{figure}

Studies of the spectral function $\chi''({\bf q}, \omega)$ at various wave vectors at
finite doping reveals that close to AF wave vector  ${\bf q \approx Q}$ the damping
$\Gamma_{\bf Q}$ is very small at  low doping and low enough temperature. In this case we
observe underdamped  spin modes characterized by sharp resonance peaks in  $\chi''({\bf
Q}, \omega) $. With increasing doping those modes evolve into overdamped (relaxation
type) spin-fluctuation modes (AF paramagnons) described by the broad spectrum as shown in
Fig.~\ref{fig4}. We found a generally good agreement with the ED data of Ref.~\cite{Prelovsek04}
in this region.
\par

\subsection{Spin dynamics in the superconducting state}
\label{sec:3c}

In the superconducting state the spin-excitation spectrum of high-$T_{\rm c}$ cuprates is dominated by a sharp magnetic peak at the  AF wave vector ${\bf Q}$ which is called the resonance mode (RM). It was discovered in the inelastic neutron scattering experiments first in the optimally doped YBa$_{2}$Cu$_{3}$O$_{y}$~(YBCO$_y$) crystal~\cite{Rossat91}
and later on, the RM was found in the Bi-2212 compounds~\cite{Sidis04},  in the
single-layer cuprates  Tl$_2$Ba$_2$CuO$_{6+x}$~\cite{He02},
HgBa$_2$CuO$_{4+\delta}$~\cite{Yu10}, and in the electron-doped
Pr$_{0.88}$LaCe$_{0.12}$CuO$_{4-\delta}$ superconductor~\cite{Wilson06}. This
demonstrates that the RM is a generic feature of the cuprate superconductors and can be related to spin excitations in a single CuO$_2$ layer.
\par
The spin-excitation dispersion close to the RM  exhibits  a peculiar ``hour-glass''-like shape  with  upward and downward dispersions. Whereas the RM energy $E_{\rm r}$  changes with doping,  no essential temperature dependence of $E_{\rm r}$ and the upward branch of the dispersion has been found. In the strongly underdoped YBCO crystal only the downward branch is suppressed  above $ T_{\rm c}$, whereas the upward dispersion and the RM are observed in the normal pseudogap state. So we can conclude that the RM and the upward dispersion are  related  to the exchange interaction $J \sim  1500$~K and for $\,T \sim T_c \ll J$  do not show a noticeable temperature dependence. The  downward branch may be explained by inhomogeneous phases related to fluctuating stripe phases with a quasi-one-dimensional order of spins and charges or to a liquid crystal state which are smeared out at $\,T \gtrsim T_c$.
\par
To prove this conclusion we consider the spin excitation damping in the superconducting state ~\cite{Vladimirov11} within the theory formulated in Sec.~\ref{sec:3a}. We can propose an alternative explanation of the magnetic RM and the upper
branch of the dispersion which are driven by the spin gap at the AF wave vector ${\bf Q}
= \pi(1,1)$ in the paramagnetic state instead of the superconducting gap $2\Delta$
proposed in the exciton spin-1 scenario (see Sec.~\ref{sec:3d}).
\par
\begin{figure}
\includegraphics[width=0.35\textwidth]{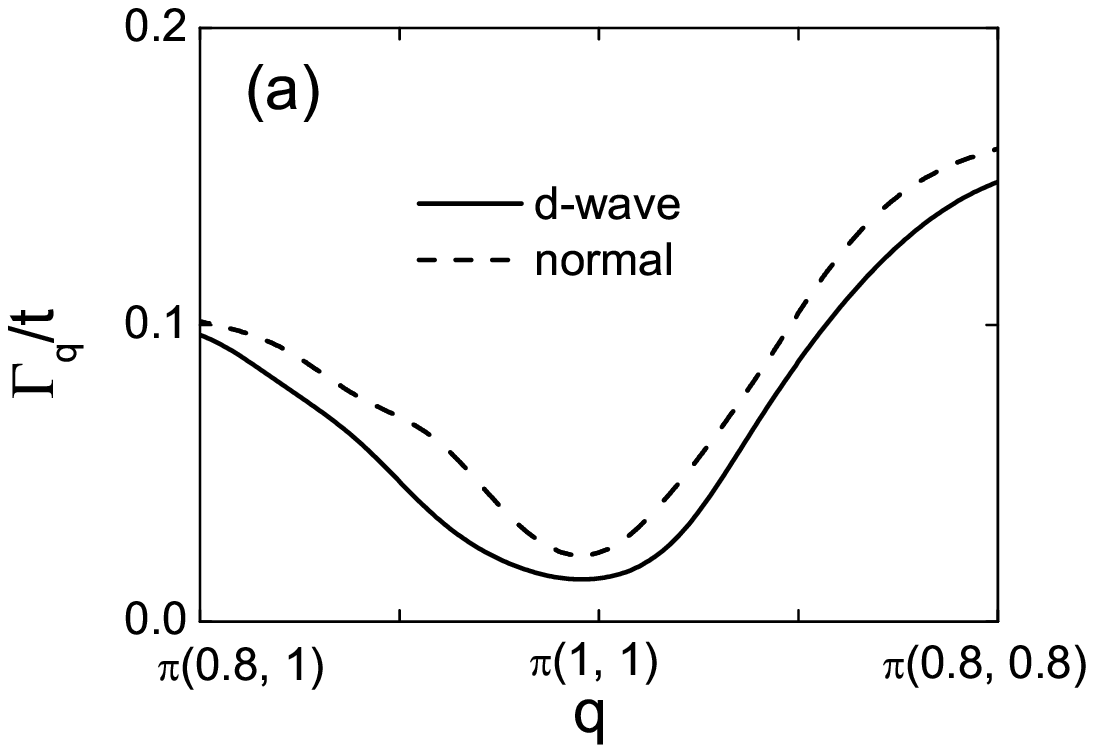}\\
\includegraphics[width=0.35\textwidth]{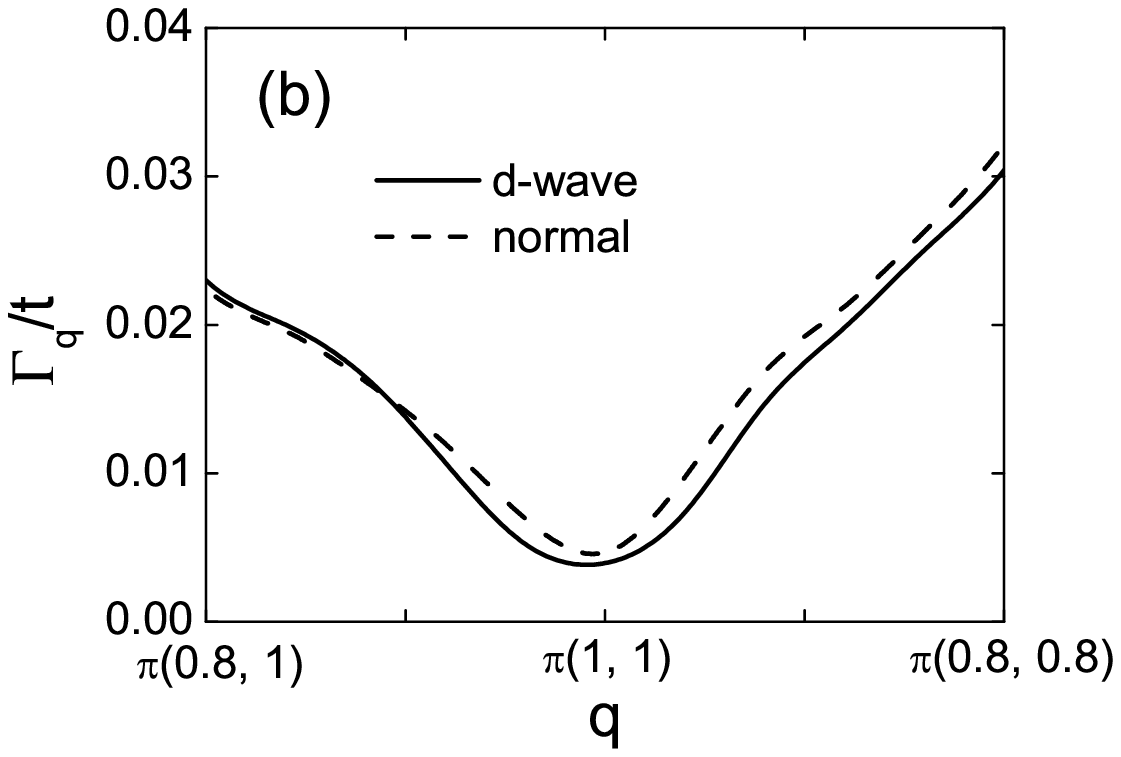}
\caption{Spin-excitation damping $\Gamma_{\bf q}$ for (a) $\delta=0.2$ and for (b)
$\delta=0.09$ at $T =0$ for  the $d$-wave pairing (solid line) and   in the normal state
(dashed line).}
 \label{fig5}
\end{figure}
As discussed in the previous section, in the normal state close to AF wave vector the damping $\Gamma_{\bf Q}$ is very small at  low doping and low enough
temperature. This results in the underdamped  spin modes characterized by sharp resonance
peaks in  $\chi''({\bf Q}, \omega) $. This property retains in the superconducting state
and the RM can be observed even above $T_{\rm c}$ in the underdoped cuprates. In
Fig.~\ref{fig5} the spin-excitation damping $\Gamma_{\bf q} = -(1/2) \,\Sigma''_t({\bf
q},\omega = \widetilde{\omega}_{\bf q})$ is  shown at $T =0$ for the overdoped case,
$\delta = 0.2$, Fig.~\ref{fig5}~(a) and for the underdoped case, $\delta = 0.09$,
Fig.~\ref{fig5}~(b). The small difference between the damping in the $d$-wave
superconducting state and the normal state  observed for the full self-energy in both the
cases, Eq.~(\ref{b13}), confirms that the superconducting gap does not play an essential
role in suppressing  the damping $\Gamma_{\bf Q}$. The damping in the underdoped
region is of the order of magnitude weaker than in the overdoped region. The sharp increase
of the damping $\Gamma_{\bf q}$ away from the AF wave-vector ${\bf Q}$ explains the
resonance character of spin excitations at ${\bf Q}$.
\par
\begin{figure}
\includegraphics[width=0.35\textwidth]{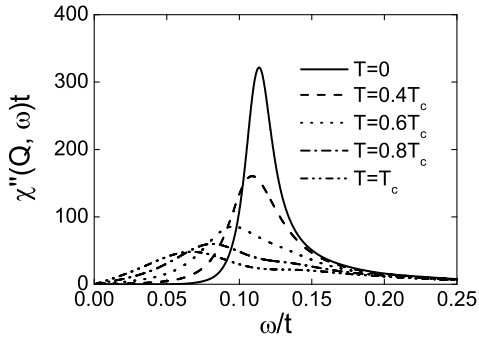}
\caption{Temperature dependence of the spectral function $\chi''({\bf Q}, \omega)$ at
$\delta=0.2$.}
 \label{fig6}
\end{figure}
\begin{figure}
\includegraphics[width=0.35\textwidth]{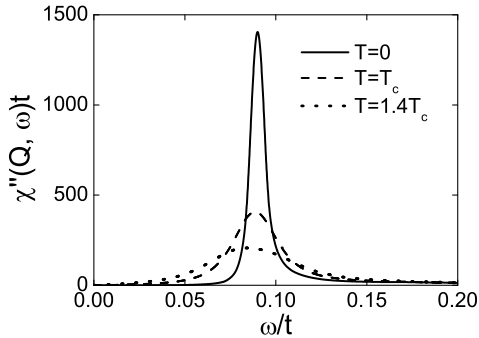}
\caption{Temperature dependence of the spectral function $\chi''({\bf Q}, \omega)$ at
$\delta=0.09$.}
 \label{fig7}
\end{figure}

The temperature dependence of the spectral function $\chi''({\bf Q}, \omega)$ in the
overdoped case  $\delta = 0.2$ is shown in Fig.~\ref{fig6}. It has high intensity at low
temperatures, but strongly decreases with temperature and becomes very broad at $T \sim
T_{\rm c}$ as found in experiments (see Ref.~\cite{Bourges98}). Note, that the shift of
the maximum of the peak to lower energies  with increasing temperature can be explained
just by the increase of the damping as this is usually observed for an overdamped
oscillator. In Fig.~\ref{fig7} the temperature dependence of the spectral function
$\chi''({\bf Q}, \omega)$ for the underdoped case $\delta = 0.09$ is plotted. Whereas the
resonance energy $E_r$ decreases with underdoping in comparison with Fig.~\ref{fig6},
the intensity of the RM greatly increases in accordance with experiments. The RM
intensity decreases with temperature but its energy $E_r$  does not change with
temperature and is still quite visible at $T = T_{\rm c}$ and even at $T = 1.4 T_{\rm
c}$.

\begin{figure}
\includegraphics[width=0.38\textwidth]{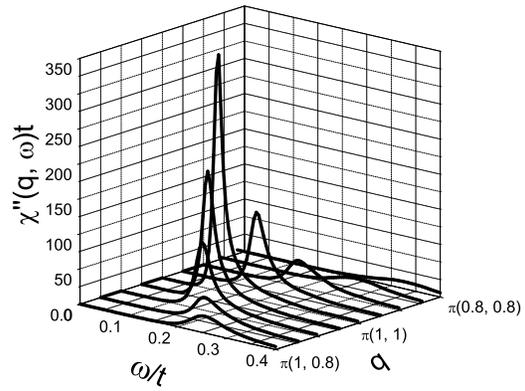}
\caption{Spectral function $\chi''({\bf q}, \omega)$ near the wave vector    ${\bf Q} =
\pi(1,1)$  at $T=0$ for $\delta=0.2$.}
 \label{fig8}
\end{figure}

The dispersion of the spectral function for $\delta = 0.2$ is shown in Fig.~\ref{fig8}
at $T=0$. A strong suppression of the spectral-function intensity away from  ${\bf Q} =
\pi(1,1)$ explains the resonance-type behavior of the function at low temperatures. This
suppression of the intensity is in accord with the  sharp increase of the damping away
from ${\bf Q}$ shown in Fig.~\ref{fig5}. However, the upper branch of the
spin-excitation spectrum reminiscent of  spin waves is still visible in Fig.~\ref{fig8}.

\begin{figure}
\includegraphics[width=0.35\textwidth]{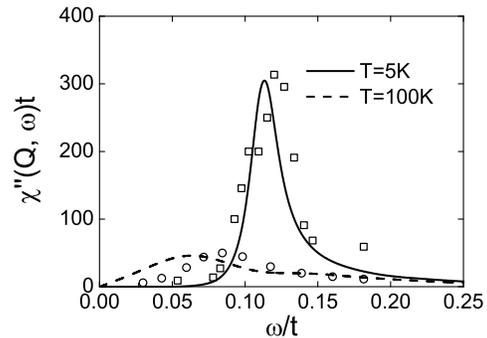}
\caption{Spectral function $\chi''({\bf Q}, \omega)$ for doping $\delta = 0.2$ compared
to experimental data  for YBCO$_{6.92}$, Ref.~\cite{Bourges98}, at $T=5K$ (squares) and
$T=100K$ (circles).}
 \label{fig9}
\end{figure}
\par
In Fig.~\ref{fig9}  we compare our results with the neutron-scattering data for the
nearly optimally doped YBCO$_{6.92}$ single crystal~\cite{Bourges98} at $T=5K$  and
$T=100K$. In this sample, $T_{\rm c} = 91$~K and the RM energy $E_{\rm r} \simeq 40$~meV$
= 5.1 k_{\rm B}\, T_{\rm c} > 2\, \Delta_0 $ (taking $2\Delta_0(\delta) = 3.52\,k_{\rm B}
T_{\rm c}(\delta)$ we have  $E_{\rm r} \simeq 2.9 \Delta_0 $). For $\delta = 0.2$,  our
calculations yield $E_{\rm r} = 0.12 t = 38$~meV$= 4.8 k_{\rm B}\, T_{\rm c} = 2.7\,
\Delta_0 $ ($t = 0.313$~eV, $k_{\rm B}\, T_{\rm c} = 0.025 t$ (see Sec.~\ref{sec:3a}).
Since the experimental data are given in arbitrary units (a. u.), our results were scaled
to fit the absolute value of the peak at $T=5K$.
\par
\begin{figure}
\includegraphics[width=0.35\textwidth]{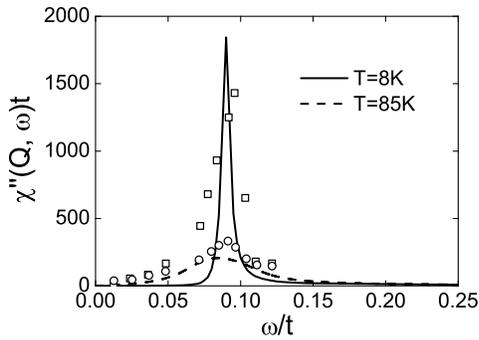}
\caption{Spectral function $\chi''({\bf Q}, \omega)$ for doping $\delta = 0.09$ compared
to experimental data  for YBCO$_{6.5}$, Ref.~\cite{Stock04},  at $T=8K$ (squares) and
$T=85K$ (circles).}
 \label{fig10}
\end{figure}
In Fig.~\ref{fig10}  our results are compared with the experimental data for  the
underdoped ortho-II YBCO$_{6.5}$ single crystal with $E_{\rm r} = 33$~meV$=6.5 \,k_{\rm
B}T_{\rm c}= 3.7\, \Delta_0 $ at $T=8K$ and $T=85K$ (see Refs.~\cite{Stock04,Stock05}).
For $\delta = 0.09$,  our theory gives $E_{\rm r} = 0.09 t = 28$~meV$ = 5.6 \, k_{\rm
B}\, T_{\rm c} = 3.2\, \Delta_0 $. Here we also scaled our results to fit the absolute
value of the peak at $T=8 K$. We note a weak temperature dependence of the RM energy
observed experimentally and obtained in our calculation. In both compounds the RM energy
is larger than the superconducting excitation energy, $ 2 \Delta_0$, while in the spin-1
exciton scenario the RM energy $E_{\rm r}$ has to be less than $ 2 \Delta_0$.  So we
obtain a good agreement of our theory with neutron-scattering experiments on YBCO
crystals both near the optimal doping and in the underdoped region.\\

\subsection{Comparison with previous theoretical studies}
\label{sec:3d}

There is a vast literature devoted to experimental and theoretical investigations of
spin-excitation spectra and  the RM in cuprates (a list of  references can be found in
reviews \cite{Bourges98,Sidis04,Sidis07,Eschrig06,Plakida10} and
Ref.~\cite{Vladimirov11}). To explain the RM  in a large number of studies the
Fermi-liquid model of itinerant electrons was assumed and the DSS was calculated within
the random phase approximation (RPA) for a one-band Hubbard model specified by the
Coulomb interaction $U$ or the AF superexchange interaction $J({\bf q})$ (see, e.g.,
Refs.~\cite{Norman00,Manske01,Eremin05}).   Summation  of the electron-hole bubble
diagrams in the  RPA for the self-energy results in a sharp RM formed below the continuum
of particle-hole excitations gapped at a threshold energy $\omega_c \leq 2\Delta({\bf
q}^*)$ determined by the superconducting $d$-wave gap $2\Delta({\bf q}^*)$ at a
particular wave vector $\,{\bf q}^*\,$ on the Fermi surface (FS). In this approach the RM
is considered as a particle-hole bound state, usually referred to as the \mbox{spin-1}
exciton.
\par
However, in the spin-1 exciton scenario, a strong temperature dependence of the RM energy
is expected below $T_{\rm c}$ due to temperature dependence of the superconducting gap
$2\Delta({\bf q}^*)$ and, hence, the threshold energy $\omega_c$. But the temperature
dependence of the RM energy was not found in experiments, in particular for the  RM at
the energy $E_{\rm r} \approx 33$~meV in the YBCO$_{6.5}$ crystal~\cite{Stock04,Stock05}.
At low temperature, $T \sim 8$~K, the RM revealed a much higher intensity than in
optimally doped crystals, and it was also seen at the same energy  with less intensity
even at $T \simeq 1.4\,T_{\rm c}$ (see Fig.~\ref{fig10}).

In the strong correlation limit the Mori projection technique in the equation of motion method for the relaxation function has been used by several groups (see, e.g.,
Refs.~\cite{Sega03,Prelovsek04,Sherman03,Sherman06,Sega06,Prelovsek06}). However, in
several studies of the $t$-$J$ model the contribution of the accompanied spin excitation
in the self-energy has been neglected~\cite{Onufrieva02} or approximated by static or
mean-field-type expressions (see, e.g., Refs.~\cite{Sega03} and ~\cite{Sherman03}) which
results in the self-energy similar to the RPA formula given by Eq.~(\ref{b16}).

\begin{figure}
\includegraphics[width=0.35\textwidth]{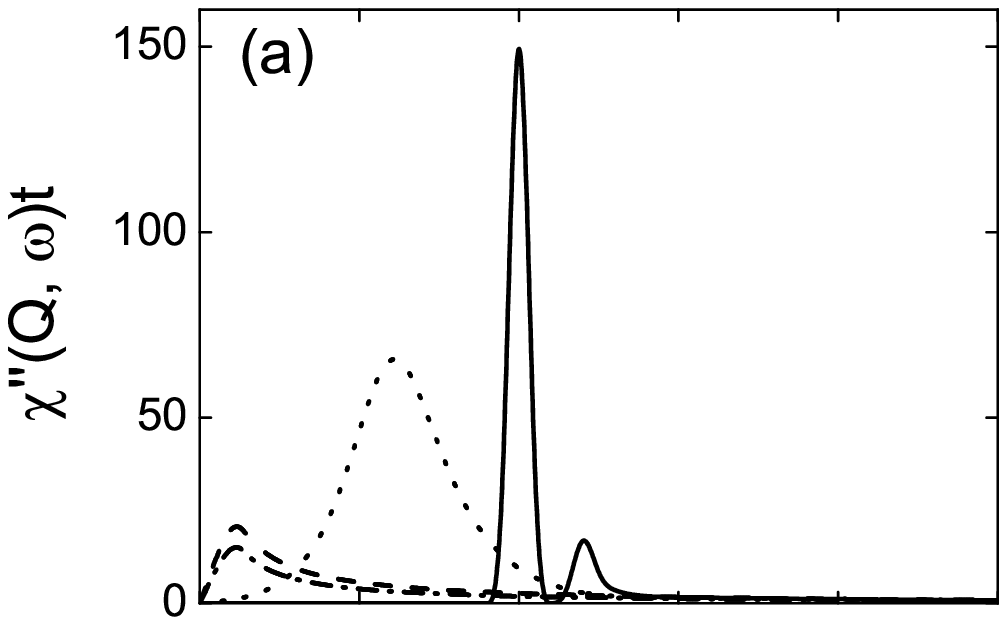}\\
\includegraphics[width=0.35\textwidth]{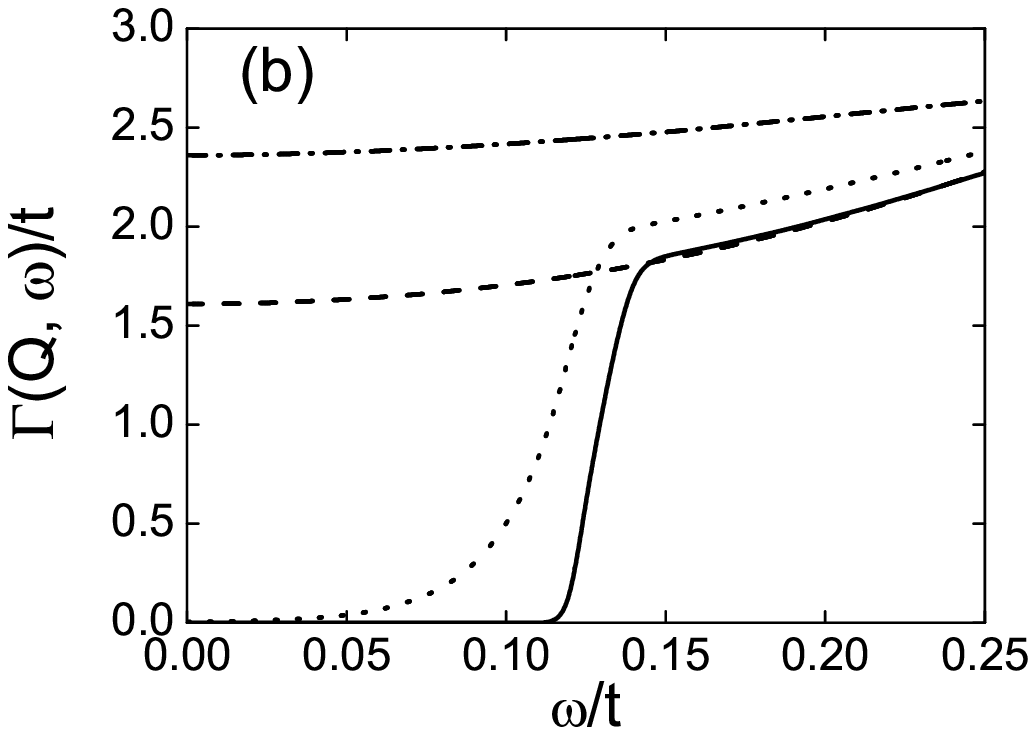}
\caption{(a) Spectral function $\chi''({\bf Q}, \omega)$ and (b) spin-excitation damping
$\Gamma({\bf Q}, \omega)$ calculated   in the particle-hole bubble approximation,
Eq.~(\ref{b16}),  at $\delta = 0.2$ for the $d$-wave pairing ($\Delta_0 = 0.1 t$ taken
from Ref.~\cite{Sega03} ) at $T = 0$ (solid line) and $T=0.4 T_{\rm c}$ (dotted line),
and for the normal state at  $T = 0$ (dashed line) and $T= T_{\rm c}$ (dash-dotted
line).}
 \label{fig11}
\end{figure}

That is, in these approximations the spin-excitation contribution was ``decoupled'' from
the particle-hole pair. We can derive the particle-hole bubble approximation from
Eq.~(\ref {b13}), if  we ignore the spin-energy contribution $\omega_2$ in comparison
with the electron-hole pair energy, or, equivalently, if in the MCA, the time-dependent
spin correlation function is approximated by its static value: $\langle S_{\bf -q}^{-}
S_{\bf q}^{+}(t)\rangle \simeq \langle S_{\bf -q}^{-} S_{\bf q}^{+}\rangle = C_{\bf q}$.
Moreover, excluding the spin-excitation wave-vector  ${\bf q}_2$ from the wave-vector
conservation law, we have ${\bf q} = {\bf q}_1 + {\bf q}_3 $. As a result of these
approximations in Eq.~(\ref {b13}), we obtain the self-energy in the form of the
particle-hole bubble approximation:
\begin{eqnarray}
&&\widetilde{\Sigma}_t''({\bf q},\omega)
 =  -\frac{\pi(2t)^4\, }
{ m({\bf q})\,\omega} \int_{-\infty}^{\infty} d\omega_1 [ n(\omega_1)- n(\omega_1
+\omega)]
 \nonumber\\
&&  \times\frac{1}{N}\,\sum_{{\bf q_1}}\,
 \big[ \widetilde{\Lambda}^N_{{\bf q_1, q -q_1}}\, A^N_{\bf
q_1}(\omega_1) \, A^N_{\bf q - q_1}(\omega_1 +\omega)
\nonumber \\
&&  - \widetilde{\Lambda}^S_{{\bf q_1, q -q_1}}\, A^S_{{\bf q_1} \sigma}(\omega_1)\,
A^S_{{\bf q- q_1} \sigma}(\omega_1 +\omega) \big].
 \label{b16}
\end{eqnarray}
where the  averaged over the spin-excitation wave vector ${\bf q_2}$ vertexes are
introduced,
\begin{eqnarray}
 \widetilde{\Lambda}^N_{{\bf q_1, q_3}}& = & \frac{1}{N}
  \sum_{{\bf q_2}} \, C_{{\bf q}_2}
 [\Lambda^2_{\bf q_1, q_2, q_3} + \Lambda^2_{\bf q_3, q_2,
   q_1}] \, ,
\label{b16n} \\
\widetilde{\Lambda}^S_{{\bf q_1, q_3}} &=& \frac{2}{N}
 \sum_{{\bf q_2}} \, C_{{\bf q}_2} {\Lambda}_{\bf q_1, q_2, q_3}
{\Lambda}_{\bf q_3, q_2, q_1} .
 \label{b16s}
\end{eqnarray}
In the approximation (\ref{b16})  only the opening of  a superconducting gap in the
particle-hole excitation can suppress the damping of spin excitations due to the decay into particle-hole pairs which may result in the RM. Figure~\ref{fig11}
shows   the spectral function $\,\chi''({\bf q}, \omega) \, $ and the damping
$\Gamma({\bf Q}, \omega)$ using Eq.~(\ref{b16}). To compare these functions with those
calculated in Ref.~\cite{Sega03}, we adopt  the electron dispersion used in
Ref.~\cite{Sega03}, $\,\varepsilon_{\bf q}^{eff} = -4\,t_{eff} \gamma_{\bf q} - 4\,
t'_{eff} \cos q_x \cos q_y  \, $ with $ t_{eff} = 0.3t $ and $ t'_{eff} = - 0.1t\,$ and
take the gap parameter $\, \Delta_0 = 0.1t$. The obtained results are quite close to
those  shown in Fig.~1 of Ref.~\cite{Sega03} (where $\chi_{zz}''({\bf q}, \omega) =
(1/2)\chi''({\bf q}, \omega)\,$ is plotted). At $T = 0$, we observe a much narrower RM,
but with a lower intensity  in comparison with the RM calculated with the full
self-energy, Eq.~(\ref{b13}), as shown in Fig.~\ref{fig6}. The energy $E_{\rm r}$ of the
RM shown Fig.~\ref{fig11}(a) noticeably decreases with increasing temperature,  contrary
to a negligible shift of the RM shown in Fig.~\ref{fig6} for $T=0.4 T_{\rm c}$. This
comparison demonstrates that in the particle-hole bubble approximation the
superconducting gap plays a crucial role in the occurrence of the RM with  $E_{\rm r}(T)
< 2 \Delta(T)$, while in the full self-energy (\ref {b13}) the superconducting gap and
details of the electron dispersion  are less important. Note that the damping in the
normal state for the reduced self-energy, Eq.~(\ref{b16}), is the order of magnitude
larger than for the full self-energy, Eq.~(\ref{b13}) (see Fig.~\ref{fig11}(b)).

This difference can be explained as follows. Whereas in the particle-hole bubble
approximation given by Eq.~(\ref{b16}) the spin excitation with the  energy $\omega$ at
the wave vector ${\bf Q}$  can decay only into a particle-hole pair with the energy
$\omega({\bf Q}) = E_{\bf Q + q} + E_{\bf q}$, in a more general process described by
Eq.~(\ref{b13}) an additional spin excitation participates  in the scattering. In the
limit $T \to 0$, the decay process in this case  is governed by another
energy-conservation law, $\omega({\bf Q}) = E_{\bf q_3} + E_{\bf q_1} +
\widetilde{\omega}_{\bf q_2}$ where the largest contribution from the spin excitation
comes from $\widetilde{\omega}_{\bf q_2} \simeq \widetilde{\omega}_{\bf Q} $ due to the
factor $\, m({\bf q}_2) \,$ (\ref{b6}) in $B({\bf q}_2)$~(\ref{b17c}) in Eq.~(\ref{b13}).
This energy-momentum conservation law strongly reduces the phase space for the decay and
suppresses the damping of the initial spin excitation with the energy $\omega({\bf Q}) $.
In fact, the occurrence of an additional spin excitation   in the scattering process with
the finite energy $\widetilde{\omega}_{\bf Q}$ plays a role similar to the
superconducting gap in the excitation of the particle-hole pair in  Eq.~(\ref{b16}).
Therefore, the damping at low temperatures ($k_{\rm B}\,T \ll \widetilde{\omega}_{\bf Q}
\sim E_{\rm r} $) appears to be small even in the normal state as demonstrated in
Fig.~\ref{fig6}. In the case of the particle-hole relaxation, the condition for the
occurrence of the RM, $\,\omega({\bf Q}) = E_{\bf q + Q} + E_{\bf q} \leq 2 \Delta({\bf
q}^*) $, imposes a strong restriction on the shape of the FS which should cross the AF
Brillouin zone to accommodate the scattering vector ${\bf Q}$ and the vector ${\bf q}^*$
on the FS.  In the case of the full self-energy, Eq.~(\ref{b13}), the energy-momentum
conservation law for three quasipartricles does not impose such strict limitations.

\section{Kinematical spin-fluctuation mechanism of pairing
in cuprates}
 \label{sec:4}

\subsection{General formulation}
\label{sec:4a}

{\bf Hamiltonian.}  In this section  we consider a more general than in Ref.~(\ref{1})
extended Hubbard model
\begin{eqnarray}
 H &= & \varepsilon_1\sum_{i,\sigma}X_{i}^{\sigma \sigma}
  + \varepsilon_2\sum_{i}X_{i}^{22} + \sum_{i\neq j,\sigma}\,
t_{ij}\,\bigl\{ X_{i}^{\sigma 0} \, X_{j}^{0\sigma}
\nonumber \\
& + &    X_{i}^{2 \sigma}X_{j}^{\sigma 2}
 + \sigma \,(X_{i}^{2\bar\sigma} X_{j}^{0 \sigma} +
  {\rm H.c.})\bigr\} + H_{c, ep},
 \label{2a}
\end{eqnarray}
which includes also electron-phonon interaction   $g_{ij}$  defined by the Hamiltonian
\begin{eqnarray}
H_{c, ep} &= & \frac{1}{2}  \sum_{i\neq j}\,V_{ij} N_i N_j + \sum_{i, j}\,g_{i j} N_i\,
u_j,
 \label{1a}
\end{eqnarray}
where $u_j$ describes  an atomic displacement on the lattice site $j$ for a particular
phonon mode. More generally, the  electron-phonon interaction can be written as a sum
$\sum_{\nu} g^{\nu}_{i,j} u_{j}^{\nu}$ over the normal modes $\nu$.

We emphasize here that  the Hubbard model (\ref{2a}) does not involve a dynamical
coupling of electrons (holes) to fluctuations of spins or charges. Its role is played by
the  kinematical interaction caused by the complicated  commutation relations (\ref{7}),
as was already noted by Hubbard~\cite{Hubbard65}. For example, the equation of motion for
the HO $\, X\sb{i}\sp{\sigma 2} = a^{\dag}_{i\sigma}a_{i\sigma}a_{i\bar\sigma}\, $ for
the model (\ref{2a}) reads:
\begin{eqnarray}
 i\frac{d}{d t}  X\sb{i}\sp{\sigma 2} &= &[X\sb{i}\sp{\sigma 2}, H] =
   (\varepsilon_1 + U)\, X_{i}^{\sigma 2}\,
\nonumber \\
  &+& \sum\sb{l,\sigma '}\! \left( t\sb{il}\sp{22}
    B\sb{i\sigma\sigma '}\sp{22} X\sb{l}\sp{\sigma ' 2} -
    \sigma t\sp{21}\sb{il} B\sb{i\sigma\sigma '}\sp{21}
    X\sb{l}\sp{0\bar\sigma '} \right)
\nonumber \\
 &-& \sum\sb{l} X\sb{i}\sp{02} \left( t\sp{11}\sb{il}
    X\sb{l}\sp{\sigma0} +  \sigma t\sp{21}\sb{il}
    X\sb{l}\sp{2 \bar\sigma} \right)
\nonumber \\
& +&  \sum\sb{l}  X_{i}^{\sigma 2}( V_{i l}\,N_{l} +
 g_{il} \, u_{l} ).
\label{8}
\end{eqnarray}
Here $B\sb{i\sigma\sigma'}\sp{\eta \zeta}$  are the Bose-like operators,
\begin{eqnarray}
  B\sb{i\sigma\sigma'}\sp{22} & = & (X\sb{i}\sp{22} +
   X\sb{i}\sp{\sigma\sigma}) \, \delta\sb{\sigma'\sigma} +
   X\sb{i}\sp{\sigma\bar\sigma} \, \delta\sb{\sigma'\bar\sigma}
\label{8a} \\
  &  = &( N\sb{i}/2 +  \sigma\, S\sb{i}\sp{z}) \, \delta\sb{\sigma'\sigma} +
    S\sb{i}\sp{\sigma} \, \delta\sb{\sigma'\bar\sigma},
\nonumber \\
  B\sb{i\sigma\sigma'}\sp{21} & = & ( N\sb{i}/2 +
   \sigma S\sb{i}\sp{z}) \, \delta\sb{\sigma'\sigma} -
   S\sb{i}\sp{\sigma}\,  \delta\sb{\sigma'\bar\sigma}.
  \label{8b}
\end{eqnarray}
We see that the hopping amplitudes here depend on the  number and spin operators due to
the kinematical interaction of electrons with spin and charge fluctuations. In
phenomenological spin-fermion models, a dynamical coupling of electrons with spin
fluctuations is specified by fitting parameters, while in Eq.~(\ref{8}) the interaction
is determined by the hopping energy $t_{ij}$ fixed by the electronic dispersion.

{\bf  Dyson equation.} To consider  superconducting pairing in the model (\ref{2a}), we
introduce the two-time thermodynamic GF~\cite{Zubarev60} expressed in terms of the
four-component Nambu operators, $\, \hat X_{i\sigma}$ and   $\, \hat
X_{i\sigma}^{\dagger}=(X_{i}^{2\sigma}\,\, X_{i}^{\bar\sigma 0}\,\, X_{i}^{\bar\sigma
2}\,\, X_{i}^{0\sigma}) \,$:
\begin{eqnarray}
 {\sf G}_{ij\sigma}(t-t') & = & -i \theta(t-t')\langle \{
 \hat X_{i\sigma}(t) ,  \hat X_{j\sigma}^{\dagger}(t')\}\rangle
 \nonumber \\
 & \equiv & \langle \!\langle \hat X_{i\sigma}(t) \mid
    \hat X_{j\sigma}^{\dagger}(t')\rangle \!\rangle,
 \label{9}
\end{eqnarray}
where $ \{A, B\} = AB + BA$,  $ A(t)= \exp (i Ht) A\exp (-i Ht)$, and $\theta(x) = 1 $
for $x > 0 $ and $\theta(x) = 0 $ for $x < 0 $. The Fourier representation in $({\bf k},
\omega) $-space is defined by the relations:
\begin{eqnarray}
{\sf G}_{ij\sigma}(t-t') &= & \frac{1}{2\pi}\int_{-\infty}^{\infty} dt e^{- i(t-t')} {\sf
G}_{ij\sigma}(\omega),
 \label{9ft}\\
{\sf G}_{ij\sigma}(\omega) & = &\frac{1}{N}\,\sum_{\bf k}\exp[i{\bf k (i-j)}] {\sf
G}_{\sigma}({\bf k}, \omega).
    \label{9fk}
\end{eqnarray}
The GF (\ref{9fk}) is convenient to write in the matrix form
\begin{equation}
{\sf G}_{\sigma}({\bf k}, \omega)=
  {\hat G_{\sigma}({\bf k}, \omega)  \quad \quad
 \hat F_{\sigma}({\bf k}, \omega) \choose
 \hat F_{\sigma}^{\dagger}({\bf k}, \omega) \quad
   -\hat{G}_{\bar\sigma}(-{\bf k}, -\omega)} ,
 \label{9m}
\end{equation}
where the normal $\hat G_{\sigma}({\bf k}, \omega)$ and anomalous (pair) $\hat
F_{\sigma}({\bf k}, \omega) $  GFs are
 $2\times 2$ matrices for  two Hubbard subbands:
\begin{equation}
\hat G_{\sigma}({\bf k}, \omega) = \langle\! \langle \left(
\begin{array}{c}
     X_{\bf k}^{\sigma2}  \\
     X_{\bf k}^{0 \bar\sigma } \\
        \end{array}\right)  \mid
  X_{\bf k}^{2\sigma} X_{\bf k}^{\bar\sigma 0}
 \rangle \! \rangle_{\omega},
 \label{9a}
\end{equation}
\begin{equation}
\hat F_{\sigma}({\bf k}, \omega)  = \langle\! \langle \left(
\begin{array}{c}
     X_{\bf k}^{\sigma2}  \\
     X_{\bf k}^{0 \bar\sigma } \\
        \end{array}\right)  \mid
  X_{-\bf k}^{\bar\sigma 2} X_{-\bf k}^{0 \sigma}
 \rangle \! \rangle_{\omega}.
 \label{9b}
\end{equation}
To calculate the GF (\ref{9})  we use the equation of motion method. Differentiating the
GF with respect to  time $t$ we obtain the equation for the Fourier representation
(\ref{9ft})
\begin{equation}
 \omega {\sf G}\sb{ij\sigma}(\omega) = \delta \sb{ij} {\sf Q} +
   \langle \!\langle
    [\hat X\sb{i\sigma},H] \mid  \hat X\sb{j\sigma}\sp{\dagger}
   \rangle \!\rangle\sb{\omega}\, .
\label{10}
\end{equation}
Here the  correlation function ${\sf Q} = \langle \{\hat X\sb{i\sigma},\hat
X\sb{i\sigma}\sp{\dagger}\}\rangle   = \hat{\tau}_{0} \times \hat Q\,$  where   $\, \hat
Q =
\left(  \begin{array}{cc} Q\sb{2} & 0 \\
      0 & Q\sb{1} \end{array}  \right)\,$
and $\hat{\tau}_{0}$ is the $2 \times 2$ unit matrix. The spectral weights of the Hubbard
subbands  in the paramagnetic state  $\, Q\sb{2} = \langle X\sb{i}\sp{22} +
X\sb{i}\sp{\sigma\sigma} \rangle = n/2 \,$ and $\, Q\sb{1} = \langle X\sb{i}\sp{00} +
X\sb{i}\sp{\bar\sigma \bar\sigma} \rangle = 1-Q\sb{2}\, $ depend on the average
occupation number of holes (\ref{3a}). In  the $\, {\sf Q}$ matrix we neglect anomalous
averages of the type $\langle X_i^{02} \rangle$ which are irrelevant for  the $d$-wave
pairing~\cite{Adam07}.
\par
To introduce the zero-order quasiparticle (QP) excitation energy we use the Mori-type
projection operator method~\cite{Mori65,Plakida11}. In this approach, the many-particle
operator $\hat Z\sb{i\sigma} = [\hat X\sb{i\sigma},H]$ in (\ref{10}) is written as  a sum
of a linear part and an irreducible part $\hat Z\sb{i\sigma}\sp{(ir)}$ orthogonal to
$\hat X\sb{j\sigma}\sp{\dagger}$:
\begin{equation}
  \hat Z\sb{i\sigma} = [\hat X\sb{i\sigma}, H] =
    \sum\sb{l}{\sf E}\sb{il\sigma} \hat X\sb{l\sigma} +
    \hat Z\sb{i\sigma}\sp{(\rm ir)}.
\label{11}
\end{equation}
From the  orthogonality condition,
 $\langle \{ \hat Z\sb{i\sigma}\sp{(\rm ir)}, \,
 \hat X\sb{j\sigma}\sp{\dagger}\} \rangle =  0$,
we obtain  the excitation  energy in the GMFA which is given by the time-independent
matrix of correlation functions:
\begin{eqnarray}
 {\sf E}\sb{ij\sigma} &=&  \langle \{ [\hat X\sb{i\sigma}, H],
    \hat X\sb{j\sigma}\sp{\dagger} \} \rangle {\sf Q}^{-1}.
 \label{12}
\end{eqnarray}
The corresponding zero-order GF reads
\begin{equation}
  {\sf G}\sp{0}\sb{\sigma }({\bf k},\omega) =
    \Bigl( \omega \tilde{\tau}\sb{0} - {\sf E}\sb{\sigma}({\bf k})
      \Bigr) \sp{-1} {\sf Q} \, ,
\label{13}
\end{equation}
where $\tilde{\tau}\sb{0}$ is the $4\times 4$ unit matrix. The spectrum of QP excitations
in the GMFA is given by the Fourier component of the matrix (\ref{12}):
\begin{eqnarray}
  {\sf E}\sb{\sigma}({\bf k})&=&
  \frac{1}{N}\sum_{\bf k}\exp[i{\bf k (i-j)}] \langle \{ [\hat X\sb{i\sigma}, H],
    \hat X\sb{j\sigma}\sp{\dagger} \} \rangle {\sf Q}^{-1}
 \nonumber\\
   &= & \left(
\begin{array}{cc}
  \hat{\varepsilon}({\bf k})  & \hat{\Delta}_{\sigma}({\bf k}) \\
     \hat{\Delta}_{\sigma}^{*}({\bf k}) &
     -\hat{\varepsilon}({\bf k})
\end{array}\right)   ,
\label{12a}
\end{eqnarray}
where $\hat{\varepsilon}({\bf k})$ and $\hat{\Delta}_{\sigma}({\bf k})$ are the normal
and anomalous parts of the energy matrix.
\par
To calculate the multiparticle GF $\,\langle \langle
    \hat Z\sb{i\sigma}\sp{(\rm ir)} (t) \mid  \hat X\sb{j\sigma}\sp{\dagger}(t')
   \rangle \rangle \,$ in (\ref{10}) we differentiate it with respect
to the second time $t'$ and apply the same projection procedure as in (\ref{11}). This
results in the equation for the GF (\ref{10}) in  the form,
\begin{equation}
 {\sf G}\sb{\sigma}({\bf k}, \omega) =
 {\sf G}\sp{0}\sb{\sigma }({\bf k},\omega) + {\sf G}\sp{0}\sb{\sigma }({\bf
 k},\omega)\,{\sf T}\sb{\sigma }({\bf k},\omega)\,
 {\sf G}\sp{0}\sb{\sigma }({\bf k},\omega),
\label{14a}
\end{equation}
where the scattering matrix
\begin{equation}
  {\sf T}\sb{\sigma}({\bf k}, \omega) =
 {\sf  Q}\sp{-1}\,\langle\!\langle {\hat Z}\sb{{\bf k}\sigma}\sp{(\rm ir)} \!\mid\!
     {\hat Z}\sb{{\bf k}\sigma}\sp{(\rm ir)\dagger} \rangle\!\rangle
      \sb{\omega}\;{\sf  Q}\sp{-1} .
\label{15a}
\end{equation}
Now we can introduce the self-energy operator ${\sf \Sigma}\sb{\sigma}({\bf q}, \omega)$
as the {\it proper} part ({\rm pp}) of the scattering matrix (\ref{15a}) which has no
parts connected by the zero-order GF (\ref{13}) according to the equation: ${\sf T} =
{\sf \Sigma} + {\sf \Sigma } \, {\sf G}\sp{0} \, {\sf T}$. The definition of the proper
part of the scattering matrix (\ref{15a}) is equivalent to an introduction of a projected
Liouvillian superoperator for the memory function in the conventional Mori
technique~\cite{Mori65}.
\par
Using the self-energy operator instead of the scattering matrix in Eq.~(\ref{14a}) we
obtain the Dyson equation for the GF (\ref{9}):
\begin{equation}
 {\sf G}\sb{\sigma}({\bf k}, \omega) =
  \left[\omega \tilde{\tau}\sb{0} - {\sf E}\sb{\sigma}({\bf k})
  -    {\sf  Q} {\sf \Sigma}_{\sigma}({\bf k}, \omega)
  \right] \sp{-1} {\sf Q},
\label{14}
\end{equation}
where the self-energy operator is given by
\begin{equation}
 {\sf  Q} {\sf \Sigma}\sb{\sigma}({\bf k}, \omega) =
    \langle\!\langle {\hat Z}\sb{{\bf k}\sigma}\sp{(\rm ir)} \!\mid\!
     {\hat Z}\sb{{\bf k}\sigma}\sp{(\rm ir)\dagger} \rangle\!\rangle
      \sp{(\rm pp)}\sb{\omega}\;{\sf  Q}\sp{-1} .
\label{15}
\end{equation}
Dyson equation  (\ref{14})  with the zero-order QP excitation energy (\ref{12a}) and the
self-energy (\ref{15}) gives an exact representation for the GF (\ref{9}).  The
self-energy takes into account processes of inelastic scattering of electrons (holes) on
spin  and charge fluctuations due to the kinematical interaction and the dynamic
intersite CI and the electron-phonon interaction.

The self-energy operator (\ref{15}) can be  written in the same matrix form as the  GF
(\ref{9m}):
\begin{equation}
 {\sf  Q} {\sf \Sigma}\sb{\sigma}({\bf k}, \omega) =  {\hat M_{\sigma}({\bf k}, \omega)
 \quad  \quad
\hat\Phi_{\sigma}({\bf k}, \omega) \choose \hat\Phi_{\sigma}^{\dagger} ({\bf k},
\omega)\quad -\hat{M}_{\bar\sigma}({\bf k}, -\omega)} {\sf Q}^{-1}  \, ,
 \label{23}
\end{equation}
where the matrices $\hat M$ and $\hat\Phi$  denote the corresponding normal and anomalous
(pair) components of the self-energy operator.

The system of equations for the $(4 \times 4)$ matrix GF (\ref{9m}) and the self-energy
(\ref{23}) can be reduced to a system of equations for the normal ${\hat G}_\sigma({\bf
k},\omega)$ and the pair ${\hat F}_\sigma({\bf k},\omega)$ matrix components. Using
representations for the energy matrix (\ref{12a}) and the self-energy (\ref{23}), we
derive  for these components  the following system of matrix equations:
\begin{eqnarray}
&&{\hat G}({\bf k},\omega) = \Bigl\{
  \hat {G}_{N}({\bf k},\omega)^{-1}
\nonumber\\
& + &  \hat{\varphi}_\sigma({\bf k},\omega)\,
  \hat{G}_{N}({\bf k},- \omega)\,\hat{\varphi}^{*}_\sigma({\bf
k},\omega)  \Bigr\}^{-1} \, \hat{Q},
 \label{24} \\
&& {\hat F}_\sigma({\bf k},\omega) = -\hat{G}_{N}({\bf k},-
\omega)\,\hat{\varphi}_\sigma({\bf k},\omega) \,
 \hat{G}({\bf k},\omega) ,
 \label{25}
\end{eqnarray}
where we introduced  the normal state  GF
\begin{eqnarray}
{\hat G}_{N}({\bf k},\omega)& = & \Bigl( \omega \hat\tau_0 - \hat{\varepsilon}({\bf k}) -
  \hat{M}({\bf k},\omega)/ \hat{Q} \Bigr)^{-1},
\label{26}
 \end{eqnarray}
and the  superconducting gap function
\begin{eqnarray}
{\hat \varphi}_\sigma({\bf k},\omega)& = & \hat{\Delta}_{\sigma}({\bf k}) +
 \hat\Phi_{\sigma}({\bf k},\omega) /\hat{Q} .
 \label{27}
\end{eqnarray}
\par
The system of equations for the normal GFs (\ref{24}), (\ref{26}), the anomalous GF
(\ref{25})  and the gap equation (\ref{27}) should be solved self-consistently  for the
multiparticle GFs in the self-energy operator (\ref{23}) as discussed below.

\subsection{Generalized mean-field approximation}
\label{sec:4b}

The superconducting pairing in the Hubbard model already occurs in  the MFA  and is
caused by the  kinetic superexchange interaction as in the $t$--$J$
model~\cite{Anderson87}. Therefore, it is reasonable to consider at first the MFA
described by the zero-order GF (\ref{13}). The energy matrix (\ref{12}) is calculated
using the commutation relations (\ref{5}) for the HOs. The normal part
$\hat{\varepsilon}({\bf k})\, $ of the energy matrix (\ref{12a}) after diagonalization
determines the QP spectrum in two Hubbard subbands in the GMFA (for detail
see~\cite{Plakida07}):
\begin{eqnarray}
{\varepsilon}_{1, 2} ({\bf k})& = & ({1}/{2}) [\omega_{2} ({\bf k}) + \omega_1 ({\bf k})]
\mp({1}/{2}) \Lambda({\bf k}),
 \label{17}\\
 {\omega}_\iota({\bf k})& = &
 4  t\,\alpha_{\iota} \gamma({\bf k})
 + 4 \,\beta_{\iota}\,t'\gamma'({\bf k})+
 4 \,\beta_{\iota}\,t''\gamma''({\bf k})
 \nonumber   \\
& + &\omega^{(c)}_\iota({\bf k}) + U \delta_{\iota,2} - \mu,
 \quad (\iota = 1, 2)
\label{17a}\\
  \Lambda({\bf k}) &= &    \{[\omega_{2} ({\bf k})
  - \omega_1 ({\bf k})]^2 + 4 W({\bf k})^2 \}^{1/2},
\nonumber\\
 W({\bf k}) & = &  4  t\,\alpha_{12} \gamma({\bf k})
 + 4 t' \,\beta_{12} \gamma'({\bf k})+
 4 t'' \,\beta_{12} \gamma''({\bf k}).
\nonumber
\end{eqnarray}
Here  the hopping parameter is defined by the expression:
\begin{eqnarray}
t_{ij} & = & (1/N)\,\sum_{\bf k}\exp[i{\bf k (i-j)}]\, t({\bf k}),
\label{17b}\\
t({\bf k})& =&  4 t \, \gamma({\bf k}) + 4 t' \,\gamma'({\bf k}) + 4 t''\, \gamma''({\bf
k}).
 \label{17c}
\end{eqnarray}
which  are equal to  $t\,$ for the nearest neighbors (nn) sites $ a_{1} = ( \pm a_{x},
\pm a_{y})$,  $\, t'\,$ -- for the next nearest neighbors (nnn) sites $a_d = \pm (a_x \pm
a_y)$, and $ \, t''\,$ -- for the  nnn sites $\,a_{2}= (\pm 2 a_{x}, \pm 2 a_{y})$. The
corresponding  ${\bf k}$-dependent functions are: $\,\gamma({\bf k})= (1/2)(\cos k_x
+\cos k_y), \; \gamma'({\bf k}) = \,\cos k_x \cos k_y \, $, and $\; \gamma '' ({\bf k})=
(1/2)(\cos 2 k_x +\cos 2 k_y) $ (the lattice constants $ a_{x}= a_{y}$ are put to unity).
\par
The kinematical interaction for the HOs results in renormalization of  the spectrum
(\ref{17}) determined by the parameters: $\, \alpha_{\iota}= Q_{\iota}[ 1 +
{C_{1}}/{Q^2_{\iota}}], \, \beta_{\iota} = Q_{\iota}[ 1 + {C_{2}}/{Q^2_{\iota}}]\,$, $\,
\alpha_{12}= \sqrt{Q_{1}Q_{2}}[ 1 - {C_{1}}/{Q_{1}Q_{2}}] ,\, \beta_{12} =
\sqrt{Q_{1}Q_{2}}[ 1 -{C_{2}}/{Q_{1}Q_{2}}]\,$. In addition to the conventional Hubbard I
renormalization given by $Q_{1}, \, Q_{2}$  parameters an essential   renormalization is
caused by the AF spin correlation functions  for the nn and the nnn sites, respectively:
\begin{equation}
C_{1} = \langle {\bf S}_i{\bf S}_{i+ a_1} \rangle, \quad C_{2} = \langle {\bf S}_i{\bf
S}_{i+ a_d (a_2) }\rangle .
 \label{18}
\end{equation}
These functions strongly depend on doping resulting in a considerable variation of the
electronic spectrum  as shown later and discussed in  detail  in Ref.~\cite{Plakida07}.
\par
The contribution from the  CI $V_{i j}$ in (\ref{17a}) is given by
\begin{equation}
\omega^{(c)}_{1(2)}({\bf k})= \frac{1}{N } \sum_{\bf q} V({\bf k -q}) N_{1(2)}({\bf q}),
 \label{17ci}
 \end{equation}
where $ N_1({\bf q}) = \langle X_{\bf q}^{0 \bar\sigma}X_{\bf q}^{\bar\sigma 0}\rangle /
Q_1\,$ and $\,N_2({\bf q})=
 \langle X_{\bf q}^{\sigma 2}X_{_{\bf q}}^{2\sigma}\rangle / Q_2\,$
are  occupation numbers in the single-particle and two-particle subbands, respectively
and  $V({\bf q})$ is the Fourier transform of intersite CI $V_{ij}$.
\par
 The anomalous component $\, \hat{\Delta}_{\sigma}({\bf
k}) \, $ of the matrix (\ref{12a}) determines the superconduction gap in the GMFA.
Considering  the singlet $d$-wave pairing, we calculate the intersite pair correlation
functions. The diagonal matrix components are given by the equations:
\begin{eqnarray}
&&   \Delta\sb{ij\sigma}\sp{22} Q_2 =  -  \sigma\, t_{ij}^{21}
   \langle X\sb{i}\sp{02} N\sb{j} \rangle  - V_{i j}
 \langle  X_{i}^{\sigma 2}\, X_{j}^{\bar\sigma 2} \rangle ,
   \label{19a}\\
&&  \Delta\sb{ij\sigma}\sp{11} Q_1 =   \sigma \, t_{ij}^{12}
   \langle N\sb{j}  X\sb{i}\sp{02} \rangle - V_{i j}
   \langle  X_{i}^{0 \bar\sigma} X_{j}^{0\sigma}  \rangle .
 \label{19b}
\end{eqnarray}
Here we used the  notation $ t_{ij}^{12}$ for the interband hopping parameters to
emphasize that the kinematical pairing $\langle X\sb{i}\sp{02} N\sb{j} \rangle$ is
mediated by the interband hopping.  In terms of  the Fermi operators  $ a_{i\sigma} =
X_{i}^{0\sigma} + \sigma X_{i}^{\bar\sigma 2}$, the pair correlation function  in
(\ref{19a}) can be written as $\, \langle X\sb{i}\sp{02} N\sb{j} \rangle  = \langle
X\sb{i}\sp{0\downarrow} X\sb{i}\sp{\downarrow 2} N\sb{j}\rangle =\langle
a\sb{i\downarrow}\, a\sb{i\uparrow} N\sb{j} \rangle $. This representation  shows that
the kinematical pairing occurs on a single lattice site but in two Hubbard
subbands~\cite{Plakida03}. The  correlation function $\, \langle X\sb{i}\sp{02} N\sb{j}
\rangle$ can be calculated directly from the    GF $\, L_{ij}(t-t') = \langle \langle
X_{i}^{02} (t) \mid N_j (t') \rangle \rangle \, $ without {\it any decoupling}
approximation as shown in Ref.~\cite{Plakida03}. In particular, under hole doping, $ n =
1 + \delta > 1$, the pair correlation function in the two-site approximation, as shown in Ref.~\cite{Plakida03}, reads:
\begin{eqnarray}
  \langle X_{i}^{02} N_j \rangle =
   -\frac{4  t_{ij}^{12} }{U}  \sigma \,
  \langle X_{i}^{\sigma2}X_{j}^{\bar\sigma2} \rangle .
 \label{20}
\end{eqnarray}
As a result, the equation for the superconducting gap in (\ref{19a})  can be written as
\begin{eqnarray}
\Delta^{22}_{ij\sigma}  =
 (J_{i j} - V_{i j})\,\langle X_{i}^{\sigma2}
 X_{j}^{\bar\sigma2}\rangle /Q_2 ,
 \label{21}
\end{eqnarray}
where $\, J_{i j} = {4\, ( t_{ij}^{12})^2}/{U }$ is the AF superexchange interaction.  A
similar equation holds  for the gap in the one-hole subband:
 $\, \Delta_{ij\sigma}^{11}= (J_{i j} -
 V_{i j})\,\langle X_{i}^{0\bar\sigma}
 X_{j}^{0\sigma} \rangle/Q_1 \,$. We thus conclude that the
pairing in the Hubbard model in the GMFA  is similar to  the superconductivity in the
$t$--$J$ model mediated by the  AF superexchange interaction $J_{i j}$ induced by the
kinematical interaction for the interband hopping as in the $t$-$J$
model~\cite{Anderson87}.

\subsection{Self-energy operator}
\label{sec:4c}

To determine  the self-energy matrix we should calculate many-particle GFs in
Eq.~(\ref{23}). Since the diagram technique for HOs is very complicated  and can be used
effectively only for the lowest order diagrams  in the Hubbard model (see, e.g.,
Refs.~\cite{Zaitsev87,Izyumov91}) we consider a more simple technique based on the
two-time decoupling of multiparticle correlation functions in the projection operator
method~\cite{Plakida11}. In this method  the SCBA is used which is equivalent to the
noncrossing approximation  in the diagram technique. In this approximation, a propagation
of Fermi-type excitations described by operators $\, X_l^{\sigma' 2}, \,$ and Bose-type
excitations described by operators $B\sb{i\sigma\sigma'}$ on different lattice sites, $\,
l \neq i$, is assumed to be independent. Therefore, the time-dependent multiparticle
correlation functions in the self-energy operators (\ref{23}) can be written as a product
of fermionic and bosonic correlation functions,
\begin{eqnarray}
 &&\langle X_{l'}^{2\sigma''} B\sb{j\sigma\sigma''}^\dag
 |B\sb{i\sigma\sigma'}(t)
X_l^{\sigma' 2}(t)\rangle
\nonumber\\
 &= &\delta_{\sigma', \sigma''}\langle X_{l'}^{2\sigma'}
X_l^{\sigma' 2}(t)\rangle \langle B\sb{j\sigma\sigma'}^\dag
 |B\sb{i\sigma\sigma'}(t) \rangle,
\label{B5} \\
&& \langle  X_{l'}^{\bar{\sigma}'' 2} B\sb{j \bar{\sigma}\bar{\sigma}''}
 | B\sb{i\sigma \sigma'}(t) X_l^{\sigma' 2}(t)\rangle
\nonumber\\
& = &\delta_{\sigma', \sigma''}
 \langle  X_{l'}^{\bar{\sigma}' 2}
  X_l^{\sigma' 2}(t)\rangle\,
  \langle   B\sb{j\bar{\sigma}\bar{\sigma}'}
  B\sb{i\sigma \sigma'}(t) \rangle \, .
 \label{B6}
\end{eqnarray}
The time-dependent correlation functions are calculated self-consistently using the
corresponding GFs.
\par
In particular, the normal and anomalous diagonal components of the self-energy for the
two-particle subband are determined by the expressions
\begin{eqnarray}
 M\sp{22}({\bf k},\omega) &= &
   \frac{1}{N} \sum\sb{\bf q}
   \int\limits\sb{-\infty}\sp{+\infty} \!\!{\rm d}z\,
   K^{(+)}(\omega,z|{\bf q },{\bf k-q})
\label{29}   \\
& \times & \Big  \{ - ({1}/{\pi}) \mbox{Im} \left[
     G\sp{22}({\bf q},z) +
   G\sp{11}({\bf q},z) \right]  \Big \} , \quad
 \nonumber
\end{eqnarray}
\begin{eqnarray}
 \Phi\sb{\sigma}\sp{22}({\bf k},\omega) & = &
 \frac{1}{N} \sum\sb{\bf q}
   \int\limits\sb{-\infty}\sp{+\infty} \!\!{\rm d}z\,
   K^{(-)}(\omega,z|{\bf q },{\bf k-q})
\label{30} \\
& \times &\Big \{ - ({1}/{\pi}) \mbox{Im} \left[
   F\sb{\sigma}\sp{22}({\bf q},z) -
 F\sb{\sigma }\sp{11}({\bf q},z)
 \right]  \Big \} ,\quad
  \nonumber
\end{eqnarray}
where $ G\sp{\alpha\alpha}({\bf q},z)$ and $F\sb{\sigma }\sp{\alpha\alpha}({\bf q},z) $
are given by the diagonal components of the matrices (\ref{24}), (\ref{25}). Similar
expressions hold for the  self-energy components $M\sp{11}({\bf k},\omega)$ and
$\Phi\sb{\sigma}\sp{11}({\bf k},\omega)$ for the one-particle subband  (see
Ref.~\cite{Plakida97}). Note, that in the paramagnetic normal state the GF (\ref{24}) and
the self-energy (\ref{29}) do not depend on the spin $\sigma$.
\par
The kernel of the integral equations (\ref{29}), (\ref{30}) has a form, similar to the
strong-coupling Migdal-Eliashberg theory~\cite{Migdal58,Eliashberg60}:
\begin{eqnarray}
&&  K^{(\pm)}(\omega,z |{\bf q },{\bf k -q })  =
\frac{1}{\pi}\int\limits\sb{-\infty}\sp{+\infty} {\rm d}\Omega
  \, \frac{1 +N(\Omega) - n(z)}{\omega - z - \Omega}
\nonumber \\
& \times &   \Big \{ |t({\bf q})|^{2} {\rm Im} \chi\sb{sf}({\bf k- q},\Omega)
\nonumber \\
& \pm & \Big[ \left( |V({\bf k -q})|^2 + | t({\bf q})|^{2}/4 \right)
 {\rm Im}\, \chi_{cf}({\bf k-q}, \Omega)
 \nonumber \\
& + & |g({\bf k -q})|^2 {\rm Im} \chi_{ph}({\bf k-q}, \Omega)\Big]\; \Big \} .
  \label{31}
\end{eqnarray}
The  spectral densities of bosonic excitations are determined by the dynamic
susceptibility for spin $( sf)$, number (charge) $( cf)$, and lattice (phonon) $( ph)$
fluctuations
\begin{eqnarray}
\chi\sb{sf}({\bf q},\omega) & = &
  -  \langle\!\langle {\bf S\sb{q} | S\sb{-q}}
\rangle\!\rangle\sb{\omega},
\label{32a} \\
 \chi\sb{cf}({\bf q},\omega) &= &
 - \langle\!\langle \delta N\sb{\bf q} | \delta N\sb{-\bf q}
   \rangle\!\rangle\sb{\omega} ,
\label{32b}\\
\chi\sb{ph}({\bf q},\omega) &= &
 - \langle\!\langle  u\sb{\bf q} | u\sb{-\bf q}
   \rangle\!\rangle\sb{\omega} ,
\label{32c}
\end{eqnarray}
which are defined in terms of the commutator GFs~\cite{Zubarev60} for the spin ${\bf S
\sb{q}} $,  number $\delta N_{\bf q} = N_{\bf q} - \langle N_{\bf q} \rangle$, and
lattice displacement (phonon) $ u\sb{\bf q} $  operators.
\par
In the SCBA (\ref{B5}), (\ref{B6}),  vertex corrections to the kinematical  interaction
$\, t({\bf q}) \,$ of electrons with spin- and charge-fluctuations (\ref{32a}),
(\ref{32b}) induced by the intraband hopping are neglected. It is assumed that the system
is far away from a charge instability or a stripe formation and charge-fluctuations give
a small contribution to the pairing. The largest contribution from spin fluctuations
comes from wave vectors close the AF wave vector  ${\bf Q} = \pi\,(1,1)$ where their
energy $\omega_{s}({ \bf Q})$  is much smaller than the Fermi energy,  $\, \omega_{s}({
\bf Q}) /E_{\rm F} \ll 1\,$ (see, e.g.,~\cite{Vladimirov09}). Therefore, vertex
corrections to the kinematical interaction should be small as in Eliashberg
theory~\cite{Eliashberg60} for electron interaction with phonons, where $\,
\omega_{ph}/E_{\rm F} \ll  1\,$. Consequently, the SCBA for the self-energy and the GFs
calculated self-consistently is quite reliable and makes it possible to examine the
strong coupling regime  which is essential in study of renormalization of the QP spectrum
and  the superconducting pairing as shown in Refs.~\cite{Plakida07,Plakida13} and
discussed later.

\section{Results and discussion}
\label{sec:5}

\subsection{Self-consistent system of equations}
\label{sec:5a}

The self-consistent system of equations for the diagonal components of normal  GFs
(\ref{24}), (\ref{26}) and the self-energy  (\ref{29}) can be written in the
form~\cite{Plakida07}:
\begin{eqnarray}
 {G}^{11(22)}_{N}({\bf k},\omega)& =&
[1 - b({\bf k})] {G}_{1(2)}({\bf k},\omega)
\nonumber\\
 &+ & b({\bf k}){G}_{2(1)}({\bf k},\omega)  ,
 \label{33}
\end{eqnarray}
where the hybridization parameter $\,  b({\bf k}) = [{\varepsilon}_{2} ({\bf k}) -
\omega_{2}({\bf k})] / [{\varepsilon}_{2} ({\bf k}) - {\varepsilon}_{1} ({\bf k})]\, $. The two subband  GFs in Eq.~(\ref{33})  in  the imaginary frequency representation,  $ i\omega_{n}=i\pi T(2n+1)$, $n = 0,\pm 1, \pm 2, ...\, $, take the form:
\begin{eqnarray}
\{G_{1(2)}({\bf k},  \omega_{n})\}^{-1}= i\omega_n  - {\varepsilon}_{1(2)}({\bf k}) -
\Sigma({\bf k},\omega_n) .
 \label{45}
\end{eqnarray}
The self-energy $\Sigma({\bf k},\omega)$  can be approximated by the same  function for two subbands:
\begin{eqnarray}
{\Sigma}({\bf k}, \omega_{n}) & = & - \frac{T}{N}\sum_{\bf q}
 \sum_{m}\lambda^{(+)}({\bf q, k-q} \mid
\omega_{n}-\omega_{m})
\nonumber\\
 &\times &
   [{G}_{1}({\bf q}, \omega_{m})+
  {G}_2({\bf q}, \omega_{m})]
\nonumber\\
 &\equiv & i\omega_{n}\,[1-Z({\bf k},\omega_n)]
  + X({\bf k},\omega_n).
 \label{45a}
\end{eqnarray}
Density of  states (DOS) is determined by
\begin{equation}
A(\omega) = \frac{1}{N } \sum_{\bf k}\, {A}({\bf k}, \omega) ,
 \label{39}
\end{equation}
where the spectral function  reads
\begin{eqnarray}
 {A}({\bf k}, \omega) & = &  [Q_1 +  P({\bf k})]
 {A}_{1}({\bf k}, \omega)
\nonumber\\
 &+ &  [Q_2 -  P({\bf k})]{A}_{2}({\bf k}, \omega),
\label{40} \\
A_{1 (2)}({\bf k}, \omega) &=&
 - ({1}/{\pi})\, {\rm Im}{G}_{1 (2)}({\bf k},\omega).
\nonumber
\end{eqnarray}
Here the hybridization parameter $\,P({\bf k}) =(n-1) b({\bf k}) - 2 \sqrt{Q_1\, Q_2}$ $[
W({\bf k})/\Lambda({\bf k})]\,$ takes into account contributions  from both the diagonal
and off-diagonal components of the GF  (\ref{26}).

To calculate $T_c$ we can use a linear approximation for the pair GF (\ref{25}). In
particular, Eq.~(\ref{27})  for the two-particle subband gap $\varphi({\bf k},\omega)=
\sigma \varphi_{2, \sigma}({\bf k},\omega)$ can be written as
\begin{eqnarray}
\varphi({\bf k}, \omega_n) & = &
  \frac{T_c}{N}\sum_{\bf q} \,  \sum_{m}\,
\{\, J({\bf k-q}) - V({\bf k-q})
\nonumber \\
& +  & \lambda^{(-)}({\bf q, k-q} \mid \omega_{n}-\omega_{m}) \}
 \label{46} \\
 &\times & \frac{[1 - b({\bf q})]^2\,
  \varphi({\bf q}, \omega_{m})}{[\omega_m Z({\bf q},\omega_m)]^2
  + [{\varepsilon}_{2}({\bf q})+
X{\bf q},\omega_m)]^2}\; .
 \nonumber
 \end{eqnarray}
The  interaction functions in (\ref{45a}) and (\ref{46}) in the imaginary frequency
representation are given by
\begin{eqnarray}
 && \lambda^{(\pm)}({\bf q },{\bf k -q } | \nu_n)  = -
|t({\bf q})|^{2} \, \chi\sb{sf}({\bf k- q},\nu_n)
\nonumber \\
&& \mp \big\{\left[ |V({\bf k -q})|^2 + | t({\bf q})|^{2}/4 \right]\chi_{cf}({\bf k-q},
\nu_n)
\nonumber \\
&& + |g({\bf k -q})|^2 \, \chi_{ph}({\bf k-q}, \nu_n)
  \,  \big\}\, .
  \label{47}
\end{eqnarray}
Thus, we have derived  the self-consistent system of equations for the normal GF
(\ref{45}), the self-energy (\ref{45a}), and the gap function (\ref{46}).
\par
In the present theory we do not perform  self-consistent computation of spin and charge
excitation spectra  but adopt certain models for the spin (\ref{32a}), charge
(\ref{32b}), and phonon (\ref{32c}) susceptibility  in Eq.~(\ref{31}) or Eq.~(\ref{47}).
Since we consider the electronic spectrum only in the normal state and calculate
superconducting transition temperature $T_c$ from the linearized gap equation (\ref{46})
the feedback effects caused by opening a superconducting gap are not essential which
justifies usage of model functions for the susceptibility. Due to a large energy scale of
charge fluctuations, of the order of several $\,t$,  in comparison with the spin
excitation energy of the order of $J$, the charge fluctuation contributions  can be
considered in the static limit for the susceptibility (\ref{32b})
\begin{eqnarray}
\chi_{cf}({\bf k}) &=& \chi_{cf}^{(1)}({\bf k})
  + \chi_{cf}^{(2)}({\bf k}), \label{59} \\
\chi_{cf}^{(\alpha)}({\bf k}) &=& - \frac{1}{N}\sum_{\bf q}
 \frac{N^{(\alpha)}({\bf q +k}) - N^{(\alpha)}({\bf q})}
 {\varepsilon_{\alpha}({\bf q +k}) -
 \varepsilon_{\alpha}({\bf q})},
 \nonumber
\end{eqnarray}
where the   occupation numbers $ N^{(\alpha)}({\bf q})$ are defined as
\begin{eqnarray}
N^{(1)}({\bf k}) &=& [Q_1 + (n-1)b({\bf k})]\, {N}_{1}({\bf k}),
\nonumber \\
N^{(2)}({\bf k})  &=& [ Q_2 - (n-1)b({\bf k})]\, {N}_{2}({\bf k}),
\nonumber \\
{N}_{\alpha}({\bf k})&=&  ({1}/{2}) + T\sum_{m} \,
   G_{\alpha}({\bf k}, \omega_{m}).
  \label{38}
\end{eqnarray}
\par
For the dynamical spin susceptibility $\, \chi_{sf}({\bf q},\omega)$ (\ref{32a}) we used
a model suggested in Ref.~\cite{Jaklic95}
\begin{eqnarray}
  && {\rm Im}\, \chi_{sf}({\bf q},\omega+i0^+) =
 \chi_{sf}({\bf q}) \; \chi_{sf}''(\omega)
 \nonumber \\
& = &\frac {\chi_{ Q}}{1+ \xi^2 [1+ \gamma({\bf q})]} \;  \tanh \frac{\omega}{2T}
\frac{1}{1+(\omega/\omega_{s})^2}\, .
 \label{51}
\end{eqnarray}
Such type of the spin-excitation spectrum was found in the microscopic theory for the
$t$-$J$ model in Ref.~\cite{Vladimirov09}. The model is determined by two parameters: the
AF correlation length $\xi$ and  the cut-off energy of spin excitations of the order of
the exchange energy $\omega_s \sim J$. The strength of the spin-fluctuation interaction
given by the static susceptibility $\chi_{ Q} = \chi_{sf}({\bf Q}) $
\begin{equation}
\chi_{ Q} = \frac{3 (1- \delta)}{2\omega_{s} }
 \left\{ \frac{1}{N} \sum_{\bf q}
 \frac{1}{ 1+\xi^2[1+\gamma({\bf q})]} \right\}^{-1} ,
 \label{52}
 \end{equation}
is defined by the normalization condition:
\begin{eqnarray}
 \frac{1}{N} \sum_{\bf q} \int\limits_{0}^{\infty}
  \frac{d \omega}{\pi}  \coth\frac{\omega}{2T} \, {\rm Im}
  \chi_{sf}({\bf q},\omega)=\langle {\bf S}_{i}^2\rangle
  = \frac{3}{4}(1- \delta). &&
  \nonumber
\end{eqnarray}
The spin correlation functions (\ref{18}) in the single-particle excitation spectrum
(\ref{17}) are calculated using the same model (\ref{51}):
\begin{eqnarray}
C_1 &= & \frac{1}{N}   \sum_{\bf q}\, C_{\bf q}\, \gamma({\bf q}),\quad
 C_2 = \frac{1}{N}   \sum_{\bf q}\, C_{\bf q}\, \gamma'({\bf q}),
 \nonumber \\
 C_{\bf q} &=& \frac{\omega_{s}}{2}\,\frac{\chi_{ Q}}{1+\xi^2[1+ \gamma({\bf q})}
 \label{52a}
\end{eqnarray}
To estimate the contribution from phonons in Eq.~(\ref{31}) we consider  a model
susceptibility for optic phonons and the electron-phonon  matrix element in the form
similar to Ref.~\cite{Lichtenstein95}:
\begin{equation}
V_{ep}({\bf q},\omega)= |g({\bf q})|^2\chi_{ph}({\bf q},\omega) = g_{ep}\,
\frac{\omega_0^2}{\omega^2_{0} - \omega^2}\,S(q),
   \label{53}
\end{equation}
where $ g_{ep}\,$ is the ``bare''  matrix element for the short-range  electron-phonon
interaction, while the momentum dependence of the electron-phonon interaction is
determined by the  vertex correction $ S(q)\,$.  It takes into account a strong
suppression of charge fluctuations at small distances (large scattering momenta $q$)
induced by electron correlations as proposed in Ref.~\cite{Zeyher96}. For the vertex
function we take the model
\begin{equation}
 S(q)= \frac{1}{\kappa_1^2 + q^2}\equiv
 \frac{\xi_{ch}^2}{1 + \xi_{ch}^2 \,q^2},
    \label{53a}
\end{equation}
where the charge correlation length $\xi_{ch} = 1/\kappa_1\,$ determines the radius of a
``correlation hole''. Taking into account that $\xi_{ch} \sim a/\delta$~\cite{Zeyher96},
we can use the relation $\xi_{ch} = 1/ (2 \delta )$ in numerical computations. This gives
$\xi_{ch} \simeq 10$ for the underdoped case ($ \delta = 0.05 $) and $\xi_{ch}\simeq 2$
for the overdoped case ($ \delta = 0.25 $). We assume a strong   electron-phonon
interaction $g_{ep} = 5\, t $ and take  $\omega_{0} = 0.1\,t$.

For the intersite CI $\,V_{ij} \,$ we consider a model for repulsion of two electrons
(holes) on neighbor lattice sites,
\begin{equation}
V({\bf q}) = 2 V\, (\cos q_x + \cos q_y )\, ,
 \label{50}
\end{equation}
with various values of  $\,  V = 0.0,\, 0.5\,t,\, 1.0\,t\,$ and $  2.0 \,t\,$. Note, that
in cuprates the  intersite CI (\ref{50}) is quite small, $\,V \lesssim 0.5\,
t$~\cite{Feiner96}. The 2D screened CI model  $\, V_c({\bf q}) =  u_c /(|{\bf q}|
  + \kappa )$  was also considered in Ref.~\cite{Plakida13}.
 Most of the calculations are done for $\, U = 8 \,t$ while several
results are presented for $\, U = 4\, t, \, 16 \,t$ and $32\, t$. The AF exchange
interaction for neighbor sites is described by the function $\,J({\bf q})= 2 J\, (\cos
q_x + \cos q_y ) $ with $J = 0.4t$. In the GMFA the CI $V_{ij}$ gives no contribution to
the exchange interaction $\, J_{i j}$ and therefore it is assumed to be the same for all
values of $V\,$ (cf. with Ref.~\cite{Plekhanov03}). In computations we use the following
hoping parameters $\, t' = - 0.2 \, t, \quad t'' = 0.10 \, t$ where $t = 0.4$~eV is the
energy unit.

\subsection{Electronic spectrum in the normal state}
\label{sec:5b}

\begin{figure}[h!]
\includegraphics[width=0.35\textwidth]{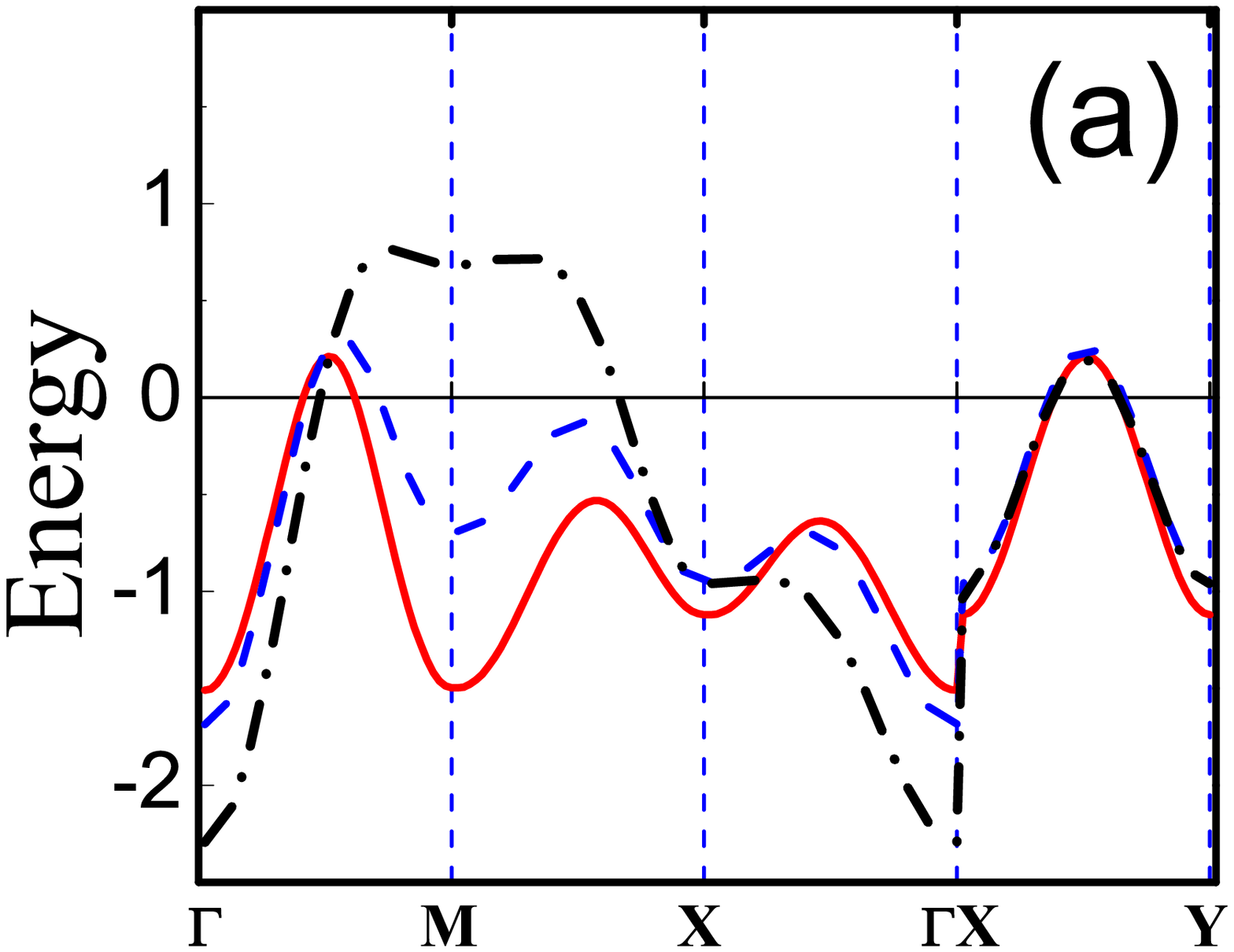}
\hspace{5mm}
\includegraphics[width=0.35\textwidth]{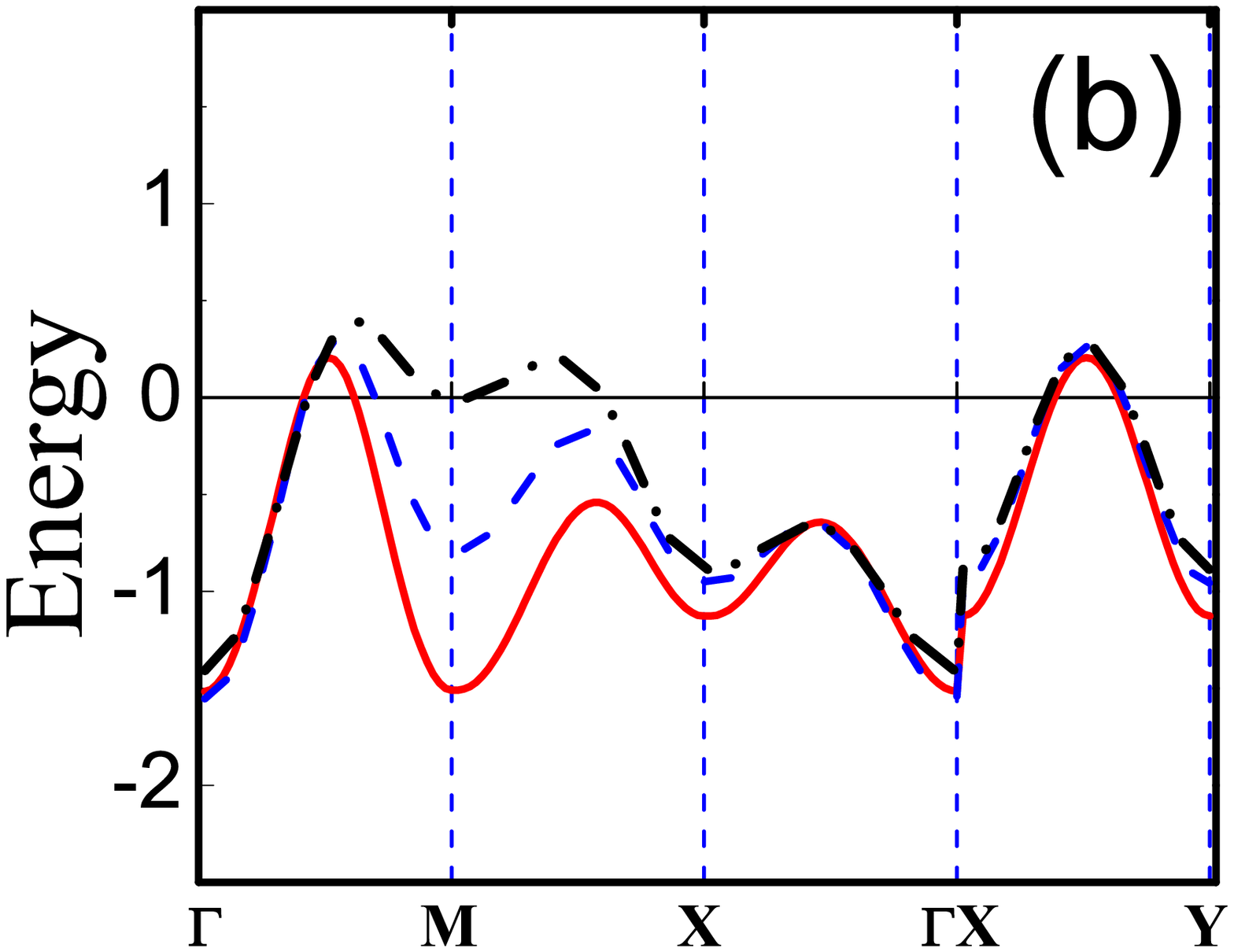}
 \caption{(Color online) Electron dispersion in the GMFA
${\varepsilon}_{2} ({\bf k})$ for (a) $V=0$ and (b)  $V=2$ at $U = 8$  along the symmetry
directions $\Gamma(0, 0)\rightarrow M(\pi,\pi) \rightarrow X (\pi, 0) \rightarrow
\Gamma(0, 0)$ and $X (\pi, 0) \rightarrow Y(0, \pi)$ for $\delta = 0.05\,$ (red solid
line), $0.10$ (blue dashed  line), and $0.25$ (black dash-dotted line). Fermi energy for
hole doping is at $\omega = 0$.}
 \label{fig12}
\end{figure}

\begin{figure}[ht!]
\includegraphics[width=0.32\textwidth]{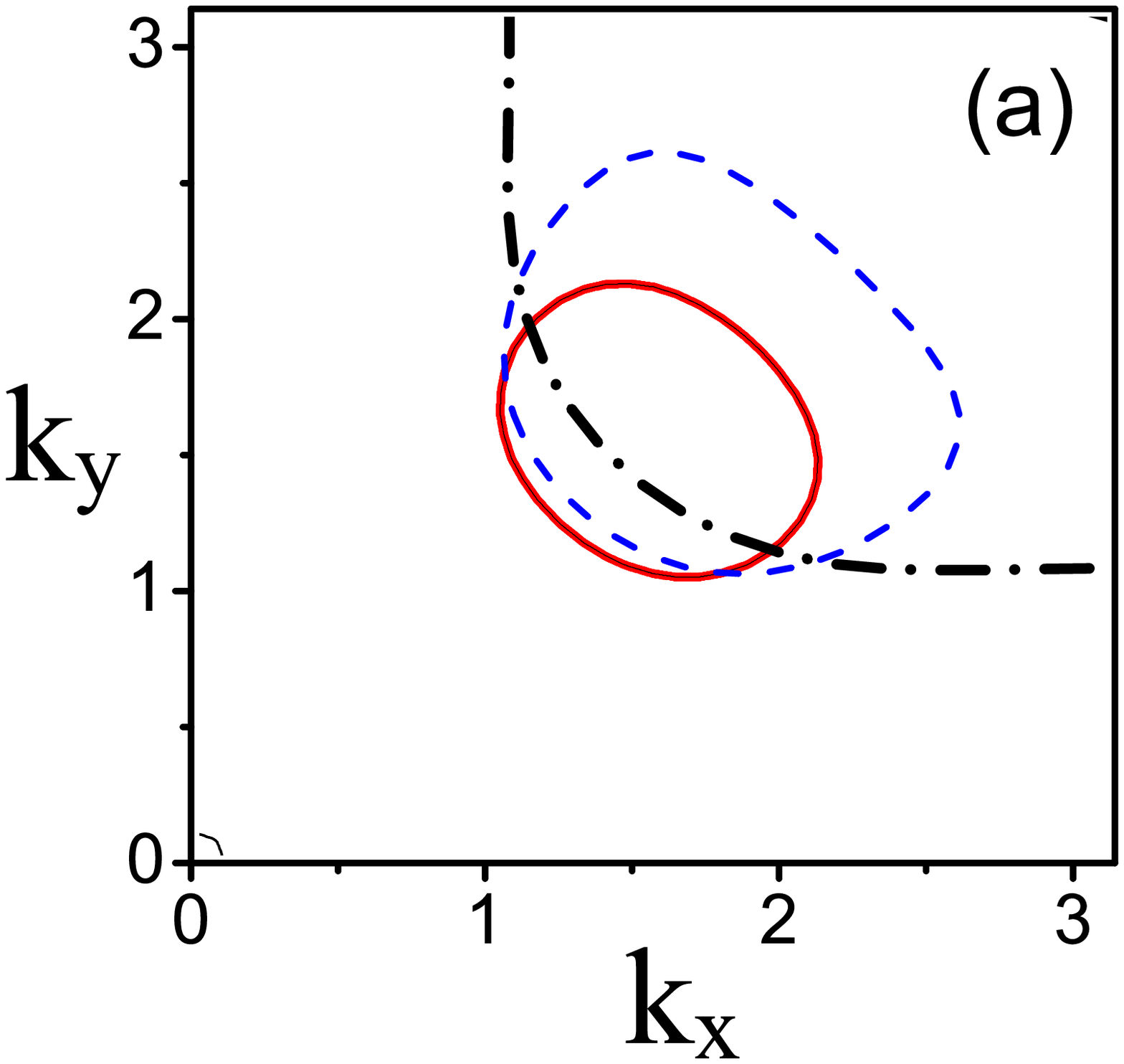}
\hspace{5mm}
\includegraphics[width=0.32\textwidth]{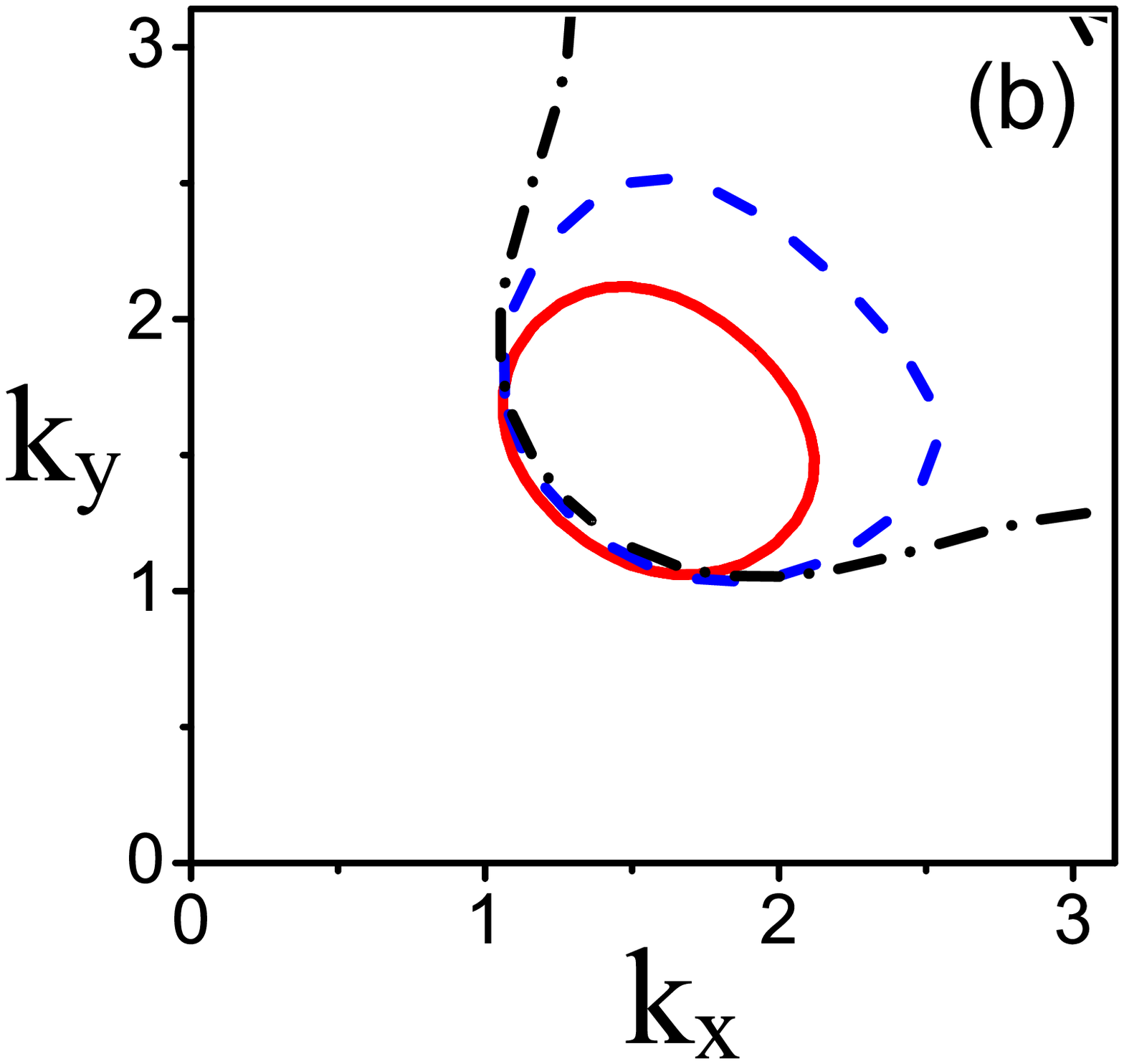}
\caption{(Color online)  Fermi surface for (a) $V=0$ and (b) $V=2$ at $U = 8$ in  the
quarter of the BZ in the GMFA at hole doping $\delta = 0.05$ (red solid line), $ 0.10$
(blue dashed line), and $ 0.25$ (black dash-dotted line).}
 \label{fig13}.
\end{figure}

First we consider  results in the GMFA for the electronic spectrum (\ref{17}). The doping
dependence of the electron dispersion for the two-hole subband  ${\varepsilon}_{2} ({\bf
k})$  along the symmetry directions in the 2D Brillouin zone (BZ) is shown in
Fig.~\ref{fig12}  for $U = 8$ for $V = 0$ (a) and for $V = 2$ (b).  The corresponding FS
determined by the equation: $\, \varepsilon_{2}({\bf k_{\rm F}}) = 0 \,$ is plotted in
Fig.~\ref{fig13}. For small doping, $\, \delta = 0.05$, the energy at the $M(\pi,\pi)$ and
$\Gamma(0, 0)$ points are nearly equal as in the AF phase. Only small hole-like FS
pockets close to the $(\pm \pi/2,\pm \pi/2)$ points emerge at this doping as shown in
Figs.~\ref{fig12}, \ref{fig13}. With increasing doping, the AF correlation length decreases
that results in increasing of the electron energy at the $M(\pi,\pi)$ point and at some
critical doping $\delta \sim 0.12$ a large FS emerges. The doping dependence  of the AF
correlation length $\xi$  on several values of   hole concentrations $\delta$ are given
in Table~\ref{Table1}.
\begin{table}
\caption{ AF correlation length $\xi$,  spin correlation functions $C_1,\, C_2$,  projected spin susceptibility $\widehat{\chi}\sb{sf}$, and
electron-phonon interaction parameter $\widehat{V}_{ep}$ for several values of hole
concentration $\delta$. }
 \label{Table1}
\begin{tabular}{crrrrc}\\
\hline
 $\quad  \delta\; $  &   $\; \xi/a  \quad $ &   $ \; C_1 \quad $ &
  $ \; C_2 \quad  $ &
   $ \; - \widehat{\chi}\sb{sf} \,\cdot t \; $ &
   $ \;  \widehat{V}_{ep} \,/ t \; $\\
\hline
 0.05 & $\;$ 3.4 $\;$ &  - 0.26 $\;$ &   0.16 $\;$& 1.32 $\;$ & 1.96 \\
 0.10 & $\;$ 2.4 $\;$ & - 0.20  $\;$  & 0.11 $\;$  &
 1.05 $\;$ & 1.4 \\
0.25  & $\;$  1.5 $\;$ & - 0.12 $\;$  &  0.05 $\;$ & 0.61 $\;$& 0.76  \\
\hline
\end{tabular}
\end{table}
A remarkable feature is that the part of the FS close to the $\Gamma(0, 0)$ point in the
nodal direction in Fig.~\ref{fig13} does not shift much with doping (or temperature) being
pinned to a large FS as observed in ARPES experiments (see, e.g.~\cite{Kordyuk05}). At
the same time, the renormalized two-hole subband width increases with doping, as e.g. for
$U=8$ and $V = 0$ from $\widetilde{W} \approx 2\,t$ at $\, \delta = 0.05\,$ to
$\widetilde{W} \approx 3\,t$ at $\, \delta = 0.25$, which, however, remains less than the
``bare'' Hubbard subband width $W = 4 t\,(1+\delta)$ where short-range AF correlations
are disregarded. Note that in the dynamical mean field theory (DMFT) this narrowing of
the subbands due to the short-range AF correlations is missed~\cite{Georges96,Kotliar06},
while they are partly taken into account  in the cluster DMFT~\cite{Haule07}. With
increasing  $V$ the subband width shrinks as seen from comparison panels (a) and (b)
in Figs.~\ref{fig12} and  in Figs.~\ref{fig13}.

\begin{figure}
\includegraphics[width=0.35\textwidth]{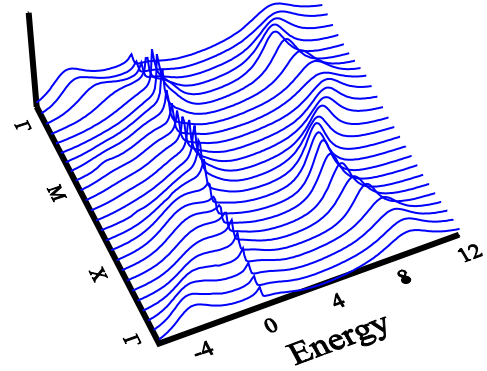}
 \caption{(Color online)  Spectral function  in the SCA along the symmetry directions $\Gamma(0, 0)\rightarrow M(\pi,\pi)
\rightarrow X (\pi, 0) \rightarrow \Gamma(0, 0)$ for hole concentration $\delta = 0.10$.}
 \label{fig14}
\end{figure}

\begin{figure}
\includegraphics[width=0.35\textwidth]{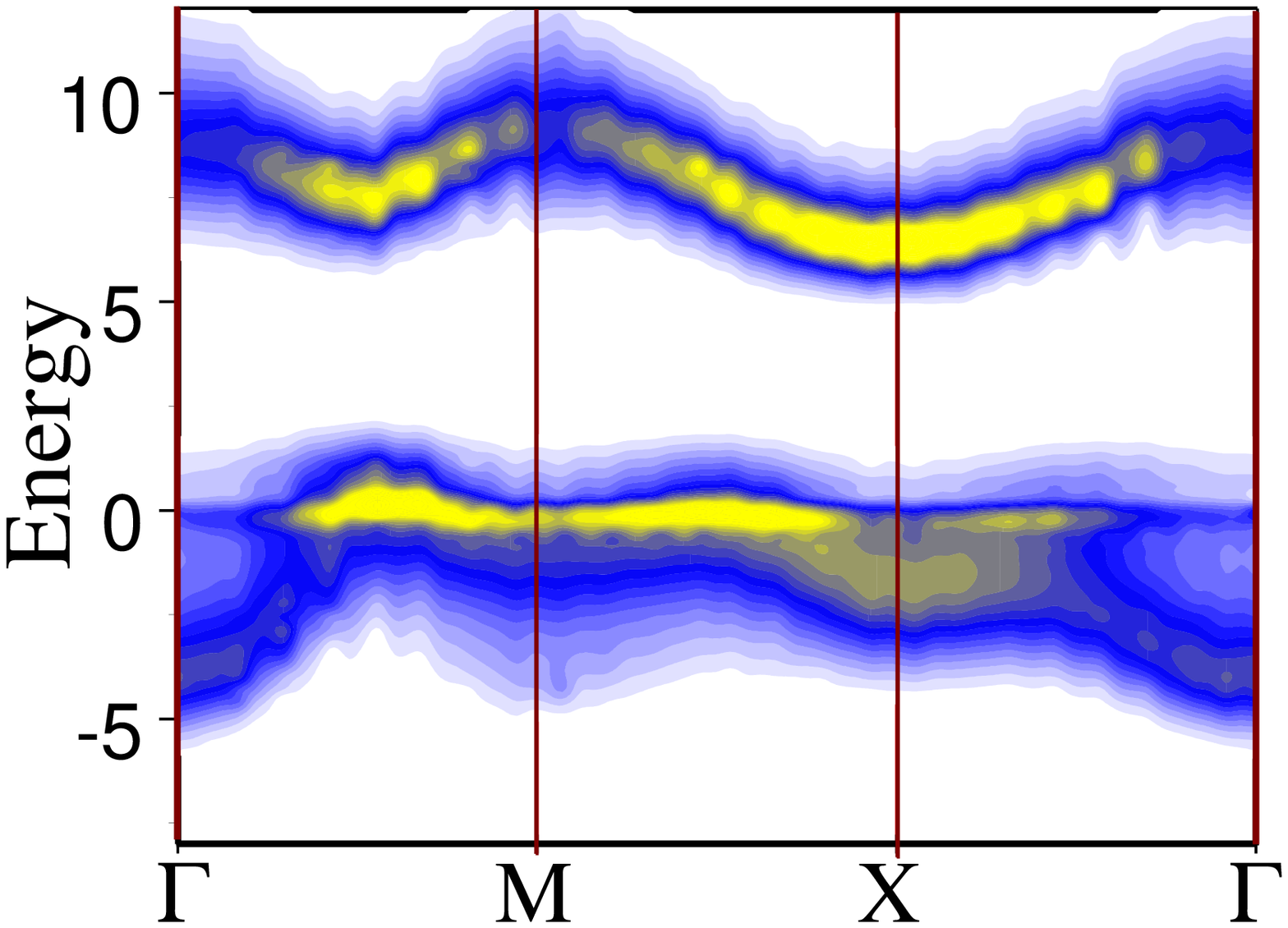}
 \caption{(Color online)  Electron dispersion curves  in the SCA along the
symmetry directions $\Gamma(0, 0)\rightarrow M(\pi,\pi) \rightarrow X (\pi, 0)
\rightarrow \Gamma(0, 0)$ for hole concentration $\delta = 0.10$.}
 \label{fig15}
\end{figure}

\begin{figure}
\includegraphics[width=0.35\textwidth]{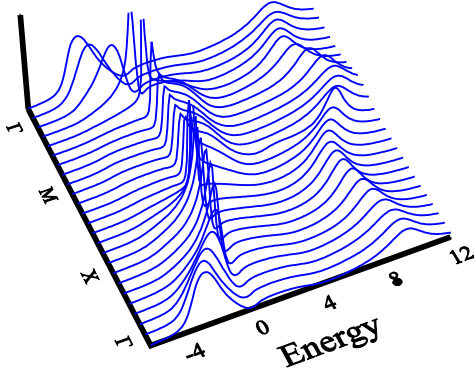}
 \caption{(Color online)  Spectral function  in the SCA along the symmetry directions  for hole concentration $\delta =
0.25$.}
 \label{fig16}
\end{figure}

\begin{figure}
\includegraphics[width=0.35\textwidth]{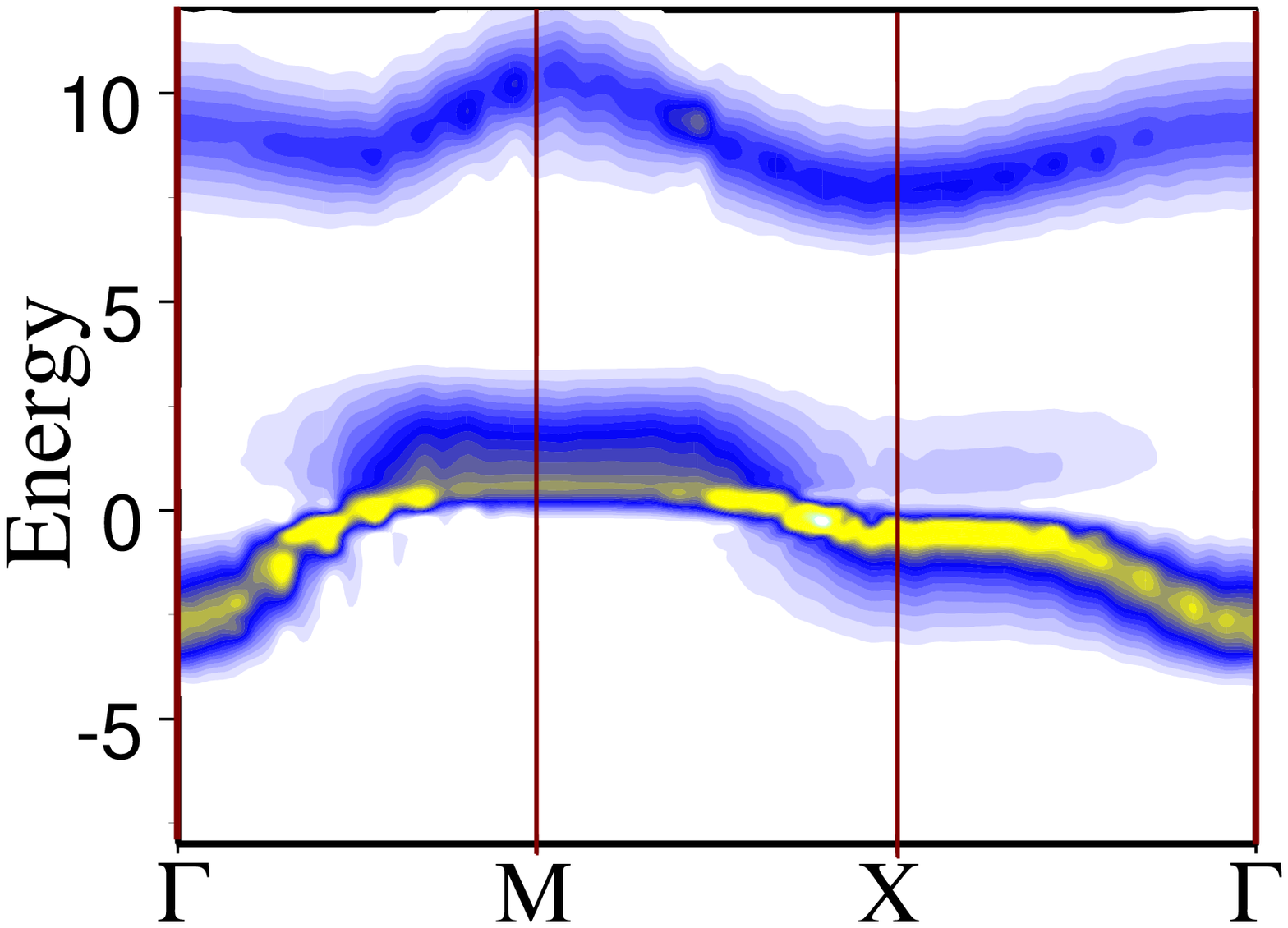}
 \caption{(Color online)  Electron dispersion curves  in the SCA along the
symmetry directions  for hole concentration $\delta = 0.25$.}
 \label{fig17}
\end{figure}

To consider the self-energy effects in  the electronic spectrum a strong coupling
approximation (SCA) should be considered by  a self-consistent solution of the system of
equations for the normal GF (\ref{45}) and the self-energy (\ref{45a}). In
Ref.~\cite{Plakida07} a detailed investigation of the normal state electronic spectrum
for the conventional Hubbard model in SCA  was performed. Here we present only
the  results of the electronic spectrum computation for the  model (\ref{2a}) which are important for further studies of superconductivity in the model. The  spectral function $\, A({\bf k}, \omega ) $ (\ref{40}) along the symmetry directions is presented  in Fig.~\ref{fig14} and the dispersion curves given by the maximum of this spectral function  is displayed in Fig.~\ref{fig15} for the  doping $\delta = 0.10$. In Fig.~\ref{fig16} and Fig.~\ref{fig17} we plot the  spectral function and the dispersion curves, respectively,   for the  doping $\delta =  0.25$.
In comparison with  the GMFA in Fig.~\ref{fig12}, a rather flat energy dispersion is found with QP peaks at the FS. In general, strong increase of the dispersion and  intensity of the QP peaks is observed in the overdoped region in  comparison with the underdoped region. This is in agreement with our detailed studies of temperature and doping dependence of the self-energy (\ref{45a}) and spectral function (\ref{40})
in~\cite{Plakida07} which have proved  strong influence of AF spin-correlations on the spectra.

\begin{figure}[!ht]
\includegraphics[width=0.32\textwidth]{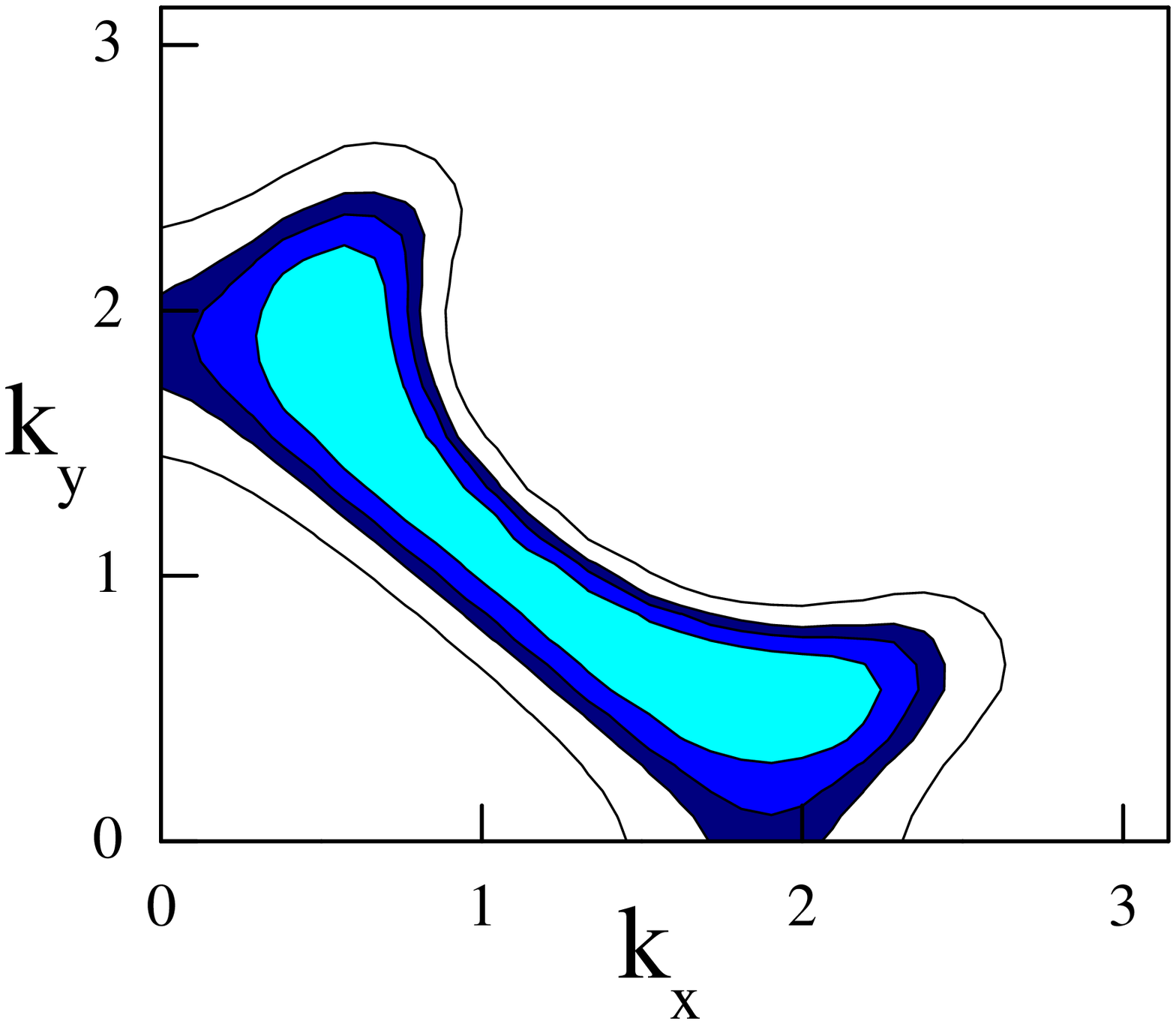}
 \caption{(Color online) $A({\bf k},\omega =0)$ on the FS at $\delta = 0.05$ at
$T=0.03t$  for $U = 4t$.}
 \label{fig18}
\end{figure}

\begin{figure}[!ht]
\includegraphics[width=0.32\textwidth]{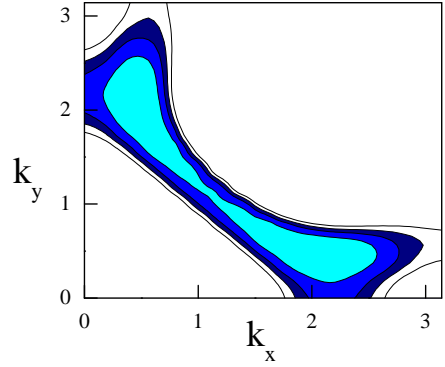}
 \caption{(Color online) $A({\bf k},\omega =0)$ on the FS
 $\delta = 0.1$  at  $T=0.03t$  for $U = 4t$.}
 \label{fig19}
\end{figure}

Noticeable  changes are observed for the FS in the SCA shown in Figs.~\ref{fig18}
and~\ref{fig19} which are determined by the equation for the spectral density (\ref{40}) at the Fermi energy, $\, A({\bf k}, \omega = 0) $. Whereas in GMFA  at small doping the FS in Fig.~\ref{fig13} shows  closed pockets, in SCA  the FS  reveals arc-type behavior detected in ARPES experiments  where spectral function $\, A({\bf k}, \omega =0 ) $  is measured (see, e.g., Ref.~\cite{Shen05}).

The arc-type behavior is related to  formation of the pseudogap in the electronic spectrum induced by the AF spin-fluctuations as was found  in the two-particle self-consistent approach (TPSC) ~\cite{Vilk95,Tremblay06}  or the  model of short-range static spin (charge) fluctuations -- the $\Sigma_{\bf k}$-model~\cite{Sadovskii01,Sadovskii05}. To prove this, let us consider  the static limit for the interaction (\ref{47}) by taking into account only  zero Matsubara frequency $i\omega_{\nu} =0$  which gives $i\omega_{m}=
i\omega_{n}$ in  (\ref{45a}). In the limit of the large  AF correlation length $\xi
\gg 1 $ the static spin susceptibility $\chi_{s}({\bf q})$ in~(\ref{51}) shows a
sharp peak close to the AF wave-vector ${\bf Q} = \pi(1,1)$ and can be expanded
over the small wave-vector ${\bf p = q - Q}$:
\begin{equation}
 \chi_{s}({\bf  q})\simeq \, \frac {\chi_Q }
 {1 + \xi^2 \,{\bf p}^2}
  \simeq \frac {A}{\kappa^{2}+{\bf p}^2} ,
 \label{ar1}
\end{equation}
where we introduced $\kappa = \xi^{-1} $ and took into account that the constant
(\ref{52}) $\chi_Q \simeq A\,\xi^2$  with $\, A=
  ({6\pi}/{\omega_{s}})   [\ln(1 + 4\pi \, \xi^2)]^{-1}$ for the square lattice.
In this limit we get the following equation for the self-energy
 (\ref{45a}):
 \begin{eqnarray}
 &&{\Sigma}({\bf k}, i\omega_{n}) \simeq A\,|t({\bf k-Q})|^{2}
 \, \frac{T}{N}\sum_{\bf p} \,\frac {1}
{{\kappa^{2}+ p^2}}
  \nonumber \\
 && \times  [{G}_{1}({\bf k - Q -p}, i\omega_{n}) +
{G}_{2}({\bf k-Q -p}, i\omega_{n})] .
 \label{ar7}
\end{eqnarray}
Expanding the QP energy $\,\varepsilon_{1 (2)}({\bf k-Q -p}) \simeq
\varepsilon_{1 (2)}({\bf k-Q}) - {\bf p \cdot v}_{1 (2), \bf k-Q}\,$ we obtain
for the GFs in (\ref{ar7}) the following representation:
\begin{eqnarray}
&&{G}_{1(2)}({\bf k - Q -p}, i\omega_{n})
 \simeq \{i\omega_n - {\varepsilon}_{1 (2)} ({\bf k- Q})
\nonumber \\
&+&
 {\bf p \cdot v}_{1 (2), \bf k-Q}-
  \Sigma({\bf k -Q},i\omega_n)\}^{-1} .
  \label{ar6}
\end{eqnarray}
The system of equations for the GFs (\ref{ar6}) and the self-energy (\ref{ar7})
is similar to those one derived in the TPSC  approach~\cite{Vilk95}) and the
$\Sigma_{\bf k}$-model~\cite{Sadovskii01,Sadovskii05,Kuchinskii05} apart from the interaction function and the two-subband system of equations in our case. The coupling constant in Eq.~(\ref{ar7}) is determined by the hopping parameter $|t({\bf k-Q})|^2 $, while in the TPSC and in  the $\Sigma_{\bf k}$-model the coupling constant is related to the Coulomb scattering, $g^2 = U^2 ( \langle n_{i\uparrow} n_{i
\downarrow}\rangle / n^2 )\langle (n_{i\uparrow} - n_{i
\downarrow})^2\rangle $. However, the values
of these constants are close: the averaged over the BZ value $
\langle\sqrt{|t({\bf k})|^2}\rangle_{{\bf k}} \sim 2t$ is comparable   with the
coupling constant  used in~\cite{Sadovskii05}.  As in the TPSC theory, in the limit
$\xi \rightarrow \infty $  the AF gap $\Delta_{AF}({\bf k}) \propto |t({\bf
k-Q})|^{2} $ in the QP spectra emerges in the subband located at the Fermi
energy. This result readily follows from the self-consistent equations for the
GF (\ref{45}) with the self-energy (\ref{ar7}) where in the right-had side the GF
(\ref{ar6}) is taken at ${\bf p} =0$. Thus, in our approach the pseudogap
formation is mediated by the AF short-range order similar to TPSC theory and the
model of short-range static spin fluctuations in the generalized
DMFT~\cite{Kuchinskii06}.

\begin{figure}[!]
\includegraphics[width=0.35\textwidth]{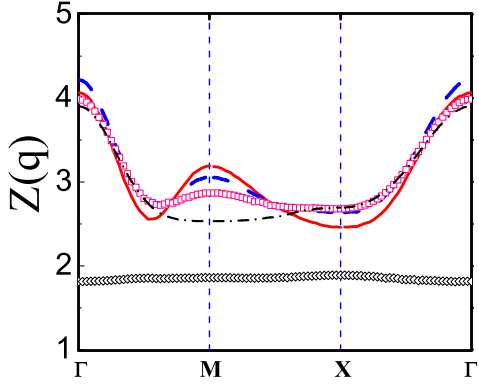}
\caption{(Color online) Doping dependence of the renormalization parameter $Z({\bf q}) $
along the symmetry directions $\Gamma(0, 0)\rightarrow M(\pi,\pi) \rightarrow X (\pi, 0)
\rightarrow \Gamma(0, 0)$ at $T \approx 140$~K for $\delta = 0.05$ (red solid line),
$\delta = 0.10$ (blue dashed line), $\delta = 0.15$ (red squares), $\delta = 0.25$ (black
dash-dotted line), and $\delta =  0.35$ (black diamonds)}
 \label{fig20}
\end{figure}

\begin{figure}[!]
\includegraphics[width=0.35\textwidth]{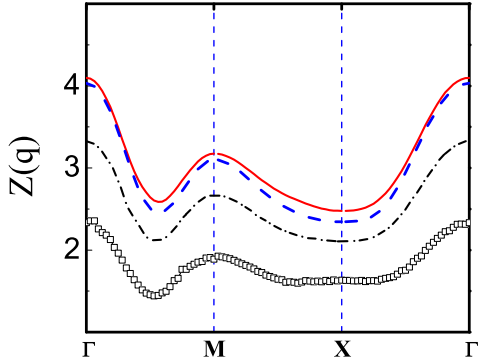}
\caption{(Color online)  Temperature  dependence of the renormalization parameter $Z({\bf
q}) $ for $\delta = 0.05$  at $T \approx 140$~K (red solid line),  $T \approx 580$~K
(black dash-dotted line), and $T \approx 1100$~K  (black  squares). Blue dashed line
shows $Z({\bf q}) $ for $\delta = 0.05$   caused only by the spin-fluctuation
contribution.}
 \label{fig21}
\end{figure}

It is important to note that the superconducting $T_c$  in the gap equation (\ref{46}) considerably depends on the quasiparticle weight determined by the renormalization  parameter $Z({\bf q}, \omega) $ in the self-energy  (\ref{45a}). We estimate it at the Fermi energy
 \begin{eqnarray}
 Z({\bf q})& = &Z({\bf q}, \omega =0)= 1 + \lambda({\bf q})
 \nonumber \\
 & =&    1 - [\, d \,{\rm Re}\,
  \Sigma({\bf q}, \omega)/{d \omega}]|_{\omega = 0},
 \label{45ba}
 \end{eqnarray}
where we introduced the coupling constant $\lambda({\bf q})$.  The doping dependence of  $Z({\bf q}) $   is shown  in Fig.~\ref{fig20}. It weakly depends on $\delta$ in the underdoped case for $\delta \lesssim 0.15$ but sharply decreases in the overdoped case for $\delta \gtrsim 0.25$. The temperature dependence of $Z({\bf q}) $ presented   in Fig.~\ref{fig21} is weak  at temperatures lower than the characteristic energy of spin fluctuations $\omega_s \sim J$. The electron-phonon interaction gives a small contribution to the coupling constant as follows from the comparison of $Z({\bf q}) $ induced by both spin-fluctuations and electron-phonon interaction contributions (red solid line) with the contribution caused only  by spin-fluctuations  (blue dashed line).

Thus, studies of the normal state electronic spectrum revealed a major role  of AF
correlations and  self-energy effects  in the renormalization of the spectrum.  The
latter show a noticeable reduction of the QP weight at low doping that strongly
suppresses superconducting pairing as we demonstrate below.

\subsection{Superconducting gap and $ \bf T_c$}
\label{sec:5c}

For a comparison of various contributions  to the superconducting gap equation
(\ref{46}),  we approximate  the interaction (\ref{47}) by   its value close to the Fermi
energy. As the result instead of the dynamical susceptibility (\ref{32a}), (\ref{32b})
the static susceptibility  $\chi({\bf q}) ={\rm Re} \, \chi({\bf q},\Omega = 0) $ appears
in the gap equation. It brings us to the BCS-type equation for the gap function
(\ref{46}) at the Fermi energy $\varphi({\bf k}) = \varphi({\bf k},\omega=0)$:
\begin{eqnarray}
&&\varphi({\bf k}) =
  \frac{1}{N}\sum_{{\bf q}} \,
  \frac{[1 - b({\bf q})]^2\,
  \varphi({\bf q})}{[Z({\bf q})]^2 \; 2\widetilde{\varepsilon}({\bf q})}
    \tanh\frac{\widetilde{\varepsilon}({\bf q})}{2T_c}
 \big\{ J({\bf k-q})
\nonumber \\
&& - V({\bf k-q})+ \big[(1/4) |t({\bf q})|\sp{2} + |V({\bf k -q})|^2\big]
\chi\sb{cf}({\bf k -q})
 \nonumber \\
&& + |g({\bf k -q})|^2 \, \chi_{ph}({\bf k-q})\,\theta(\omega_{0} -| \tilde{\varepsilon}({\bf
q})|)
\nonumber \\
&& -  |t({\bf q})|\sp{2}\; \chi_{sf}({\bf k -q})\theta(\omega_{s} -| \tilde{\varepsilon}({\bf
q})|) \big\}\, ,
     \label{55}
\end{eqnarray}
where $\,  \widetilde{\varepsilon}({\bf q}) = \varepsilon_{2}({\bf q})/Z({\bf q}) $ is
the renormalized energy. Whereas for the exchange interaction and CI there are no
retardation effects and the pairing occurs for  all electrons in the two-particle
subband, the electron-phonon interaction and spin-fluctuation contributions are
restricted  to the range of energies  $\,\pm \omega_0\, $ and  $\,\pm \omega_s$,
respectively,  near the FS as determined by the $\theta$-functions.
\par
To estimate various contributions in the gap equation (\ref{55}) we consider a model
$d$-wave gap function, $\varphi({\bf k}) = (\Delta/2) \,\eta({\bf k})$ where $ \eta({\bf k}) = (\cos k_x - \cos k_y)$. Integrating Eq.~(\ref{55}) with the function
$\eta({\bf k})$ over $ {\bf k}$ we obtain the  gap equation  in the form:
\begin{eqnarray}
 1 &=& \frac{1}{N } \sum_{\bf q}\frac{[1 - b({\bf q})]^2\,
  [\eta({\bf q})]^2}{[Z({\bf q})]^2 \;
  2\widetilde{\varepsilon}({\bf q})}
  \tanh\frac{\widetilde{\varepsilon}({\bf q})}{2T_c}
  \big\{ J - V
 \nonumber \\
 & + &\widehat{V}\sb{cf}  +(1/4)\,|t({\bf q})|\sp{2} \widehat{\chi}\sb{cf}
  + \widehat{V}_{ep}\,\theta(\omega_{0} -| \widetilde{\varepsilon}({\bf q})|)
\nonumber \\
& - &  |t({\bf q})|^{2}\, \widehat{\chi}\sb{sf}
   \theta(\omega_{s} -| \widetilde{\varepsilon}({\bf q})|) \big\}.
\label{56}
\end{eqnarray}
In this equation only $l=2$ components of the static susceptibility  and CI give
contributions
\begin{eqnarray}
 \widehat{V}\sb{cf} & = & \frac{1 }{N}\sum_{\bf k}|V({\bf k})|^2
 \, \chi\sb{cf}({\bf k})\, \cos k_x ,
\label{57b} \\
\widehat{\chi}\sb{cf} & = &\frac{1}{N}\sum_{\bf k}
 \chi\sb{cf}({\bf k}) \cos k_x ,
\label{57c} \\
   \widehat{V}_{ep} & = &  \frac{g_{ep}}{N}\sum_{\bf k}
 S({\bf k})\cos k_x,
 \label{57d} \\
  \widehat{\chi}\sb{sf} & = &  \frac{1}{N}\sum_{\bf k}
 \chi\sb{sf}({\bf k})\cos k_x \, .
\label{57e}
\end{eqnarray}
The contribution from the charge fluctuations $\,\widehat{\chi}\sb{cf} \,$ (\ref{57c})
weakly depends on $U$ and $V$ and is very small: $\, \widehat{\chi}\sb{cf}\sim
10^{-3}\,(1/t)\, - 10^{-2}\,(1/t)\, $ for  hole concentrations $\delta = 0.05 - 0.10$,
respectively. For the averaged over the BZ vertex $ \;\overline{|t({\bf q})|^2} =
(1/N)\sum_{\bf q}|t({\bf q})|^2 \simeq 4\,t^2 $ the contribution induced by the
kinematical interaction is equal to $\overline{|t({\bf q})|^2}\,\widehat{\chi}\sb{cf}
\lesssim 0.04\, t \,$ and can be neglected.  The charge fluctuation contribution $\,
\widehat{V}_{cf}\,$ (\ref{57b}) from the intersite CI (\ref{50}) for the hole
concentration $\delta = 0.05$ is  also small, $\; \widehat{V}_{cf} \lesssim 5 \cdot
10^{-2}\, t\; $ and $\, \widehat{V}_{cf} - V <0\,$ for all values of   $\,U $ and $\, V$
and consequently,  the $d$-wave pairing induced only by the charge fluctuations cannot occur.

The electron-phonon interaction contribution (\ref{57d}) for the electron-phonon
interaction model (\ref{53}) even for a strong electron-phonon interaction coupling
$g_{ep} = 5 t $ is quite small for the $d$-wave pairing, $\,\widehat{V}_{ep} = 2\,t
\,(0.8\, t)\,$ for $\delta = 0.05 \,( 0.25)$,  as shown in Table~\ref{Table1}. The
electron-phonon interaction contribution to the $s$-wave pairing is given by the $l = 0$
component $ S_0 =(1/N)\sum_{\bf q}\, S(q)= 0.31\; (0.57)\,$ for $\xi_{ch} = 2\; (10)$,
respectively. The ratio of the $d$-wave $S_d$ and the $s$-wave $ S_0$ components of the
electron-phonon  matrix elements is equal to $\,(S_d/ S_0) = 0.43\; ( 0.60)$ for
$\xi_{ch} =2 \; (10)$, respectively. This shows that at small hole concentrations
$\delta$ (large charge correlation lengths $\xi_{ch} = 1/ 2\delta $)  the electron-phonon
interaction for the both components are comparable, while for the overdoped case the
$d$-wave component $S_d$ becomes considerably smaller than the $s$-wave component in
agreement with the results of Ref.~\cite{Zeyher96}.

The spin-fluctuation contribution $\, \widehat{\chi}\sb{sf}\,$ (\ref{57e}) is calculated
for the model $\chi_{sf}({\bf q})$ in Eq.~(\ref{51}). Since the spin susceptibility has a
maximum at the AF wave vector ${\bf Q} = \pi\,(1,1)$ the integral over ${\bf k}$ in
(\ref{57e}) results in the negative value for $\, \widehat{\chi}\sb{sf}\,$ which strongly
depends on hole doping as shown in Table~\ref{Table1}. Using the averaged over BZ vertex
$\, \overline{|t({\bf q})|^2} \simeq 4\,t^2\,$ we can estimate an effective
spin-fluctuation coupling constant as $\, g_{sf} \simeq - 4\,t^2 \,\widehat{\chi}\sb{sf}
= 5\,t - 2\, t $ for $\delta = 0.05 -0.25$.  Thus, the spin-fluctuation contribution to
the pairing in Eq.~(\ref{56}) with the coupling constant $\, g_{sf} $ appears to be the
largest.

\begin{figure}
\includegraphics[width=0.35\textwidth]{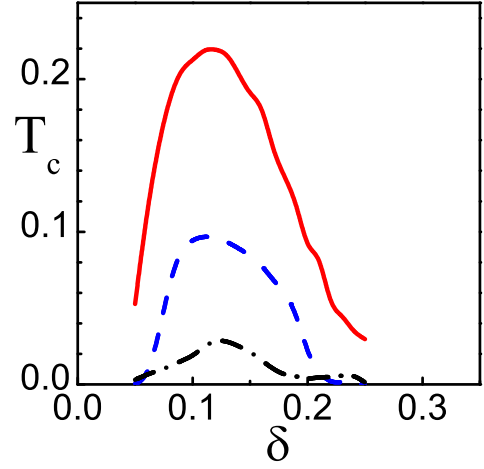}
\caption{(Color online) $T_c(\delta)$ in the WCA  induced by all interactions (red solid
line) and only by the spin-fluctuation contribution
 (blue dashed line) or only by the electron-phonon interaction  $\, \widehat{V}\sb{ep}\,$ (black dash-dotted line).}
 \label{fig22}
\end{figure}

First we calculate  doping dependence of $T_c$ in the weak-coupling approximation (WCA)
for $ Z({\bf q})= 1$ in Eq.~(\ref{56})  by taking into account the exchange interaction
$J$, the Coulomb repulsion $V$, and the contributions from the self-energy
$\, \widehat{V}\sb{ep}$, $\widehat{\chi}\sb{sf}$, and  $\, \widehat{V}\sb{cf} $,
neglecting the small contribution $\widehat{\chi}\sb{cf}$. Results of the calculation is
shown in Fig.~\ref{fig22}. The highest $T_c \approx 0.22\, t $ is found when all the
contributions are taken into account. The spin-fluctuation pairing results in
superconducting $T^{sf}_c \approx 0.1\, t $ much larger than $T^{ep}_c \approx 0.012 t$
mediated by the electron-phonon interaction. For the ${\bf k}$-independent
electron-phonon interaction $\, (S({\bf k}) =1)\,$  there is no contribution to the $d$-wave pairing. The
doping dependence of $T_c$ is qualitatively agree with experiments in cuprates but its
value is an order of magnitude higher.

\begin{figure}
\includegraphics[width=0.36\textwidth]{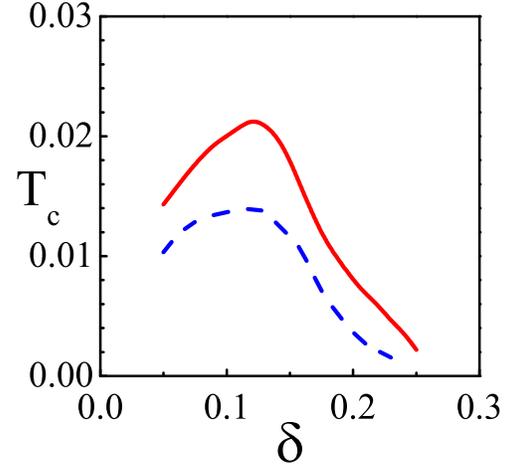}
\caption{(Color online) $T_c(\delta)$  in the SCA  induced by all interactions (red solid
line) and only by the spin-fluctuation contribution
 (blue dashed line).}
 \label{fig23}
\end{figure}

The high values for $T_c$ found in  the WCA are explained by neglecting the reduction of the QP weight caused by the renormalization factor $Z({\bf q},\omega_m)$  in the gap equation (\ref{46}). In the SCA  the gap equation (\ref{46}) is convenient to write in the form:
\begin{eqnarray}
&&\varphi({\bf k},\omega_n) =
  \frac{T_c}{N}\sum_{\bf q} \,  \sum_{m}\,
\big\{ J({\bf k -q}) - V({\bf k -q})
\nonumber \\
& & + V_{ep}({\bf k-q},\omega_{n}- \omega_{m})
 - V_{sf}({\bf q, k-q},\omega_{n}- \omega_{m}) \big\}
\nonumber\\
&&\times  \frac{[1 - b({\bf q})]^2\,
  \varphi({\bf q}, \omega_{m})}{[\omega_m Z({\bf q},\omega_m)]^2
   + [{\varepsilon}_{2}({\bf q})+ X({\bf q},\omega_m)]^2}\, .
 \label{64}
\end{eqnarray}
For  $ V({\bf k -q})$ we take the nearest-neighbor  CI (\ref{50}). Since the
charge  fluctuations $\chi_{cf}({\bf k-q}, \nu_n)$ gives a much weaker contribution than the spin-fluctuations and
electron-phonon interactions (see Table~\ref{Table1}), we neglect them in the interaction function (\ref{47}). Contributions induced by spin-fluctuations and the electron-phonon interaction are described by the functions
\begin{eqnarray}
&& V_{sf}({\bf q, k-q}, \omega_{\nu})=  |t({\bf q})|^{2}\,
 \chi_{sf}({\bf k-q})
  F_{sf}(\omega_{\nu}),\quad
 \label{65a} \\
&& V_{ep}({\bf  k-q}, \omega_{\nu})= g_{ep}\, \frac{\omega_0^2}{\omega^2_{0} +
\omega_\nu^2}\,S({\bf k - q}),
  \label{65b}
\end{eqnarray}
where  the spectral function for spin fluctuations reads:
\begin{equation}
F_{sf}(\omega_\nu)=\frac{1}{\pi} \int_{0}^{\infty}\frac {2x dx}{x^2 +
(\omega_\nu/\omega_s)^2}\frac{\tanh ( x\,\omega_{s}/ 2T)}{1+x^2} \, .
 \label{66}
\end{equation}
To calculate $T_c$ and to find out the energy- and ${\bf k}$-dependence of the gap $\,
\varphi({\bf k},\omega)$, Eq.~(\ref{64}) was solved by a direct diagonalization in $({\bf
k },  \omega_n)$-space. Since the largest contribution in Eq.~(\ref{64}) comes from
energies close to the FS, we have used the renormalization parameters at the Fermi energy
$Z({\bf q})$ (\ref{45ba}) and $X({\bf q}) $  instead of the energy dependent ones. The
results for $T_c(\delta)$ is shown in  Fig.~\ref{fig23}. The highest $T_c \sim 0.021 t
\sim 100$~K is found when all the contributions are taken into account, though pairing
induced only by spin-fluctuations also results in  high $T^{sf}_c \sim 0.014 t \sim
65$~K. The $d$-wave pairing induced only by the electron-phonon interaction is rather
weak and does not displayed in   Fig.~\ref{fig23}. The value of $T_c$ is reduced by an
order of magnitude in comparison with the WCA in  Fig.~\ref{fig22}  due to a suppression
of the QP weight by the factor $[1/Z({\bf q})]^2$. The maximum value of $T_c$ is found at
lower value of doping $\delta_{opt} \approx 0.12$ than in experiments,
$\delta^{exp}_{opt} = 0.16$.
\par
\begin{figure}
\includegraphics[width=0.32\textwidth]{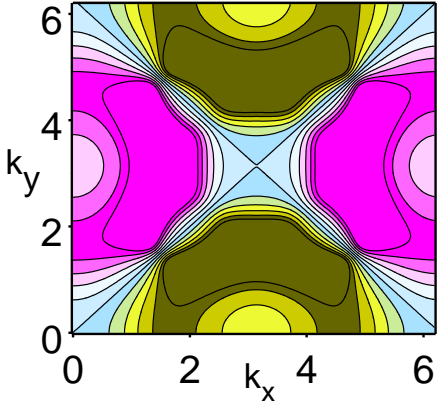}
 \caption{(Color online)  2D plot of the SC gap
 $\varphi({\bf k},\omega \simeq 0)$. }
 \label{fig24}
\end{figure}
\begin{figure}
\includegraphics[width=0.36\textwidth]{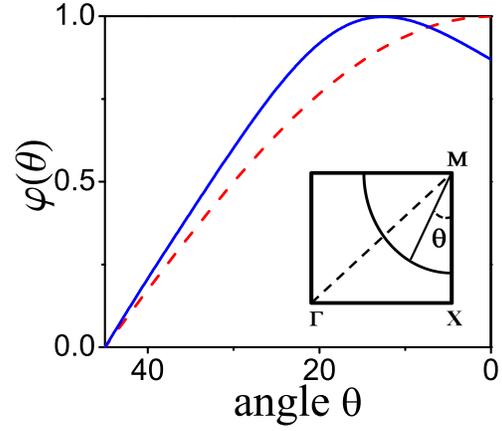}
 \caption{(Color online) Angle dependence of the SC gap $\varphi(\theta) $ on the FS (blue bold line) in
comparison with the model $d$-wave dependence $\varphi_d(\theta) = \cos 2\theta$ (red
dashed lines).}
 \label{fig25}
\end{figure}
\begin{figure}
\includegraphics[width=0.36\textwidth]{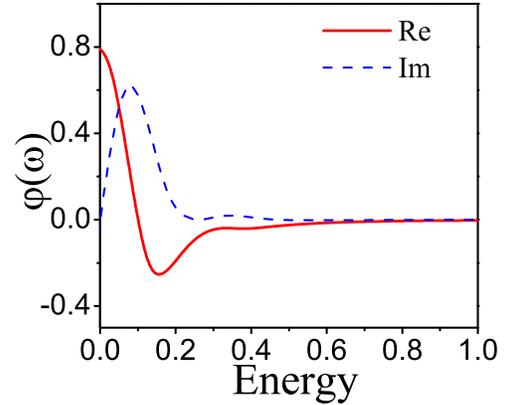}
\caption{(Color online)  Energy  dependence of the real, Re~$\varphi({\bf k},\omega)$, and imaginary,
Im~$\varphi({\bf k},\omega)$, parts of  the SC gap in arbitrary units.}
 \label{fig26}
\end{figure}

The ${\bf k}$-dependence of the gap function  $\varphi({\bf k}, \omega \simeq 0)\,$ at
doping $\delta = 0.13$ for $\, (0 \leq k_x,\, k_y \leq 2\pi)$ is plotted in
Fig.~\ref{fig24}. The gap reveals a distinct $d$-wave symmetry with maximum values in the
vicinity of  the FS.  As shown in  Fig.~\ref{fig25}, its angle dependence on the FS is
close to the model $d$-wave dependence $\varphi_d(\theta) = \cos 2\theta$.
Energy dependence  (in units of $\,t \,$) of  the gap function $\varphi({\bf k},\omega)$,
the real and imaginary parts, is presented in Fig.~\ref{fig26} at ${\bf k}\approx
(0,\pi/2)$ and $\delta = 0.13$. Since the gap function was obtained as a solution of the
linear equation at $T = T_c$ the  value of the gap is given in arbitrary units. The
energy  variation of the gap occurs in the region of  $\omega \lesssim 0.4 t $, of the
order of the characteristic  spin-fluctuation energy $\omega_s = J = 0.4 t$.
\par
\begin{figure}
\includegraphics[width=0.35\textwidth]{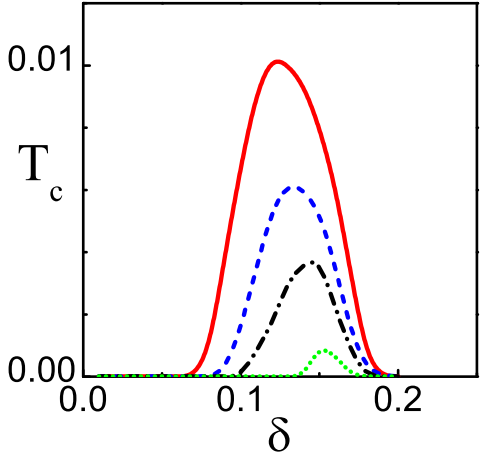}
 \caption{(Color online) $T_c(\delta)$ for $V = 0.0$
(bold red line), $V = 0.5 $ (blue dashed line), $V = 1.0 $ (black dash-dotted line), and
$V = 2.0 $ (green dotted line) for $U = 8 \,$.  }
 \label{fig27}
\end{figure}
\begin{figure}
\includegraphics[width=0.35\textwidth]{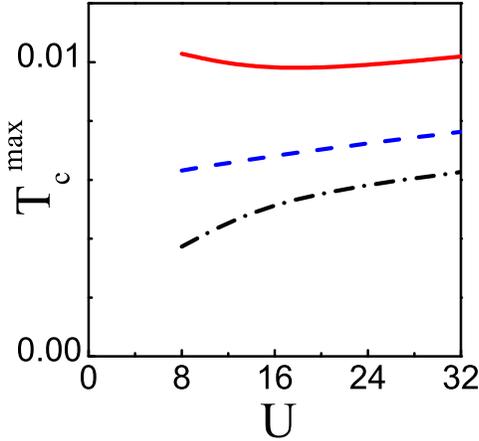}
 \caption{(Color online)  Maximum $T^{\rm max}_c$ at
the optimal doping as a function of $U \,$ for $V = 0.0$  (bold red line), $V = 0.5 $
(blue dashed line), and $V = 1.0 $ (black dash-dotted line).}
 \label{fig28}
\end{figure}

Dependence of the superconducting temperature $T_c$  on the intersite CI  is shown in Fig.~\ref{fig27}
for $\, V = 0.0,\, 0.5 ,\, 1.0 $, and $\, 2 \,$.  Increasing of  $V$ suppresses $T_c$
which becomes small only for high values of $V = 2t - 3t$ comparable with the
spin-fluctuation coupling $\, g_{sf}$ and much larger than the exchange interaction $\, J
= 0.4 t$.  The maximum $T^{\rm max}_c$ at the optimal doping as a function of $U$ and $V$
is shown in  Fig.~\ref{fig28}. It is remarkable, that $T^{\rm max}_c$ only weakly
depends on $\,U \, $ that supports the kinematical mechanism of pairing where the
coupling constant $\,  |t({\bf q})|^2 \,$ does not depend on $U$. A weak increase of
$T^{\rm max}_c$ with $U$ is  explained by narrowing of the electronic band as seen in
Figs.~\ref{fig12},~\ref{fig13} and corresponding increase of the density of state.

\par
In the current approach one can also consider  the $s$-wave pairing. For the extended
$s$-wave gap function, $\varphi_s({\bf k}) = (\Delta/2) \,\eta_s({\bf k})$ where $
\eta_s({\bf k}) = (\cos k_x + \cos k_y)$, a similar to (\ref{55}) equation for $T_c$ can
be derived. Solution of this equation reveals a finite and quite high $T_c$  as in
Refs.~\cite{Zaitsev04, Valkov11} where  superconducting pairing in the Hubbard model was found to be  robust in respect to the intersite Coulomb interaction $V$. However, the $s$-wave pairing  violates  the well known constraint of ``no double occupancy'' in strongly correlated systems. First, it was pointed out in
Refs.~\cite{Plakida89,Yushankhai91} for the $t$-$J$ model and then in
Ref.~\cite{Plakida14} for the Hubbard model. This constraint can be formulated in terms
of a specific  relation for the anomalous (pair) correlation function for the Hubbard
operators. It is easy to verify that the product of two HOs for the singly occupied
subband equals zero: $\, X_i^{0 \sigma }  X_i^{0 \bar{\sigma}} = a_{i\sigma}(1-
N_{i\bar\sigma})\, a_{i\bar\sigma}(1- N_{i\sigma}) = 0$. Therefore,   the corresponding
single-site pair correlation function should vanish:
\begin{equation}
F_{ii,\sigma} =
  \langle  X_i^{0 \sigma }   X_i^{0 \bar{\sigma}} \rangle =
    \frac{1}{N} \sum_{{\bf q}}\, \langle  X_{\bf q}^{0 \sigma }
    X_{-{\bf q}}^{0 \bar{\sigma}} \rangle  \equiv 0.
\label{67}
\end{equation}
The symmetry of the Fourier-component of the  pair correlation function
$\,F_{\sigma}({\bf q}) = \langle  X_{\bf q}^{0 \sigma } X_{-{\bf q}}^{0 \bar{\sigma}}
\rangle  \,$ has the symmetry of the superconducting order parameter, i.e., the gap
function. For instance, in the quasiparticle approximation  we have
 \begin{equation}
 F_{\sigma}({\bf q})  =
 \frac{\varphi(\bf q)}
  {[Z({\bf q})]^2 \; 2\widetilde{\varepsilon}({\bf q})}
    \tanh\frac{\widetilde{\varepsilon}({\bf q})}{2T_c} .
     \label{w2}
\end{equation}
  For the tetragonal lattice for the $d$-wave pairing $ \,F_{\sigma}(q_x, q_y) =
- F_{\sigma}(q_y, q_x)\,$ and the condition (\ref{67}) after integration over $q_x, q_y$
is fulfilled. For the $s$-wave pairing $ \,F_{\sigma}(q_x, q_y) =  F_{\sigma}(q_y,
q_x)\,$ and the condition (\ref{67}) is violated. The same condition holds for the pair
correlation function for the second Hubbard subband, $\langle  X_i^{\sigma 2}
X_i^{\bar{\sigma}2} \rangle = 0$. Therefore, the $s$-wave pairing in the both Hubbard
subbands is prohibited in the limit of strong correlations.

To overcome the restriction (\ref{67}) in Refs.~\cite{Valkov03}  it was proposed  to
consider the modified time-dependent pair correlation function:
\begin{equation}
\tilde{F}_{ii,\sigma} (t) =
  \int_{- \infty}^{+\infty} d \omega \, {\rm e}^{i\omega t}\,
  \tilde{J}_{ii,\sigma}(\omega) ,
\label{68}
\end{equation}
where
\begin{equation}
  \tilde{J}_{ii,\sigma}(\omega) = {J}_{ii,\sigma}(\omega)-
 \delta(\omega) \;\int_{- \infty}^{+\infty} d \omega_1 \,  J_{ii,\sigma}(\omega_1)  .
\label{69}
\end{equation}
The  spectral density $\,{J}_{ii,\sigma}(\omega)\,$ determines the original correlation
function
\begin{equation}
{F}_{ii,\sigma} (t) =
  \langle  {X_{i}^{0\sigma}}(t)  {X_{i}^{0 \bar{\sigma}}} \rangle =
  \int_{- \infty}^{+\infty} d \omega \, {\rm e}^{i\omega t}\,
 {J}_{ii,\sigma}(\omega) .
\label{70}
\end{equation}
For the modified spectral density  (\ref{69}) the condition  (\ref{67}) is trivially
satisfied for any spectral function $\, J_{ii,\sigma}(\omega) \,$ and the restriction on
the $s$-wave pairing seems to be  lifted. However, the spectral density (\ref{69})
results in the nonergodic behavior~\cite{Kubo57}  of the pair correlation function
(\ref{68}):
\begin{eqnarray}
C_{ii,\sigma} =  \lim_{t \to \infty} \tilde{F}_{ii,\sigma} (t)
 = - \frac{1}{N} \sum_{{\bf q}}\, \langle X_{\bf q}^{0 \sigma }
    X_{-{\bf q}}^{0 \bar{\sigma}} \rangle ,
\label{71}
\end{eqnarray}
where the conventional pair correlation function decays in the limit ${t \to \infty} $
due to finite life-time effects
\begin{equation}
 \lim_{t \to \infty} {F}_{ii,\sigma} (t)  =
\int_{- \infty}^{+\infty} d \omega \, {\rm e}^{i\omega t}\, J_{ii,\sigma}(\omega) = 0.
\label{72}
\end{equation}
The nonergodic behavior of the modified pair correlation function (\ref{68})  contradicts to
the basic  properties of physical systems and appears for some pathological models with
local integrals of motion~\cite{Suzuki71,Huber77}. In that case, the nonergodic constant
can be found from $1/\omega$ poles of the anticommutator or causal Green functions, as
described for the Hubbard model for spin or charge excitations in
Refs.~\cite{Mancini03,Avella07} contrary to the arbitrary definition (\ref{69}).
Therefore, the statement given in Refs.~\cite{Valkov03}: ``The inclusion of a singular
contribution to the spectral intensity of the anomalous correlation function regains the
sum rule and remove the unjustified forbidding of the $s$-symmetry order parameter in
superconductors with
strong correlations''  cannot be accepted.\\

\subsection{Isotope effect}
\label{sec:5d}

The  observation  of  the isotope  effect (IE) in conventional superconductors, i.e., the
dependence of $T_{\rm c}$ on the mass $ M $ of the lattice ions, $T_{\rm c} \propto
M^{-\alpha}$, with the  isotope exponent $\alpha =  -d \ln T_{\rm c}/ d \ln M \simeq 0.5
$ was a direct evidence of the electron-phonon pairing mechanism  in these materials. In
cuprate superconductors  a weak isotope shift with $\alpha \leq 0.1 $ was found at
optimal doping. In the underdoped region isotope exponent increases and can be even
larger than in conventional superconductors, $\alpha \sim 1 \, $  (see, e.g., reviews
~\cite{Zhao01,Khasanov04,Muller07}). Small values of the isotope
exponent at optimal doping suggests nonphononic, e.g., magnetic, mechanism of pairing.
However, it is difficult to explain the IE on $T_c$ within the spin-fluctuation pairing
mechanism. By taking into account the electron-phonon interaction within the extended
Hubbard model (\ref{2a}) we can observe the IE in qualitative agreement with experiments.

\begin{figure}
\includegraphics[width=0.35\textwidth]{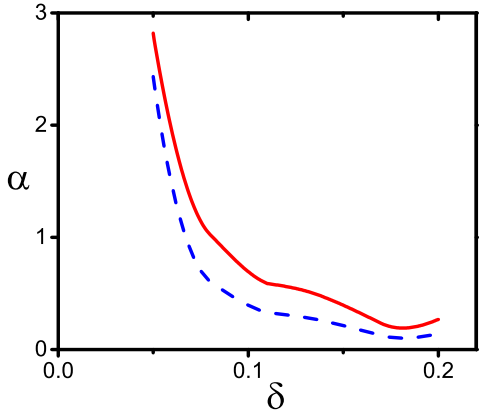}
 \caption{(Color online) Isotope exponent $\alpha(\delta)$
for $g_{ep} = 5\, t$ (red solid line) and  $g_{ep} = 2.5\, t$ (blue dashed  line).}
 \label{fig29}
\end{figure}

To study the isotope effect on $T_c$ using the gap equation (\ref{56}) we consider the
mass-dependent phonon frequency $\omega_{0}$ in the $\theta$-function in (\ref{56}),
$\omega_{0} = \omega_{0}^{(0)}\sqrt{ {M_0}/(M_0 +\Delta M)}= \omega_{0}^{(0)}(1 -
\beta)$. For instance, for the oxygen isotope shift $^{16}$O $\rightarrow ^{18}$O we have
$\,\beta = 0.057$. We neglect the polaronic effect for the $d$-wave electron-phonon
coupling constant $\,g_{ep}({\bf q}) = |g({\bf q})|^2 \, \chi_{ph}({\bf q}) \,$ assuming
it  to be mass-independent (for a discussion see Ref.~\cite{Plakida11a}).  The result for the isotope exponent found by  numerical solution of the gap equation
(\ref{56}) is shown in Fig.~\ref{fig29} for two values of
electron-phonon interaction $g_{ep} = 5\, t$ and $g_{ep} = 2.5\, t\,$ in Eq.~(\ref{57d})
for $Z({\bf q}) = 5$.  The doping dependence of the
exponent agrees with experiments, it is quite small, $\, \alpha=  0.09 - 0.18\,$, at
doping close to the optimal, while drastically increases in the underdoped case, $\,
\alpha = 0.38 - 0.68\,$ at $\delta = 0.1$ for $g_{ep} = 2.5\, t - 5\, t\,$, respectively.
Similar
results were  obtained in Ref.~\cite{Shneyder09} for the $t$-$J$ model with the
electron-phonon interaction.

\subsection{Comparison with previous theoretical studies}
\label{sec:5e}

In studies of superconductivity in the framework of the conventional Hubbard model in the limit of strong correlations the  MFA has been often used (see, e.g.,
Refs.~\cite{Beenen95,Avella97a,Stanescu00,Mancini03,Avella07,Avella07b}). As was shown in Sec~\ref{sec:4b},  superconducting pairing in MFA is induced by the exchange interaction $\, J\sb{ij} = {4 t\sp{2}}/{U}$ (see Eq.~(\ref{21})). The
exchange interaction vanishes in the limit $\, U \to \infty\,$, a feature which explains the disappearance of the pairing at large $U$ observed in
Refs.~\cite{Beenen95,Avella97a,Stanescu00}. To obtain nonzero pairing in the limit $U \gg t $ the self-energy contribution induced by the kinematical interaction should be taken into account which does not depend on $U$ as shown in Fig.~\ref{fig28}.
\par
In the limit of strong correlations various numerical methods were extensively used. Here
we refer to numerical simulations for finite clusters (see
reviews~\cite{Dagotto94,Bulut02,Scalapino07,Senechal12r}), the DMFT (see
reviews~\cite{Georges96,Kotliar06}), the dynamical cluster approximation
(DCA)~\cite{Maier05,Maier06} and the cluster DMFT (see, e.g.,
Refs.~\cite{Haule07,Kancharla08}). More accurate results have been obtained within the
DCA and cluster DMFT methods where short-range AF correlations are partially taken into
account. Extensive numerical studies for finite clusters have revealed a tendency to the
$d$-wave pairing in the Hubbard model, though a delicate balance between
superconductivity and other instabilities (AF, spin-density wave, charge-density wave,
etc.) was found (see, e.g.,
Refs.~\cite{Bulut02,Scalapino07,Maier05,Maier06,Kancharla08}). In
Ref.~\cite{Macridin09}),  using the DCA with the quantum Monte Carlo method, the
superconducting $d$-wave pairing and the isotope effect similar to  observed in cuprates
were found for the Hubbard-Holstein model. However, in several publications an appearance
of the long-range superconducting order has not been confirmed (see, e.g.,
Ref.~\cite{Aimi07}).

As  discussed in Sec.~\ref{sec:5c}, the intersite Coulomb repulsion $V$ is detrimental
for pairing induced by the on-site CI $U$ in the Hubbard model or higher-order
contributions from $V$  in the weak correlation limit. Here we would like to comment on
several studies of this problem  in the strong correlation limit and to compare them with
our analytical results for the $d$-wave pairing. Following the original idea of
Anderson~\cite{Anderson87}, it is commonly believed that the exchange interaction $\,J =
4t^2/U \,$ induced by the interband hopping in the Hubbard model plays the major role in
the $d$-wave superconducting pairing. Since the excitation energy of electrons in the
interband hopping $\, U\, $ is much larger than their intraband kinetic energy $\,  W \,$
the exchange pairing has no retardation effects contrary to the electron-phonon pairing
where large Bogoliubov-Tolmachev logarithm~\cite{Bogoliubov58} diminishes the Coulomb
repulsion $V \rightarrow V/[1 + \rho_c\, \ln(\mu/\omega_{ph})]$ where $\rho_c = N(0)\, V
$ and $\omega_{ph}$ is the phonon energy. Consequently, without the retardation effects
the  Coulomb repulsion $V$ should destroy the exchange  pairing for $\,V > J\,$.
\par
Our  conclusion concerning  importance of the kinematical mechanism of pairing is
supported by the studies in Ref.~\cite{Plekhanov03}. Using the variational Monte Carlo technique the superconducting $d$-wave  gap was observed for the extended Hubbard model with a weak exchange interaction $J = 0.2 \, t$  and a repulsion $V \leq 3\, t $ in a broad range of $\,0 \leq U \leq 32$. It was found that the gap decreases with increasing $V$ at all $U$ and can be suppressed for  $V > J $ for small $\, U $. But for large $ \,U \gtrsim U_c \sim 6\, t\,$ the gap becomes  robust and exists up to  large values of $V \sim 10\, J = 2\, t $ which was explained by effective enhancement of   $J$. At the same time, the gap does not show notable variation with $U$ for large $U = 10 - 30$ though it should depend  on the conventional exchange interaction in the Hubbard model $J = 4t^2/U$ (or $J = 4t^2/(U - V) $). We can suggest another explanation of these results by pointing
out that at large $U \gtrsim U_c \,$ concomitant decrease of the bandwidth (as shown in
Fig. 3~b in Ref.~\cite{Plekhanov03}) results in the splitting of the Hubbard band into
the upper and lower subbands and the emerging kinematical interaction induces the
$d$-wave pairing in one Hubbard subband. In that case the second subband for large $U$
gives a small contribution which results in $U$-independent pairing as shown in
Fig.~\ref{fig28}. It can be suppressed by the repulsion $V$ only larger than the
kinematical interaction, $V \gtrsim 4t$.
\par
In Ref.~\cite{Raghu12a} the extended Hubbard model is considered in the weak or
intermediate correlation limits as in Ref.~\cite{Raghu12} and in the strong correlation
limit within the slave-boson representation in the MFA. In the strong correlation limit a
small value of $V = J$  suppresses the $d$-wave superconducting gap.  The kinetic energy term  described by the projected electron operators, $t\, \hat{c}_{i\sigma }^\dag \hat{c}_{j \sigma } = t\, {c}_{i\sigma }^\dag (1 -  n_{i - \sigma }){c}_{j \sigma }(1 - n_{j - \sigma })\equiv t\, X_{i}^{\sigma 0} X_{j}^{ 0\sigma} $,  was approximated by the conventional fermion (spinon) operators, $t\,\delta \, f_{i\sigma }^\dag f_{j \sigma } $ where the slave-boson contribution in the MFA was described by the hole concentration $\delta$. In this approximation the most important contribution from the kinematical interaction given by $\widehat{\chi}\sb{sf}$ (\ref{57e}) in Eq.~(\ref{56})  was lost in the resulting BCS-type gap equation (13) in Ref.~\cite{Raghu12a}.  Thus, the slave-boson theory
treated in MFA fails to describe superconducting pairing in the limit of strong
correlations.
\par
In Refs.~\cite{Feng03,Feng04,Feng06} the slave-boson representation was considered beyond the MFA within the extended $t$-$J$ model. A kinetic-energy driven mechanism of superconductivity in the effective fermion-spin theory was proposed  where the pairing of fermions is induced  by spin excitations described by  slave bosons. A special procedure of the charge-spin separation and then the  charge-spin recombination   have  been  used in calculation of the electronic GFs.  Similar to our kinematic spin-fluctuation pairing theory, the Eliashberg-type system of equations was obtain where the coupling constant is given by the hopping parameter. However, as in the conventional slave-boson theory the local constraint of no double occupancy cannot be treated rigorously contrary to the HO technique used in our theory. Nevertheless, this approach yields many results such as doping dependence of $T_c$,  the normal state pseudogap, electromagnetic response, charge transport which are in a broad agreement with experiments in cuprates ~\cite{Feng15}.

\section{Conclusion}
\label{sec:6}

In the present review  we present the microscopic theory of spin excitations and
superconductivity for strongly correlated electronic systems as cuprates. Studies of the spin excitations  in the normal state  have shown  a crossover from well-defined spin-wave-like excitations at low doping and temperatures to relaxation-type spin-fluctuation excitations (AF paramagnon) with increasing hole doping.  This results  agree  with inelastic neutron-scattering experiments, RIXS and numerical simulations for finite clusters.
We propose a new explanation for the magnetic RM observed in superconducting state.  As was shown in Sec.~\ref{sec:3d}, a weak damping of spin excitations close to    the AF wave vector ${\bf Q}$  is the reason for appearance  of the RM. The weak damping of the RM  is explained by a contribution from  spin excitations to the decay process besides a particle-hole pair usually considered in the spin-1 exciton scenario. In this case the temperature independent RM at $ E_{\rm r}$ appears since
 the damping essentially depends on the gap $\widetilde{\omega}_{\bf Q} \simeq E_{\rm r}$ in the spin-excitation spectrum  and  opening of the superconducting gap $2\Delta(T)$ below $T_{\rm c}$ is less important.
\par
The theory of superconducting pairing was developed within the extended  Hubbard model (\ref{2a}) in the limit of strong electron correlations. Using the Mori-type projection technique we  obtained a self-consistent system of equations for the normal and anomalous (pair) GFs and for the self-energy calculated in the SCBA.  From our study we can draw the following  conclusion concerning the mechanism of pairing in the extended Hubbard model.
Solution of the gap equation  shows that for the $d$-wave pairing  relevant contributions come from the $l=2$ angular momentum component of interactions. This results in a considerable reduction of the intersite Coulomb repulsion  and electron-phonon interaction. In particular, a momentum independent electron-phonon interaction  give no contribution to the $d$-wave pairing. At the same time, a strong local electron-phonon interaction may induce a noticeable polaronic effect observed in the magnetic penetration depth~\cite{Khasanov04}. Thus we conclude that the electron-phonon interaction plays a secondary role in achieving high-$T_c$ though it should be taken into account to explain  the weak isotope effect on $T_c$. The largest contribution to the $d$-wave pairing comes from the electron coupling to spin fluctuations induced by the strong kinematical interaction $ |t({\bf q})|$ which brings about the superconductivity with high $T_c$ observed in cuprates.
\par
It is important to point out that the kinematical interaction induced by the kinetic
energy  of electrons  moving in the Hubbard subband is responsible both for  the damping of spin excitations at finite doping and the spin-fluctuation superconducting pairing. The kinematical interaction is characteristic for systems with strong electron correlations   which is absent in the fermionic models. Therefore, we believe that the spin-fluctuation magnetic mechanism of superconducting pairing in the Hubbard model in the limit of strong correlations is a relevant mechanism of high-temperature
superconductivity in the copper-oxide materials.\\

{\bf Acknowledgments}\\

The author thanks S. Adam, G. Adam, D. Ihle, V. Oudovenko, and A. Vladimirov  for the fruitful collaboration.   The author is grateful to the
MPIPKS, Dresden, for the hospitality  during his stay at the Institute, where a part of these investigations  has been done. The work was  supported   by the Heisenberg--Landau Program of JINR.\\

\end{document}